\documentclass[prx,aps,superscriptaddress,floatfix,tightenlines,showpacs,twocolumn,longbibliography]{revtex4-1}
\usepackage{graphicx,amsfonts,amssymb,amsmath,hyperref,hypcap,enumerate}

\usepackage{amsthm}
\usepackage{multirow}
\usepackage{graphicx}
\usepackage{dcolumn}   
\usepackage{bm}        
\usepackage{amssymb}   
\usepackage{amsmath}
\usepackage{url}
\usepackage{layout}
\usepackage{epsfig}
\usepackage{graphicx}
\usepackage{epstopdf}
\usepackage{booktabs}
\usepackage{float}
\usepackage{color}

\DeclareMathAlphabet{\mathitb}{OT1}{cmr}{bx}{sl}

\begin{document}
\title{Quantum Multicriticality in Disordered Weyl Semimetal}

\author{Xunlong Luo}
\affiliation{International Center for Quantum Materials, Peking University, Beijing 100871, China}
\affiliation{Collaborative Innovation Center of Quantum Matter, Beijing 100871, China}

\author{Baolong Xu}
\affiliation{International Center for Quantum Materials, Peking University, Beijing 100871, China}
\affiliation{Collaborative Innovation Center of Quantum Matter, Beijing 100871, China}

\author{Tomi Ohtsuki}
\affiliation{Department of Physics, Sophia University, Chiyoda-ku, Tokyo 102-8554, Japan}

\author{Ryuichi Shindou}
\email{rshindou@pku.edu.cn}
\affiliation{International Center for Quantum Materials, Peking University, Beijing 100871, China}
\affiliation{Collaborative Innovation Center of Quantum Matter, Beijing 100871, China}

\date{\today}

\begin{abstract}
In electronic band structure of solid state material, two band touching points with linear dispersion appear in pair in the momentum space. When they annihilate with each other, the system 
undergoes a quantum phase transition from three-dimensional 
Weyl semimetal (WSM) phase to a band insulator phase such as 
Chern band insulator (CI) phase. The phase transition is described by a new critical theory with 
a `magnetic dipole' like object in the momentum space. In this paper, we reveal that the critical theory 
hosts a novel disorder-driven quantum multicritical point, which is encompassed by three quantum phases, 
renormalized WSM phase, CI phase, and diffusive metal (DM) phase. 
Based on the renormalization group argument, we first clarify scaling properties around 
the band touching points at the quantum multicritical point  
as well as all phase boundaries among these three phases. Based on numerical calculations of 
localization length, density of states and critical conductance distribution, we next prove  
that a localization-delocalization transition between the CI phase with a finite zero-energy 
density of states (zDOS) and DM phase belongs to an ordinary 3D unitary class. Meanwhile, 
a localization-delocalization transition between the Chern insulator phase with zero 
zDOS and a renormalized Weyl semimetal (WSM) phase turns out to be a direct phase 
transition whose critical exponent $\nu=0.80\pm 0.01$. We interpret these numerical 
results by a renormalization group analysis on the critical theory. 
\end{abstract}


\maketitle

\section{Introduction}
Novel quantum matters called Weyl semimetal~\cite{1,2} have been recently discovered 
in a number of experimental materials~\cite{3,4,5,6}, which show non-trivial magnetotransport 
properties~\cite{7,8,9} as a consequence of the chiral anomaly in 3+1 dimension. Besides its 
magnetotransport property, a Weyl semimetal with a linearly dispersive band-touching point 
(`Weyl' node) exhibits an intriguing quantum phase transition in the 
presence of moderate disorder strength; a quantum phase 
transition between renormalized Weyl semimetal (WSM) and diffusive metallic (DM) phases~\cite{10,11,12,13,14,15,16,17,18,19,20,21,22,23,24,25,26,27,28,29,30,31,32}.
To clarify a critical nature of this quantum phase transition, 
scaling properties of  density of states (DOS) around the Weyl 
node~\cite{10,12,15,20,21,22,25,27,31} and zero-energy 
conductivity~\cite{15,21,25} as well as their interplay 
with rare-event states~\cite{17,30} have been extensively discussed. 

The celebrated Nielsen-Ninomiya theorem dictates that single Weyl node cannot exist by itself 
in the momentum space of solid state material. Instead, two Weyl nodes with opposite 
magnetic charges always appear in pair in the first Brillouin zone. When a pair of the two 
nodes annihilate with each other in the momentum space, 
the system undergoes a quantum phase transition from WSM phase 
to a three-dimensional Chern band insulator (CI) phase. The quantum 
phase transition between these two phases is described 
by a new critical theory, which we call in this paper as a ``magnetic dipole" model, 
\begin{eqnarray}
{\cal H}_{\rm eff} &= \int d^2{\bm x}_{\perp} dx_3 \!\ \psi^{\dagger}({\bm x}) 
\Big\{ -iv \big(\partial_{1} {\bm \sigma}_1 
+ \partial_2 {\bm \sigma}_2 \big) \nonumber \\
&\ \ \ + \big((-i)^2 b_2 \partial^2_3 - m \big){\bm \sigma}_3  
\Big\} \psi({\bm x}), \label{eff0}  
\end{eqnarray} 
with ${\bm x}_{\perp} \equiv (x_1,x_2)$. ${\bm \sigma}_{\mu}$ are 
2 by 2 Pauli matrices ($\mu=1,2,3$) and $\psi({\bm x})$ denotes 
a two-component slowly-varying 
fermion field. A mass term ($m$) 
separates the WSM phase ($m>0$) from the CI phase ($m<0$). 
For $m<0$, the effective Hamiltonian has a finite energy gap (CI phase). 
For $m>0$, the Hamiltonian has a pair of the Weyl points in the momentum 
space, $(p_1,p_2,p_3)=(0,0,\pm \sqrt{m/b_2})$. At the massless point ($m=0$), 
the model supports a linearly dispersive energy band within a plane that  
is subtended by coordinates $x_1$ and $x_2$, and a quadratic 
dispersion along a dipole direction (along the coordinate 
$x_3$). A tree-level renormalization group argument dictates that the 
critical phase at the massless point is robust against an infinitesimally small 
randomness (disorder), and persists up to a certain critical disorder strength. 
This suggests the presence of a new type of disorder-driven 
quantum phase transition between the critical phase near the clean limit 
and a DM phase above the critical disorder strength.

\begin{figure}[b]
	\centering
	\includegraphics[width=1\linewidth]{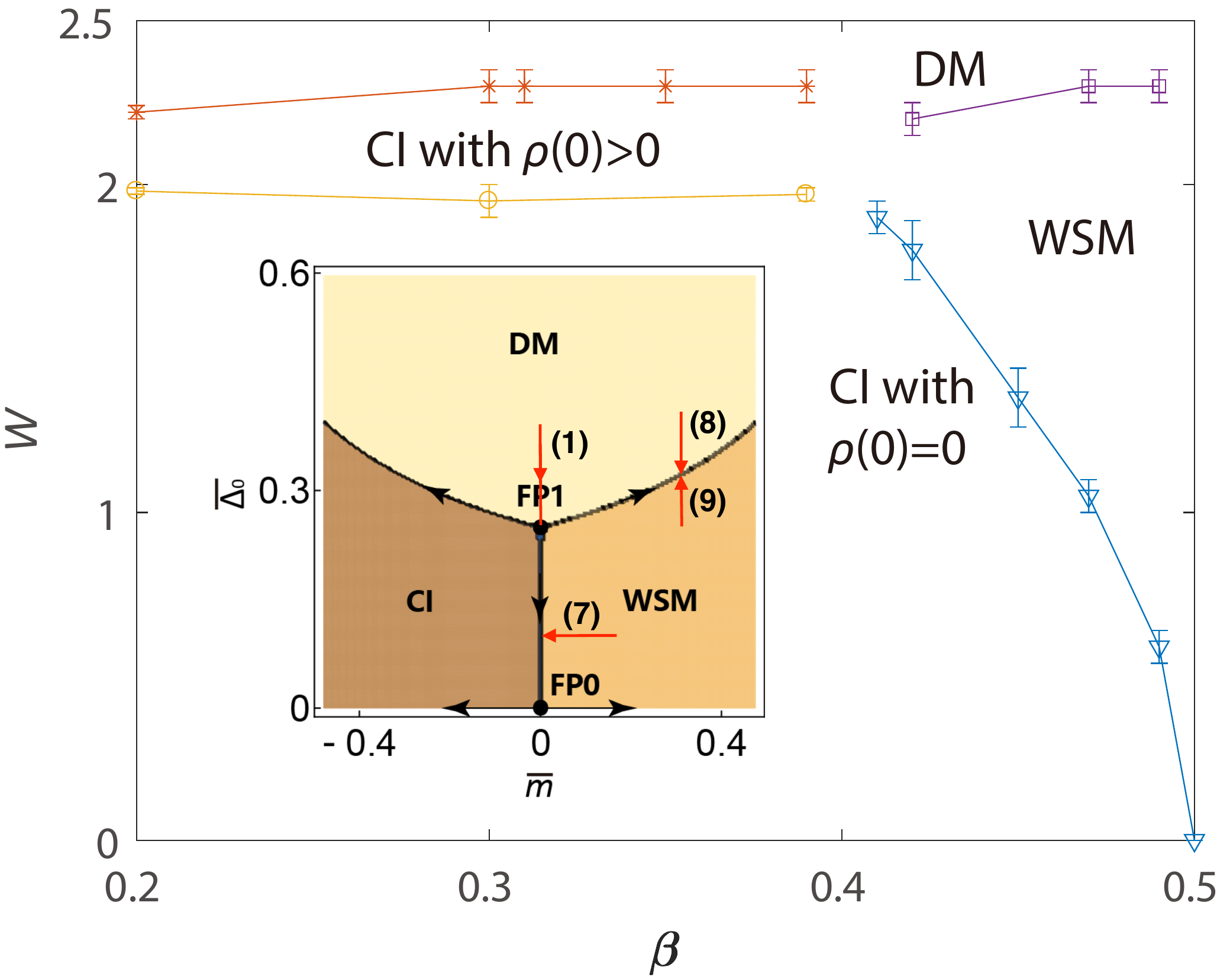}
	\caption{(color online) Phase diagram for disordered Weyl semimetal. 
The horizontal axis indicates the coupling between the layers, while the vertical 
axis indicates the strength of disorder. Red, purple and blue 
lines are boundaries between CI and DM, between WSM 
and DM and between CI and WSM, respectively. These boundaries are determined 
by the localization length calculation. Yellow line is a 
boundary between CI phase with zero zero-energy density of states (zDOS) and CI phase 
with finite zDOS, which is determined by the density of states calculation. 
Inset: Phase diagram determined by perturbative renormalization group analyses (Ref.~\cite{31}. See also Appendix \ref{sec:B} of this paper). Red lines with arrow and number ((1), (7), (8), (9)) describe how the system approaches the critical lines or critical point in those cases that are considered in Table~\ref{table:scaling0}.}
	\label{fig:1}
\end{figure}

In this paper, we reveal that critical natures of this new type of the quantum phase 
transition is controlled by a novel quantum multicritical point (QMCP), from which three 
quantum phase boundaries are branching out; WSM-CI, WSM-DM, and CI-DM phase boundaries. 
Using a renormalization-group (RG) scaling argument, we first clarify unconventional 
DOS and conductivity scalings at Weyl nodes around QMCP as well as their crossover 
behavior among different critical regimes. We especially show that, around a quantum critical 
line between WSM and CI phases as well as QMCP, the conductivity and diffusion 
constant along the in-plane direction and those along the dipole direction 
follows different universal scaling functions with different exponents! For the phase boundary 
between CI and DM phases, we numerically prove that a mobility edge and a band edge 
(red and yellow curves in Fig.~\ref{fig:1}, respectively) are distinct in the phase diagram, 
dictating the presence of a CI phase with finite zero-energy density of states (zDOS).  
A localization-delocalization transition between the CI with finite zDOS 
and DM phase belongs to the ordinary 3D unitary class~\cite{33,34,35,36,37,38}.
Around a transition point between CI with zero zDOS and CI with finite zDOS, 
DOS for different disorder strengths and single-particle energies are well fitted into a single universal scaling function. Both critical 
exponent and dynamical exponent associated with this universal function 
are evaluated to be around one, respectively. For the phase boundary between CI and 
WSM phases, we numerically confirm that the transition between these two 
is direct. The associated critical exponent $\nu$ is estimated as 
0.80 [0.79,0.81] by the localization length calculation. 
We interpret these results from a viewpoint of renormalization group 
analyses for a disordered magnetic-dipole model.  

\begin{table*}[ht]
  \centering
    \begin{tabular}{p{6mm} |c|c|c}
\hline 
& $\rho(0)$ or $\rho({\cal E})$ & $\sigma_3(0)$ or $\sigma_3({\cal E})$ & $\sigma_{\perp}(0)$ 
or $\sigma_{\perp}({\cal E})$   \\ \hline \hline 
(1) & $\delta \overline{\Delta}_0^{\frac{2d-1-2z}{2y_{\Delta}}}$ 
& $\delta \overline{\Delta}_0^{\frac{2d-3}{2y_{\Delta}}}$ & $\delta \overline{\Delta}_0^{\frac{2d-5}{2y_{\Delta}}}$ \\
(2) & $|{\cal E}|^{d-\frac{3}{2}}$ & $|{\cal E}|^{d-\frac{3}{2}}$  & $|{\cal E}|^{d-\frac{5}{2}}$ \\ 
(3) & $|\delta \overline{\Delta}_0|^{\frac{2d-1}{2} \frac{1-z}{y_{\Delta}}} |{\cal E}|^{d-\frac{3}{2}}$ 
& $|\delta \overline{\Delta}_0|^{\frac{2d-3}{2} \frac{1-z}{y_{\Delta}}}|{\cal E}|^{d-\frac{3}{2}}$ 
& $|\delta \overline{\Delta}_0|^{\frac{2d-5}{2} \frac{1-z}{y_{\Delta}}}|{\cal E}|^{d-\frac{5}{2}}$ \\  
(4) & $|{\cal E}|^{\frac{d-z'}{z'}}$ & {$|{\cal E}|^{\frac{d-2}{z'}}$} &same as $\sigma_3$ \\ 
(5) & $m^{\frac{2d(z'-z)-z'}{2z'y_m}} |{\cal E}|^{\frac{d-z'}{z'}}$ 
& $m^{\frac{2d(z'-z)+4z-3z'}{2z'y_m}} |{\cal E}|^{\frac{d-2}{z'}}$ 
& $m^{\frac{2d(z'-z)+4z-5z'}{2z'y_m}} |{\cal E}|^{\frac{d-2}{z'}}$  \\ 
(6) & $|{\cal E}|^{\frac{2d-1-2z}{2z}}$ & $|{\cal E}|^{\frac{2d-3}{2z}}$ & $|{\cal E}|^{\frac{2d-5}{2z}}$ \\
(7) & $m^{-\frac{1}{2}} |{\cal E}|^{d-1}$ & $m^{d-\frac{3}{2}}$ & $m^{d-\frac{5}{2}}$  \\
(8) & $\delta \overline{\Delta}_0^{\frac{d-z'}{y^{\prime}_{\Delta}}}$ 
&{$\delta \overline{\Delta}_0^{\frac{d-2}{y^{\prime}_{\Delta}}}$} &same as $\sigma_3$\\
(9) &$|\delta \overline{\Delta}_0|^{-\frac{dz'-d}{y^{\prime}_{\Delta}}} |{\cal E}|^{d-1}$
&{$|\delta \overline{\Delta}_0|^{\frac{d-2}{y^{\prime}_{\Delta}}}$}& same as $\sigma_3$\\  \hline 
 \end{tabular} 
 \caption{Scalings of DOS and conductivities near Weyl nodes. ${\cal E}$ denotes a single-particle energy. 
$\rho(0)$ and $\rho({\cal E})$ are DOS for the zero-energy states and for finite (but small) energy states, 
respectively. $\sigma_{\mu}(0)$ and $\sigma_{\mu}({\cal E})$ are the conductivities 
along the $\mu$ direction ($\mu=3, \perp$) for 
the zero-energy and the finite energy states, respectively. 
The number such as (1), (7), (8), and (9)  in the first column specifies  
a route along which the system approaches the quantum multicritical point ((1)) or quantum critical lines 
((7), (8), (9)). The routes with the same number are depicted as a red line with arrow 
in Fig.~\ref{fig:1}. In (2), (4) and (6), the system is on a quantum critical line between CI and WSM 
phases, on that between DM and WSM phases, and on the quantum multicritical point, respectively. 
In (3) and (5), the system approaches QMCP along 
(2) and (4). $\delta\overline{\Delta}_0$ denotes the disorder strength measured 
from a critical disorder strength of respective critical point (see also the text and Appendix). 
$m$ denotes the mass term. $y_{\Delta}$, $y_m$ and $z$ are scaling dimensions of the disorder 
strength $\delta \overline{\Delta}_0$, the mass term $m$, and the dynamical exponent 
at the quantum multicritical point (QMCP), respectively. $y^{\prime}_{\Delta}$ and $z^{\prime}$ are 
scaling dimension of the disorder strength and dynamical exponent around a fixed point with finite 
disorder in disordered single Weyl node model 
respectively. See also Table~\ref{table:scaling} in Appendix \ref{sec:C}.}\label{table:scaling0}
\end{table*}

\section{Lattice Model, Magnetic Dipole Model and Multicritical Quantum Phase Transition} 

Let us first introduce a lattice model which supports the magnetic dipole model as its low-energy 
effective electronic Hamiltonian. In this paper, we numerically study a two-orbital tight-binding 
model on a cubic lattice~\cite{24,25,26,27}. The model is a layered Chern insulator (LCI), which 
consists of two-dimensional Chern insulator in the $x_1$--$x_2$ plane with 
interlayer couplings along the $x_3$-direction (see Appendix \ref{sec: A}). In the two-orbital model, 
we set a half electron filling (one band occupied, the other empty in the clean limit), and  
change an interlayer coupling strength (which we call $\beta$ henceforth, see the Appendix \ref{sec: A}).
When the coupling strength exceeds a critical value $\beta_c$, the two energy 
bands form pairs of linearly dispersive band-touching points 
(`Weyl' nodes) at the Fermi level. An electronic phase 
for $\beta>\beta_c$ is referred to as Weyl semimetal  (WSM) phase, 
and a  phase with an energy gap for $\beta<\beta_c$ as 
Chern band insulator (CI) phase.  
At the critical point, the low-energy effective electronic Hamiltonian takes 
the form of Eq.~(\ref{eff0}), where $\psi^{\dagger}({\bm x})$ and $\psi({\bm x})$ 
therein denote a two-component slowly-varying fermion field obtained from 
the ${\bm k}\cdot {\bm p}$ expansion around degenerate Weyl points at $\beta=\beta_c$. 
$m$ corresponds to the interlayer coupling strength, $m=\beta-\beta_c$.

Effects of various types of disorders on the critical theory at the massless point of Eq.~(\ref{eff0}) can 
be captured by 
renormalization group (RG) analyses (Ref.~\cite{31} and Appendix \ref{sec:B}). A tree-level scaling argument 
dictates that a scaling dimension of the most relevant disorder is $-(d-5/2)$, where $d$ is 
the spatial dimension, $d=3$. This suggests the presence of 
a non-trivial fixed point with a finite critical disorder strength, below which the 
disorder is irrelevant, while above which the disorder blows up to a larger value. A 
perturbative RG equation takes a form of coupled equations of the mass term $m$ and 
properly normalized (chemical-potential type) 
disorder strength $\overline{\Delta}_0$;
\begin{align}
\frac{dm}{dl} &= m(1-2\overline{\Delta}_0), \label{m-rg1} \\ 
\frac{d\overline{\Delta}_0}{dl} &= -\frac{1}{2} \overline{\Delta}_0 + 2\overline{\Delta}^2_0,  \label{m-rg2} 
\end{align}    
with $dl$ the scale change. The equation has a trivial fixed point in the clean limit $(\overline{\Delta}_0,m)=(0,0)$ (FP0) and 
the nontrivial fixed point at finite critical disorder strength $(\overline{\Delta}_0,m)=(\Delta_c,0)$ 
with $\Delta_{c}=1/4$ (FP1). The critical phase described by Eq.~(\ref{eff0}) is 
stable up to the critical disorder strength ($\overline{\Delta}_{0}<\Delta_c$), above which a 
system enters a diffusive metallic (DM) phase ($\overline{\Delta}_0>\Delta_c$). The 
critical phase in $0<\overline{\Delta}_0<\Delta_c$ and $m=0$ intervenes WSM and CI phases. The 
fixed point at $(\overline{\Delta}_0,m)=(\Delta_c,0)$ plays role of a 
quantum multicritical point among WSM, CI and DM phases (inset of Fig.~\ref{fig:1}). 
In the phase diagram, there are three phase boundaries that branch out
from the multicritical point. They are a boundary between CI and WSM phases, 
that between WSM and DM phases, and that between CI and DM phases. 
In the following section, we first employ the RG argument to clarify a rich variety of  
scaling properties of DOS, diffusion constant and conductivity around the 
quantum critical line between CI and WSM phases including the QMCP.   

\section{DOS and Conductivity Scalings around QMCP} 

According to the RG equation, the critical properties of the quantum phase boundary between CI and WSM phases are controlled by the critical theory at the massless point of the 
magnetic dipole model (FP0). The theory has a quadratic dispersion along the dipole direction, 
and a linear dispersion along the in-plane direction. This naturally leads to a spatially anisotropic 
scaling around FP0;
\begin{align}
x^{\prime}_3 &= b^{\frac{1}{2}} x_3, \label{scaling-3} \\ 
{\bm x}^{\prime}_{\perp} &= b {\bm x}_{\perp}, \label{scaling-12}
\end{align} 
where $b \equiv e^{-dl}<1$ counts how many times we carry out the renormalization; 
$-\ln b \equiv dl$ is same as $dl$ in the left hand side of Eqs.~(\ref{m-rg1}) and (\ref{m-rg2}).  
Here and henceforth, quantities with prime and quantities without prime 
denote those after the renormalization and before the renormalization, respectively. Under the anisotropic 
scaling, a volume is scaled by $b^{d-\frac{1}{2}}$. Thus, a total number of single-particle states below 
a given energy ${\cal E}$ per unit volume $N({\cal E},\overline{\Delta}_0,m)$ is scaled by 
$b^{-(d-\frac{1}{2})}$ under the renormalization;
\begin{align}
N^{\prime}({\cal E}^{\prime},\overline{\Delta}^{\prime}_0,m^{\prime}) = b^{-(d-\frac{1}{2})} N({\cal E},\overline{\Delta}_0,m),  
\label{N-scaling}
\end{align}
with ${\cal E}^{\prime}=b^{-\overline{z}}{\cal E}$, $\overline{\Delta}^{\prime}_0=b^{-\overline{y}_{\Delta}}\overline{\Delta}_0$, 
and $m^{\prime}=b^{-\overline{y}_m}m$. $\overline{z}$, $\overline{y}_{\Delta}$ and $\overline{y}_m$ are dynamical 
exponent, scaling dimensions of the disorder strength $\overline{\Delta}_0$, and mass term $m$ around 
the fixed point in the clean limit (FP0). In terms of the dynamical exponent, the density of states is scaled by 
$b^{-(d-\frac{1}{2}-\overline{z})}$ under the renormalization, 
\begin{eqnarray}
\rho^{\prime}({\cal E}^{\prime},\overline{\Delta}^{\prime}_0,m^{\prime}) = b^{-(d-\frac{1}{2}-\overline{z})} \rho({\cal E},\overline{\Delta}_0,m) 
\label{rho-scaling}
\end{eqnarray}
with $\rho({\cal E}) \equiv dN({\cal E})/d{\cal E}$. The critical theory in the clean limit has its dynamical exponent 
to be one; $\overline{z}=1$. Around FP0, the disorder is an irrelevant scaling variable, 
$\overline{y}_{\Delta}=-(d-\frac{5}{2})$, while the mass term is a relevant scaling variable, $\overline{y}_m=1$. 
To see how DOS near Weyl nodes is scaled by the single-particle energy near the quantum 
critical line between CI and WSM phases, we take $m$ to be tiny and $\overline{\Delta}_0<\Delta_c$. Let 
us renormalize the system many times, such that $m^{\prime}=b^{-\overline{y}_m}m = 1$. 
Solving $b$ in favor for small $m$, we reach,
\begin{eqnarray}
\rho({\cal E},\overline{\Delta}_0,m) = m^{d-\frac{3}{2}} \rho^{\prime}(m^{-1}{\cal E},m^{-\overline{y}_{\Delta}}\overline{\Delta}_0,1). \label{rho-scaling-2}
\end{eqnarray}
When $m$ is tiny, the second argument in the right hand side 
can be replaced by zero, because $\overline{y}_{\Delta}<0$. 
This leads to the following universal DOS scaling around 
the CI-WSM branch:
\begin{eqnarray}
\rho({\cal E},\overline{\Delta}_0,m) = m^{d-\frac{3}{2}} \Phi(m^{-1}{\cal E}), \label{rho-scaling-a}
\end{eqnarray} 
for small $m$ and arbitrary $\overline{\Delta}_0<\Delta_c$. 

The anisotropic scaling in the magnetic dipole model also 
leads to unconventional scalings of the diffusion constant near Weyl nodes. 
To see this, we consider a mean square displacement of single-particle states 
of energy ${\cal E}$ at a time $s$ as a function of the two scaling variables:
\begin{align}
g_{3}({\cal E},s,\overline{\Delta}_0,m) &\equiv \langle x^2_3 \rangle ({\cal E},s,\overline{\Delta}_0,m), \label{g3} \\
g_{\perp}({\cal E},s,\overline{\Delta}_0,m) &\equiv \langle {\bm x}^2_{\perp} \rangle ({\cal E},s,\overline{\Delta}_0,m). \label{g12} 
\end{align} 
Under the scaling, the displacement along the dipole direction ($g_3$) and that along the in-plane direction 
($g_{\perp}$) are scaled with different exponents;
\begin{align}
b^{-1} g^{\prime}_3({\cal E}',s',\overline{\Delta}^{\prime}_0,m') &= g_3({\cal E},s,\overline{\Delta}_0,m), \label{g3-s} \\
b^{-2} g^{\prime}_{\perp}({\cal E}',s',\overline{\Delta}^{\prime}_0,m') &= g_{\perp}({\cal E},s,\overline{\Delta}_0,m), \label{g12-s} 
\end{align}
and $s' = b^{z} s$ ($z=1$). This leads to the following universal scaling forms of the mean square 
displacements,
\begin{align}
g_{3}({\cal E},s,\overline{\Delta}_0,m) &= m^{-1} \Psi_3 (m^{-1}{\cal E},m s), \label{g3-ss} \\
g_{\perp}({\cal E},s,\overline{\Delta}_0,m) &= m^{-2} \Psi_{\perp} (m^{-1}{\cal E},m s), \label{g12-ss} 
\end{align}
for small $m$ and arbitrary $\overline{\Delta}_0 < \Delta_c$. For small $m$, we may further expand 
the right hand side with respect to small $ms$ for an arbitrary time $s$, to obtain the following 
universal scaling forms of the diffusion constants at Weyl nodes,
\begin{align}
D_{3}({\cal E},\overline{\Delta}_0,m) &= f_3(m^{-1}{\cal E}), \label{d3-s} \\
D_{\perp}({\cal E},\overline{\Delta}_0,m) & = m^{-1}f_{\perp}(m^{-1}{\cal E}). \label{d12-s}
\end{align}
Here $D_{3}$ and $D_{\perp}$ denote a diffusion constant along the dipole direction ($x_3$) and that 
within the in-plane direction (${\bm x}_{\perp}$), respectively, which are linear coefficients of 
the mean square displacements in time $s$; 
$g_{\mu}(\cdots,s,\cdots) = D_{\mu}(\cdots) s + {\cal O}(s^2)$ ($\mu=3,\perp$).   

For finite positive $m$, the low-energy electronic band structure comprises of the 
pair of the two Weyl nodes. Thereby, we regard that, in the WSM phase, 
($m>0$ and $\overline{\Delta}_{0} <\Delta_c$), $\rho({\cal E}) \propto {\cal E}^{d-1}$ and 
$D_{\mu}({\cal E}) \propto {\cal E}^{-(d-1)}$ for ${\cal E}<m$. 
When combined with Eqs.~(\ref{rho-scaling-a}), (\ref{d3-s}), and (\ref{d12-s}), 
this requires 
\begin{eqnarray}
\left\{\begin{array}{c} 
\rho({\cal E},\overline{\Delta}_0,m) \propto m^{-\frac{1}{2}} {\cal E}^{d-1}, \\ 
D_3({\cal E},\overline{\Delta}_0,m) \propto m^{d-1} {\cal E}^{-(d-1)}, \\ 
D_{\perp}({\cal E},\overline{\Delta}_0,m) \propto m^{d-2} {\cal E}^{-(d-1)}, \\ 
\sigma_3({\cal E},\overline{\Delta}_0,m) \propto m^{d-\frac{3}{2}}, \\ 
\sigma_{\perp}({\cal E},\overline{\Delta}_0,m) \propto m^{d-\frac{5}{2}},  \\ 
\end{array}\right. \label{scaling-imp1}
\end{eqnarray}
for small $m$ and arbitrary $\overline{\Delta}_0 <\Delta_c$. Here 
$\sigma_3$ and $\sigma_{\perp}$ are conductivity along the dipole direction ($x_3$) and 
that within the in-plane direction (${\bm x}_{\perp}$), that are related with diffusion constants  
and the density of states via the Einstein relation, 
$\sigma_{\mu}({\cal E}) \equiv e^2 \rho({\cal E})D_{\mu}({\cal E})$ ($\mu=3,\perp$). 
At non-zero single-particle energy, both DOS and diffusion constants are finite 
even at the critical line ($m=0$). When combined with Eqs.~(\ref{rho-scaling-a}), (\ref{d3-s}), and (\ref{d12-s}), 
this tells how DOS and diffusion constants as well as conductivities are scaled by the single-particle 
energy ${\cal E}$ at $m=0$, 
\begin{eqnarray}
\left\{\begin{array}{c} 
\rho({\cal E},\overline{\Delta}_0,m=0) \propto {\cal E}^{d-\frac{3}{2}}, \\ 
D_3({\cal E},\overline{\Delta}_0,m=0) \propto {\cal E}^{0}, \\ 
D_{\perp}({\cal E},\overline{\Delta}_0,m=0) \propto {\cal E}^{-1}, \\ 
\sigma_3({\cal E},\overline{\Delta}_0,m=0) \propto {\cal E}^{d-\frac{3}{2}}, \\ 
\sigma_{\perp}({\cal E},\overline{\Delta}_0,m=0) \propto {\cal E}^{d-\frac{5}{2}},  \\ 
\end{array}\right. \label{scaling-imp2}
\end{eqnarray}
for arbitrary $\overline{\Delta}_0 \!\ <\Delta_c$. 

According to the RG equations given by Eqs.~(\ref{m-rg1}) and (\ref{m-rg2}), 
the scaling around QMCP (FP1) has two relevant scaling variables, 
$\delta\overline{\Delta}_0 \equiv \overline{\Delta}_0-\Delta_c$ and 
the mass term $m$. Under the renormalization, DOS and mean square displacements 
are scaled as follows, 
\begin{align}
b^{d-\frac{1}{2}-z} \rho^{\prime}(b^{-z}{\cal E},b^{-y_{\Delta}}\delta\overline{\Delta}_0,b^{-y_m}m) 
&= \rho({\cal E},\delta\overline{\Delta}_0,m), \nonumber \\
b^{-1} g^{\prime}_3(b^{-z}{\cal E},b^{z} s,b^{-y_{\Delta}}\delta\overline{\Delta}_0,b^{-y_m}m) 
&= g_3({\cal E},s,\delta\overline{\Delta}_0,m), \nonumber \\
b^{-2} g^{\prime}_{\perp}(b^{-z}{\cal E},b^z s,b^{-y_{\Delta}}\delta \overline{\Delta}_0,b^{-y_m}m) 
&= g_{\perp}({\cal E},s,\delta\overline{\Delta}_0,m), \nonumber 
\end{align}
where $z$, $y_{\Delta}$ and $y_m$ are dynamical exponent, scaling 
dimensions of $\delta \overline{\Delta}_0$ and $m$ around FP1. Within large-$n$ RG analysis 
(see Ref.~\cite{31} and Appendix \ref{sec:B}), they are evaluated as 
$z=1+\frac{1}{n}+{\cal O}(n^{-2})$, $y_m=1-\frac{1}{n}+{\cal O}(n^{-2})$, and $y_{\Delta}=\frac{1}{n}+{\cal O}(n^{-2})$ 
with $n=2$. 

When the system approaches QMCP along $m=0$, the DOS and diffusion constants near Weyl nodes 
respect the following universal scaling forms,
\begin{align}
\rho({\cal E},\delta\overline{\Delta}_0,m=0) &= |\delta \overline{\Delta}_0|^{\frac{d-\frac{1}{2}-z}{y_{\Delta}}} 
\Psi(|\delta \overline{\Delta}_0|^{-\frac{z}{y_{\Delta}}}{\cal E}), \label{qmc-r-s} \\
D_3({\cal E},\delta\overline{\Delta}_0,m=0) &= |\delta \overline{\Delta}_0|^{\frac{z-1}{y_{\Delta}}} 
f_z(|\delta \overline{\Delta}_0|^{-\frac{z}{y_{\Delta}}}{\cal E}), \label{qmc-d3-s} \\
D_{\perp}({\cal E},\delta\overline{\Delta}_0,m=0) &= |\delta \overline{\Delta}_0|^{\frac{z-2}{y_{\Delta}}} 
f_{\perp}(|\delta \overline{\Delta}_0|^{-\frac{z}{y_{\Delta}}}{\cal E}), \label{qmc-d12-s}
\end{align} 
In the DM phase side ($\delta \overline{\Delta}_0>0$) with finite zero-energy DOS and 
diffusion constant, this leads to 
\begin{eqnarray}
\left\{\begin{array}{c}
\rho({\cal E}=0,\delta\overline{\Delta}_0>0,m=0) \propto |\delta \overline{\Delta}_0|^{\frac{d-\frac{1}{2}-z}{y_{\Delta}}}, \\
D_3({\cal E}=0,\delta\overline{\Delta}_0>0,m=0) \propto |\delta \overline{\Delta}_0|^{\frac{z-1}{y_{\Delta}}}, \\
D_{\perp}({\cal E}=0,\delta\overline{\Delta}_0>0,m=0) \propto |\delta \overline{\Delta}_0|^{\frac{z-2}{y_{\Delta}}}, \\ 
\sigma_3({\cal E}=0,\delta\overline{\Delta}_0>0,m=0) \propto |\delta \overline{\Delta}_0|^{\frac{d-\frac{3}{2}}{y_{\Delta}}}, \\
\sigma_{\perp}({\cal E}=0,\delta\overline{\Delta}_0>0,m=0) \propto |\delta \overline{\Delta}_0|^{\frac{d-\frac{5}{2}}{y_{\Delta}}}.
 \\ 
\end{array}\right. \label{qmc-1}
\end{eqnarray} 
In the side of the quantum critical line between CI and WSM phases 
($\delta \overline{\Delta}_0<0$), Eqs.~(\ref{qmc-r-s}), (\ref{qmc-d3-s}), and (\ref{qmc-d12-s}) 
in combination with Eq.~(\ref{scaling-imp2}) leads to, 
\begin{eqnarray}
\left\{\begin{array}{c}
\rho({\cal E},\delta \overline{\Delta}_0<0,m=0) \propto |\delta \overline{\Delta}_0|^{\frac{2d-1}{2}\frac{1-z}{y_{\Delta}}} 
{\cal E}^{d-\frac{3}{2}}, \\ 
D_{3}({\cal E},\delta \overline{\Delta}_0<0,m=0) \propto |\delta \overline{\Delta}_0|^{\frac{z-1}{y_{\Delta}}},  \\
D_{\perp}({\cal E},\delta \overline{\Delta}_0<0,m=0) \propto |\delta \overline{\Delta}_0|^{\frac{2(z-1)}{y_{\Delta}}}  
{\cal E}^{-1}, \\
\sigma_{3}({\cal E},\delta \overline{\Delta}_0<0,m=0) \propto 
|\delta \overline{\Delta}_0|^{\frac{2d-3}{2}\frac{1-z}{y_{\Delta}}}  {\cal E}^{d-\frac{3}{2}},  \\
\sigma_{\perp}({\cal E},\delta \overline{\Delta}_0<0,m=0) \propto 
|\delta \overline{\Delta}_0|^{\frac{2d-5}{2}\frac{1-z}{y_{\Delta}}}  {\cal E}^{d-\frac{5}{2}}. \\
\end{array}\right. \label{qmc-2}
\end{eqnarray}

When the system approaches QMCP along $\delta \overline{\Delta}_{0}=0$, the DOS and diffusion constants 
respect the following universal scaling forms, 
\begin{align}
\rho({\cal E},\delta \overline{\Delta}_0=0,m) &= |m|^{\frac{d-\frac{1}{2}-z}{y_{m}}} 
\Psi(|m|^{-\frac{z}{y_{m}}}{\cal E}), \label{qmc-r-ss} \\
D_3({\cal E},\delta \overline{\Delta}_0=0,m) &= |m|^{\frac{z-1}{y_{m}}} 
f_z(|m|^{-\frac{z}{y_{m}}}{\cal E}), \label{qmc-d3-ss} \\
D_{\perp}({\cal E},\delta \overline{\Delta}_0=0,m) &= |m|^{\frac{z-2}{y_{m}}} 
f_{\perp}(|m|^{-\frac{z}{y_{m}}}{\cal E}). \label{qmc-d12-ss}
\end{align} 
In the side of the quantum critical line between DM and WSM phases ($m>0$), the DOS and diffusion 
constant are scaled by finite small energy ${\cal E}$ as 
\begin{align}
\rho({\cal E},\delta\overline{\Delta}_0=0,m>0) & \propto |{\cal E}|^{\frac{d-z^{\prime}}{z^{\prime}}}, \label{zp-1} \\ 
D_{\mu}({\cal E},\delta\overline{\Delta}_0=0,m>0) & \propto |{\cal E}|^{\frac{z^{\prime}-2}{z^{\prime}}}, \label{zp-2}
\end{align}
with $\mu=3,\perp$ (see Ref.~\cite{15,25} and Appendix). Here $z'$ is 
dynamical exponent at the semimetal-metal quantum phase transition in disordered single 
Weyl node~\cite{10,12,15,20,21,22,25,27}. 
This in combination with Eqs.~(\ref{qmc-r-ss}), (\ref{qmc-d3-ss}), and (\ref{qmc-d12-ss}) 
gives 
\begin{eqnarray}
\left\{\begin{array}{c}
\rho({\cal E},\delta \overline{\Delta}_0=0,m>0) \propto |m|^{\frac{2d(z'-z)-z'}{2z'y_{m}}} 
{\cal E}^{\frac{d-z'}{z'}}, \\ 
D_{3}({\cal E},\delta \overline{\Delta}_0=0,m>0) \propto |m|^{\frac{2z-z'}{z'y_{m}}} {\cal E}^{\frac{z'-2}{z'}}, \\
D_{\perp}({\cal E},\delta \overline{\Delta}_0=0,m>0) \propto 
|m|^{\frac{2(z-z')}{z'y_{m}}} {\cal E}^{\frac{z'-2}{z'}}, \\
\sigma_{3}({\cal E},\delta \overline{\Delta}_0=0,m>0) \propto |m|^{\frac{2d(z'-z)+4z-3z'}{2z'y_{m}}} 
{\cal E}^{\frac{d-2}{z'}}, \\
\sigma_{\perp}({\cal E},\delta \overline{\Delta}_0=0,m>0) \propto |m|^{\frac{2d(z'-z)+4z-5z'}{2z'y_{m}}} 
{\cal E}^{\frac{d-2}{z'}}, \\
\end{array}\right. \label{qmc-3}
\end{eqnarray}
for small positive $m$. On QMCP ($\delta \overline{\Delta}_0=0$, $m=0$), DOS and 
diffusion constants are scaled by finite small energy ${\cal E}$ as,  
\begin{eqnarray}
\left\{\begin{array}{c}
\rho({\cal E},\delta\overline{\Delta}_0=0,m=0) = |{\cal E}|^{ \frac{d-\frac{1}{2}-z}{z}}, \\
D_3({\cal E},\delta\overline{\Delta}_0=0,m=0) =  |{\cal E}|^{\frac{z-1}{z}}, \\
D_{\perp}({\cal E},\delta\overline{\Delta}_0=0,m=0) = |{\cal E}|^{\frac{z-2}{z}}, \\
\sigma_3({\cal E},\delta\overline{\Delta}_0=0,m=0) = |{\cal E}|^{\frac{d-\frac{3}{2}}{z}}, \\
\sigma_{\perp}({\cal E},\delta\overline{\Delta}_0=0,m=0) = |{\cal E}|^{\frac{d-\frac{5}{2}}{z}}. \\ 
\end{array}\right. \label{qmc-4}
\end{eqnarray}
The DOS and conductivities scalings around QMCP as well as the quantum 
critical lines for CI-WSM branch and that for WSM-DM branch are summarized in Table~\ref{table:scaling0}. 

As for transition between CI and DM phases, a situation becomes more involved. Firstly, our numerical 
studies below reveal that, unlike in the WSM-DM phase boundary, a mobility edge and a band edge 
are distinct for the CI-DM phase boundary, indicating the presence of the CI phase with 
finite zero-energy density of states (zDOS). Let us call the CI phase with finite zDOS as 
CI$^{\prime}$ phase. In the following, we first show that 
localization-delocalization transition between the CI$^{\prime}$ phase 
and DM phase belongs to the ordinary 3D unitary class. In Sec.~\ref{dos-ci-dm}, 
we show that, on increasing the disorder strength, the gapped Chern band insulator 
phase acquires a finite zDOS below the localization-delocalization transition. 

\section{Critical Exponent and Critical Conductance for CI-DM Branch}

\begin{figure}
	\centering
	\includegraphics[width=1\linewidth]{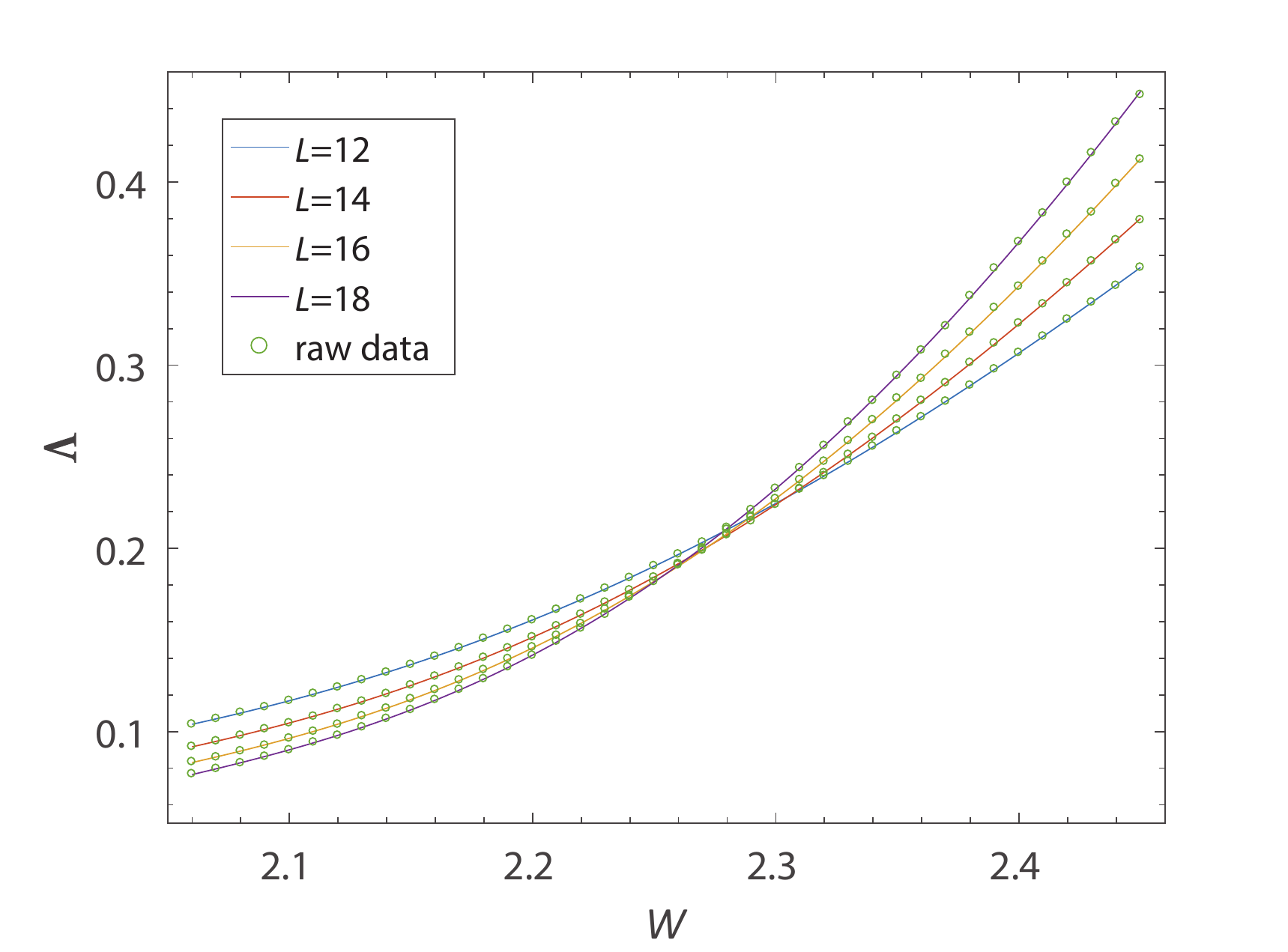}
	\caption{(color online) $\Lambda\equiv\lambda/L$ as a function of disorder strength 
for $\beta=0.2$ with a cross section size $L$ = 12, 14, 16, 18. The circle is the raw data 
of $\Lambda$ and the four curves are the fitting result with the Taylor-expansion orders: 
$(m_{1},m_2,n_1,n_2)=(2,0,3,1)$ (see the text).}\label{fig:2}
\end{figure}

To clarify the nature of the phase transition between 
CI and DM phases, we include in LCI a short-ranged on-site disorder potential, 
whose strength is denoted by $W$ henceforth, and calculate a quasi-one dimensional localization length $\lambda$ of 
eigenstates at the Fermi level ($E=0$) by transfer matrix method. 
We use a quasi-one-dimensional geometry with a square cross-section $L_x=L_y=L$ in $x_1$--$x_2$ plane, to 
calculate the localization length along the $x_3$-direction with variable length $L_z$. 
For longer $L_z$, smaller is the standard deviation $\sigma_{\Lambda}$ of 
a normalized localization length $\Lambda \equiv \lambda/L$~\cite{M81,P81}. 
We terminate the transfer matrix calculation, when a precision $\sigma_{\Lambda}/\Lambda$ 
reaches 0.1\% for $L=12,14$ and 0.2 \% for $L=16$, $18$. 
Fig.~\ref{fig:2} shows $\Lambda$ as a function of $W$ for different $L$. 
The scale-invariant behavior of $\Lambda$ indicates a localization-delocalization phase 
transition induced by $W$. 

\begin{table}[t]
  \centering
    \begin{tabular}{cccccc}
      \hline
$m_{1}$ &$m_{2}$ & GOF 
&$W_{c}$	&$\nu$ &$-y$  \\ \hline
2	&0	&0.27  &2.21[2.19,2.23] &1.34[1.23,1.53] &2.6[2.0,3.4] \\
3	&0	&0.82  &2.22[2.18,2.23] &1.27[1.18,1.48] &3.0[2.0,4.0] \\
2	&1	&0.29  &2.22[2.19,2.23] &1.31[1.22,1.52] &3.0[2.1,3.9] \\
3	&1	&0.81  &2.22[2.18,2.23] &1.26[1.18,1.48] &2.9[2.0,3.8] \\
\hline
 \end{tabular}
 \caption{Polynomial fitting results for the localization-delocalization transition between CI and DM phases 
at $\beta=0.2$ with $(n_1,n_2)=(3,1)$. The goodness of fit (GOF), critical disorder $W_{c}$, critical exponent $\nu\equiv1/\alpha_1$, 
largest irrelevant exponent $y$ are shown for different orders of the Taylor expansion ($m_1,m_2$). 
A value of $\Lambda$ at the critical point is $0.14\!\ [0.11,0.16]$. Error bar in the square bracket 
denotes the 95\% confidence interval.}\label{table}
\end{table} 

To determine a critical disorder strength $W_c$ and critical exponent of the phase transition, 
we use polynomial fitting method. We assume that the normalized localization length $\Lambda$ depends on 
disorder strength $W$ and system size $L$ through a universal function 
of only two scaling variables, $F(\phi_1,\phi_2)$~\cite{40,41}.
$\phi_1$ is a relevant scaling variable with 
positive exponent $\alpha_1\equiv 1/\nu$ and $\phi_2$ is an irrelevant scaling variable with the largest negative 
exponent $\alpha_2 \equiv y$. $\Lambda$ depends on $W$ and $L$ through the scaling variables, 
$\phi_j \equiv u_j(\omega) L^{\alpha_j}$ where $\omega \equiv (W-W_c)/W_c$. 
By definition, $u_1(\omega=0)=0$ and $u_2(\omega=0) \ne 0$. Assuming that $\omega$ is 
small and $L$ is large enough, we Taylor-expand the universal function in small $\phi_1$ and $\phi_2$ 
and $u_j(\omega)$ in small $\omega$;
$\Lambda=F(\phi_1,\phi_2)=\sum_{j_{1}=0}^{n_{1}}\sum_{j_{2}=0}^{n_{2}}a_{j_{1},j_{2}}\phi_{1}^{j_{1}}\phi_{2}^{j_{2}}$,  
$u_{i}(\omega)=\sum_{j=0}^{m_{i}}b_{i,j}\omega^{j}$. 
For a given set of $(n_1,n_2,m_1,m_2)$, we minimize $\chi^{2} \equiv \sum_{j=1}^{N_{D}}(F_{j}-\Lambda_{j})^{2}/\sigma_{j}^{2}$ 
in terms of $W_c$, $\alpha_1 \equiv \nu^{-1}$, $\alpha_2 \equiv y$, 
$a_{i,j}$, and $b_{l,m}$ with $a_{1,0}=a_{0,1}=1$ and $b_{1,0}=0$. 
$\Lambda_j$ and $\sigma_j$ ($j=1,\cdots,N_D$) are values of $\Lambda$ and 
$\sigma_{\Lambda}$ in Fig.~\ref{fig:2} for a given $L$ and $W$. $j$ specifies 
a data point, while $N_D$ is a number of those data points that are 
used for the fitting. $F_j$ is a value of $\Lambda$ from the polynomial fitting 
at the same $L$ and $W$. Fittings are carried out for several different 
$(n_1,n_2,m_1,m_2)$ with $n_1,n_2,m_1\le 3$, $m_2\le 2$. Those fitting results 
with goodness of fit (GOF) greater than 0.1 is shown in Table~\ref{table}. We also carry 
out the same fitting for 1000 synthetic data sets of $\{\Lambda_1,\cdots,\Lambda_{N_D}\}$, 
to determine 95\% confidence interval~\cite{41}.

A fit with the smallest number of parameters gives the exponent 
$\nu = 1.34 \!\ [1.23, 1.53]$. The fitting is stable against the increase of the 
expansion orders (Table~\ref{table:scaling0} and Appendix~\ref{sec: E}). 
The value is consistent with $\nu_{\rm 3d,u}\approx1.443[.437, .449]$~\cite{35} for the ordinary 
3D unitary class, suggesting that the localization-delocalization transition 
between CI and DM phases belong to the same universality class as the ordinary 3D 
unitary class~\cite{33,34,35,36,37,38}.  

\begin{figure}
	\centering
	\includegraphics[width=1\linewidth]{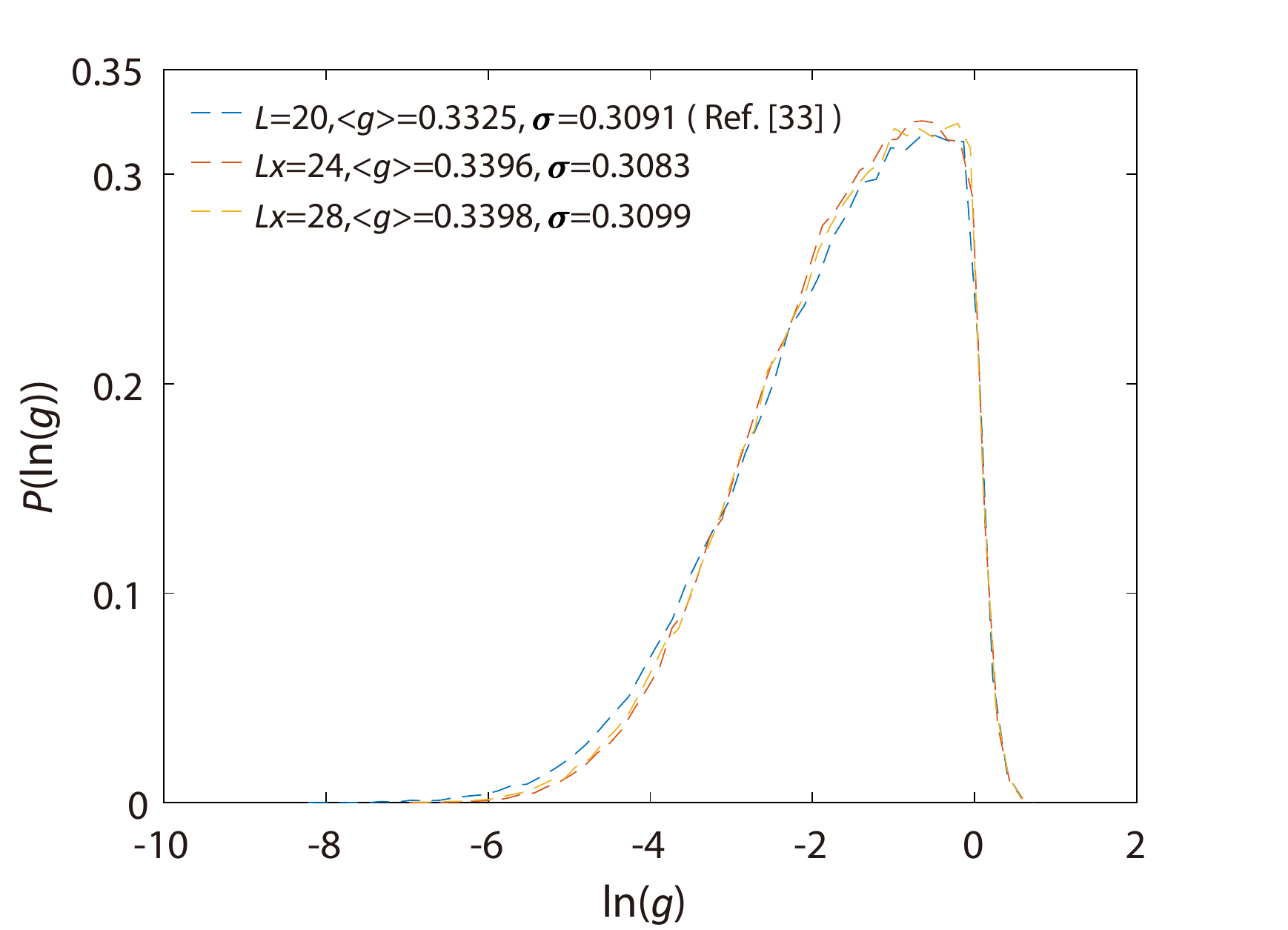}
	\caption{(color online) Critical conductance distributions (CCD) along the $x_3$-direction of LCI model  
at $W=W_{c}=2.21, \beta=0.2$ with $L_x=L_y=4L_z$ (periodic boundary conditions in $x_1$ and $x_2$ directions). 
$L_x=24$ (red broken line) and $L_x=28$ (orange). CCD of the reference model (called as a 
cubic U(1) tight-binding model~\cite{33}) (blue) is compared with them. Each distribution include 
$10^{5}$ samples. $g$ is the two terminal conductance $G_z$ in unit of $e^{2}/h$.}
	\label{fig:3}
\end{figure}

To reinforce the above result, we calculate a critical 
conductance distribution (CCD) and compare this with CCD in a reference 
tight-binding model whose Anderson transition is known to belong to the ordinary 3D universality 
class~\cite{33}. 
CCD generally depends only on the symmetry class and system's geometry, 
but free from a system size~\cite{42}. The geometry of a system is determined by mean values
of two-terminal conductance at the critical point along $x_1$, $x_2$ and $x_3$ directions, 
i. e. $G_x$, $G_y$ and $G_z$. In our model, $G_x=G_y\equiv G_{\perp}$, while out-plane 
conductance $G_z$ differs from the in-plane conductance $G_{\perp}$. 
We fine-tune $L_z$ with fixed $L_x=L_y =L$, such that $G_z$ and $G_{\perp}$
at the critical point become identical (very close) to those calculated from the reference 
tight-binding model~\cite{33}. 
For the critical point shown in Fig.~\ref{fig:2}, $L/L_z=4$ realizes this. 
Fig.~\ref{fig:3} shows a distribution of $G_{z}$ at the critical point with different 
$L$ with fixed $L/L_z=4$. The distributions are almost identical to the CCD in the 
reference tight-binding model. This in combination with the critical exponent concludes that 
the localization-delocalization transition for the CI-DM branch belong to the ordinary 
3D unitary class.

\section{Density of States for CI-DM Branch} \label{dos-ci-dm}

\begin{figure}[t]
	\centering
	\includegraphics[width=1\linewidth]{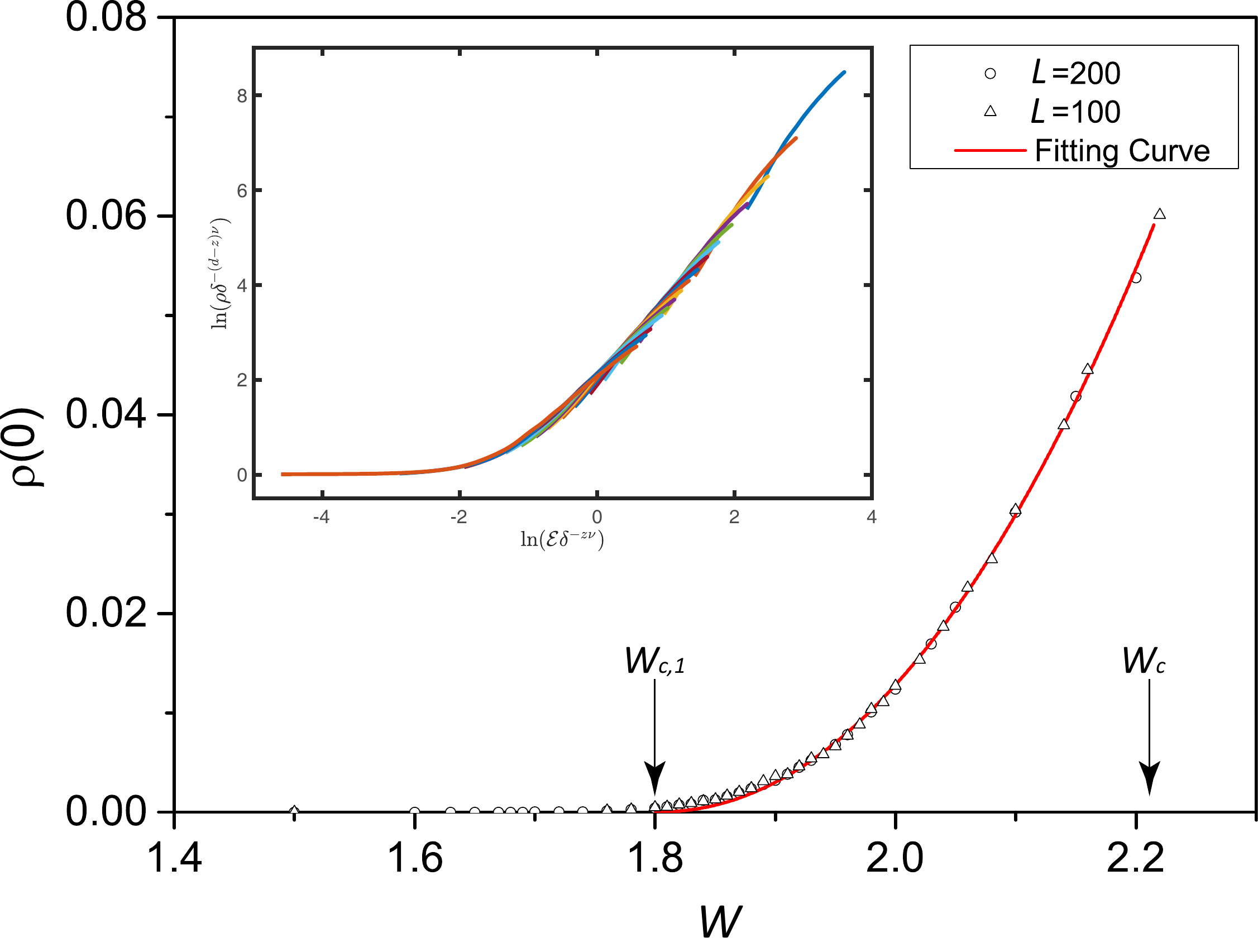}
	\caption{(color online) zDOS $\rho(0)$ as a function of disorder for $\beta=0.2$. 
The data points are fitted by a curve of $\rho(0)\propto\delta^{\phi}$ with $\phi=1.99\pm0.03$. Inset: 
Single-parameter scaling of DOS $\rho({\cal E})$ for $\beta=0.2$, different $W\!\ (>W_{c,1})$ and single-particle 
energy ${\cal E}$ with $\rho({\cal E})>0.008$. We use $\rho({\cal E})=\delta^{(d-z)\nu} \Psi({\cal E}\delta^{-z\nu})$ with 
$\delta\equiv|W-W_{c,1}|/W_{c,1}$, $d=3$, $z=1$ and $\nu=1$~\cite{15}.}
	\label{fig:4}
\end{figure}

The above observation indicates that, unlike in the critical line between DM and 
WSM phases~\cite{10,11,12,13,14,15,16,17,18,19,20,21,22,23,24,25,26,27,28,29,30}, 
zDOS is finite at the localization-delocalization transition between DM and CI phases. To confirm this, 
we calculate DOS in disordered LCI with cubic geometry ($L_x=L_y=L_z=L$) 
by kernel polynomial expansion method~\cite{43}.
We set $L$ to be 100 and/or 200 and the order of 
the Chebyshev polynomial expansion to a few thousands. 
Considering a self-averaging nature of DOS, we average the calculated 
DOS over two ($L=200$) and four ($L=100$) different disorder realizations. On increasing 
the disorder strength $W$, we found that 
the gapped (Chern) band insulator phase acquires a finite zDOS $\rho(0)$ {\it below} the 
localization-delocalization transition point, $\rho(0) \propto |W-W_{c,1}|^{\phi}$ with $\phi=1.99 \pm 0.03$ and 
$W_{c,1}<W_c$ (Fig.~\ref{fig:4} and Fig.~\ref{fig:1}). The zero-energy eigenstates are localized at $W_{c,1}<W<W_c$. 
Firstly, this concludes that zDOS at the 
localization-delocalization transition point for 
the CI-DM branch is finite and dynamical exponent $z$ 
associated with this transition is $3$. 
Note also that, within the self-consistent Born approximation, the exponent $\phi$ associated with 
$\rho(0)$ is evaluated as $1/2$~\cite{25}, while our numerical observation 
($\phi\simeq 2$) clearly deviates from this. Meanwhile, the DOS for different (but small) single-particle 
energy ${\cal E}$ and disorder strength $W>W_{c,1}$ can be well fitted into the same single-parameter 
scaling function (inset of Fig.~\ref{fig:4}). 

The critical properties for the CI-DM branch is distinct from that for the WSM-DM branch. 
This is consistent with the RG flow around the non-trivial fixed point with finite disorder 
(`FP1' in the inset of Fig.~\ref{fig:1}). Around FP1, the mass term is a relevant scaling variable, so that 
the critical properties for the CI-DM branch are controlled by saddle-point fixed points at finite 
$m \!\ (<0)$ and that for the WSM-DM branch is by a saddle-point fixed point at finite $m \!\ (>0)$. Especially, 
the single-parameter scaling in Fig.~\ref{fig:4}, 
$\rho({\cal E})\sim \delta^{(d-z)\nu} \Psi({\cal E}\delta^{-z\nu})$~\cite{15,25}, 
suggests that the DOS scaling around the phase 
boundary between CI with zero zDOS and CI with finite zDOS is controlled by a saddle-point fixed point 
at finite $m<0$. From the fitting, we evaluate $\nu\simeq1$ and 
$z\simeq1$ for this (postulated) saddle-point fixed point, which are distinct from those of the conventional 
3D unitary class ($\nu_{\rm 3d,u}\simeq 1.44$, $z_{\rm 3d,u} = 3$). This indicates that 
there exist two saddle-point fixed points between QMCP at $m=0$ and the two-dimensional 
limit with $\beta=0$. One saddle-point fixed point controls the DOS scaling around the transition 
between CI phase with zero zDOS and CI phase with finite zDOS. The other fixed point (conventional 
3D unitary class) controls critical properties around a localization-delocalization transition between 
CI with finite zDOS and DM phases (See Appendix \ref{sec:C}).

\begin{figure}[t]
	\centering
	\includegraphics[width=1\linewidth]{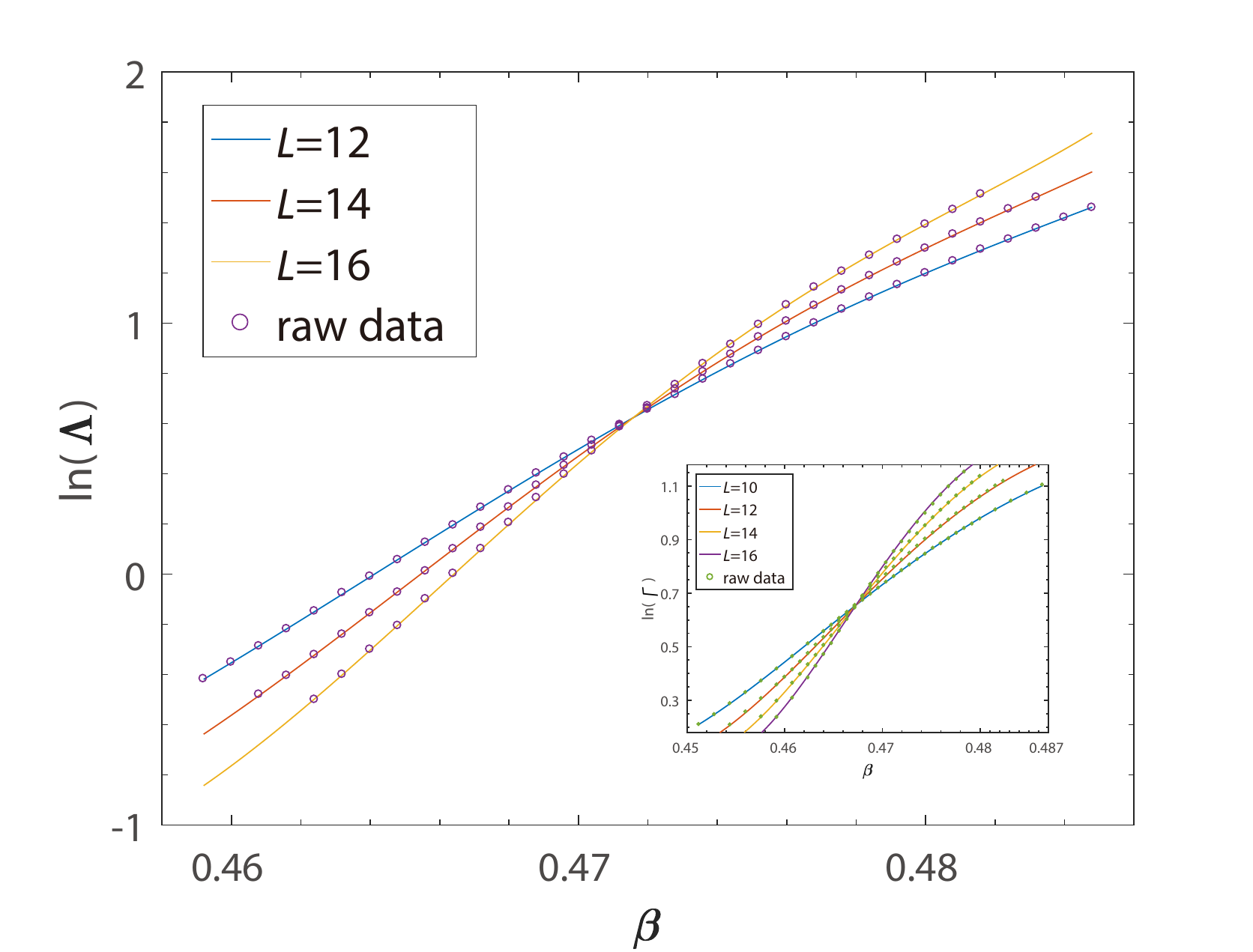}
	\caption{(color online) $\ln(\Lambda)$ as a function of $\beta$ at $W=1$ for the 
CI-WSM transition with cross-section size $L$ = 12, 14, 16. The circle is the raw data and the curves are the fitting result with the expansion orders $(n,m) = (4,2)$ (see the text). Inset: $\ln(\Gamma)$ as a function of 
$\beta$ at $W=1$ for CI-WSM transition with cross-section size $L$ = 10, 12, 14, 16. 
$\Gamma \equiv \lambda/\sqrt{L}$ and $\Lambda \equiv \lambda/L$ where $\lambda$ is 
a localization length along the $x_3$ direction (see the text). }
	\label{fig:5}
\end{figure}

\begin{table}[t]
	\centering
	\begin{tabular}{ccccc}
		\hline
		$n$   & $m$ & GOF & $\beta_{c}$ & $\nu$\\
		\hline
		4   &2	&0.79	&0.4716[0.4715,0.4716]	&0.8008[0.7947,0.8068]\\
		5   &2	&0.95	&0.4716[0.4715,0.4716]	&0.8002[0.7943,0.8069]\\
		4	&3	&0.81	&0.4716[0.4715,0.4717]	&0.8097[0.7964,0.8225]\\
		5	&3	&0.97	&0.4716[0.4715,0.4716]	&0.8124[0.7990,0.8261]\\
		\hline
	\end{tabular}
	\caption{Polynomial fitting results with goodness of fit greater than 0.1 are shown 
with the 95\% confidence interval for critical $\beta$ ($\beta_c$) 
and critical exponent $\nu$.}\label{table_w1}
\end{table} 

\section{Critical Exponent for CI-WSM Branch}
To clarify the critical nature of phase transition between CI with zero zDOS 
and WSM phases, we calculate a localization length along the $x_3$-direction 
at $E=0$. We fix $W$ and calculate 
$\lambda$ by changing $\beta$. The calculation is 
terminated when the precision of $\Lambda$ reaches 0.2\% for $L$=12, 14 and 
0.4\% for $L$=16. As shown in Fig.~\ref{fig:5}, curves of $\Lambda$ for different $L$ 
intersect almost at the same critical $\beta$. We thus use the polynomial fitting 
without the finite-size correction; $\Lambda = F(\phi_{1})$. 
Table~\ref{table_w1} shows a list of the critical exponent $\nu$ and critical mass $\beta_c$ 
with different expansion orders; 
$F(\phi_1) = \sum^{n}_{j=0} a_{j} \phi^j_1$ with $\phi_{1}(\omega) = \sum^{m}_{j=1} b_j \omega^j$. 
The fitting with the smallest number of parameters gives $\nu=0.80$ $[0.79,0.81]$, which is 
stable against the increase of the Taylor expansion orders.  

According to the RG analyses, the critical property for CI-WSM branch is determined by 
the fixed point in the clean limit (`FP0' in inset of Fig.~\ref{fig:1}). Around FP0, the mass 
term is a relevant scaling variable with its scaling dimension +1. The critical 
theory given by Eq.~(\ref{eff0}) at $m=0$ has a linear dispersion along the $x_1$--$x_2$ plane, 
and a quadratic dispersion along the $x_3$-direction. This leads to the anisotropic scaling, 
Eqs.~(\ref{scaling-3}) and (\ref{scaling-12}), dictating that a characteristic length 
scale along $x_3$-direction and that along the in-plane direction are scaled with the mass 
term differently, $\xi_3 = m^{-\frac{1}{2}}$, $\xi_{\perp}=m^{-1}$. Meanwhile, 
$\xi_3 = m^{-\frac{1}{2}}$ clearly contradicts with our precise determination of $\nu$. 

The discrepancy may be attributed 
to effect of finite disorders. Though the critical line between CI and WSM phases is 
controlled by the clean-limit fixed point, any finite small disorder could average out the spatial 
anisotropy in the  scaling property. When a characteristic length scale $\xi$ is defined 
as a cubic root of a characteristic volume, the scaling exponent of such 
$\xi$ with respect to $m$ can be around our numerical observation, 
$\nu_{\rm th}=\frac{1}{3}(\frac{1}{2}+1+1)=0.833\cdots$. Nonetheless, we cannot 
exclude a possibility that the discrepancy simply stems from a crossover phenomenon  
between FP1 and FP0 either. In fact, the perturbative RG equations given by 
Eqs.~(\ref{m-rg1}) and (\ref{m-rg2}) dictate that the exponent $\nu$ at  
FP1 and FP0 is +1 and  +1/2, respectively, while our numerical evaluation places 
between these two values.  

It is also completely legitimate to analyze the obtained numerical data in terms of another 
scaling theory which stands for the spatially anisotropic scaling dicated by 
Eqs.~(\ref{scaling-3}) and (\ref{scaling-12}). In the new scaling theory, the localization length along the 
$x_3$ direction, $\lambda$, is normalized by $\sqrt{L}$ instead of $L$;
\begin{eqnarray}
\Gamma \equiv \frac{\lambda}{\sqrt{L}}.  \label{Gamma1}
\end{eqnarray}
Namely, $L$ is a linear dimension within the $x_1$-$x_2$ plane and $\lambda$ is a length 
scale along the $x_3$ direction, so that they are scaled by $b$ and $b^{\frac{1}{2}}$ respectively 
under the anisotropic scaling given by Eqs.~(\ref{scaling-3}) and (\ref{scaling-12});
$\lambda^{\prime}=b^{\frac{1}{2}}\lambda$ and $L^{\prime}=bL$.  
Thus, $\Lambda \equiv \lambda/L$ after and before renormalization are related 
with each other by 
\begin{eqnarray}
\Lambda^{\prime}(m',L^{\prime}) = b^{-\frac{1}{2}}\Lambda(m,L) \label{Gamma2}
\end{eqnarray}
where $m'=b^{-1}m$ and $L^{\prime}=bL$. This leads to 
\begin{eqnarray}
\Lambda(m,L) = m^{\frac{1}{2}}\Phi(mL). \label{Gamma3}
\end{eqnarray}
The scaling form dictates that the conventional normalization, $\Lambda \equiv \lambda/L$, 
does not give a function only of $mL$. Instead, the new normalization, $\Gamma \equiv \lambda/\sqrt{L}$, 
does give a function only of $mL$;  
\begin{align}
\Gamma(m,L) &= \Lambda(m,L) L^{\frac{1}{2}}  \nonumber \\ 
&= (mL)^{\frac{1}{2}}\Phi(mL) = f(mL). \label{Gamma4}
\end{align}
Our fitting based on Eqs.~(\ref{Gamma1}) and (\ref{Gamma4}) shows a scaling invariant point at 
$\beta=\beta^{\prime}_c<\beta_c$. Meanwhile, an exponent $\nu'$ defined as 
$\Gamma\equiv f(m^{\nu'}L)$ is evaluated to be around 0.75 at $W=1$ 
rather than the expected value (1).

\section{Conclusion and Outlook} 
In this paper, we reveal a novel disorder-driven quantum multicriticality in disordered Weyl semimetal, 
by clarifying a rich variety of universal scalings of low-energy density of states (DOS), diffusion constants, 
and conductivities around the quantum multicritical point (QMCP) that is encompassed by three quantum 
phases; CI, WSM and DM phases. We show that, around a quantum critical line between CI and 
WSM phases, conductivities as well as diffusion constants in one spatial direction obey 
different universal scaling with different exponents from those in the other spatial 
directions. As for a phase boundary between CI and DM phases, 
we numerically demonstrated that a mobility edge and band edge are distinct in a phase diagram. We further 
show that the mobility edge belongs to the conventional 3D unitary class, while the DOS scaling around the 
band edge is controlled by another fixed point with different critical and dynamical exponents. We also 
confirmed numerically that the phase transition between CI and WSM is direct, and an associated critical 
exponent and scaling behaviour are rather consistent with the RG analyses than inconsistent.

The usage of the disordered magnetic dipole model and its RG analyses are 
not limited to the QMCP encompassed by CI, WSM and DM phases, though our simulation 
focused on this particular critical point. Being derived by the 
${\bm k}\cdot{\bm p}$ expansion around  high symmetric points, the dipole model does  
not carry any information of global ${\bm k}$-space topology 
in its gapped phase side. Thus, scaling properties revealed in this paper can be 
equally applicable to the other QMCP that is encompassed by oridinary 
three-dimensional band insulator, WSM and, DM phases~\cite{24,25,26,27}.  

Layered organic conductors may offer an experimental platform 
for exploring conductivity scalings discussed in 
this paper~\cite{45,46,46a}. Being soft, organic materials under an uniaxial pressure 
acquire substantial changes in hopping integrals among molecular orbitals, which sometimes 
result in a formation of a pair of two-dimensional gapless Dirac fermions in their elecrotonic 
band structures~\cite{45,46}. Thereby, the in-plane uniaxial pressure continuously 
changes the location of the gapless points in the ${\bm k}$-space, and it is 
predicted that the pair of the two Dirac points annihilate with each 
other above a certain critical pressure~\cite{45,46,47,48}. In such layered conductors, 
the out-of-plane uniaxial pressure can also enhance interlayer hopping 
strengths, which may lead to a formation and pair annihilation 
of three-dimensional gapless Dirac or Weyl points 
with or without the relativistic spin-orbit interaction~\cite{49,1}.  
Further detailed studies in conjuction with the ab-initio band calculation 
along this direction are awaited.

\appendix 
\section{Tight-Binding Cubic-Lattice Model for Weyl Semimetal}{\label{sec: A}}
We use the same two-orbital tight-binding model on a cubic lattice as studied previously~\cite{24,25,26,27,31,32}. The model 
consists of an $s$ orbital and a $p= p_{x}+ip_{y}$ orbital on each cubic lattice site (${\bm x}$);  
\begin{align}
{\cal H} = & \sum_{\bm {x}} \Big\{
\big(\epsilon_s + v_s({\bm x})\big) c^{\dagger}_{{\bm x},s} c_{{\bm x},s}
+ \big(\epsilon_p + v_p({\bm x})\big) c^{\dagger}_{{\bm x},p} c_{{\bm x},p} \Big\}   \nonumber \\
&+ \sum_{{\bm x}} \Big\{ - \sum_{\mu=1,2} \big(
t_s c^{\dagger}_{{\bm x} + {\bm e}_{\mu},s} c_{{\bm x},s}
- t_p c^{\dagger}_{{\bm x} + {\bm e}_{\mu},p} c_{{\bm x},p}\big)  \nonumber \\
& +  t_{sp} 
(c^{\dagger}_{{\bm x}+{\bm e}_1,p}
- c^{\dagger}_{{\bm x} - {\bm e}_1,p}  -  i c^{\dagger}_{{\bm x}+{\bm e}_2,p}
+ i c^{\dagger}_{{\bm x} - {\bm e}_2,p})  \!\ c_{{\bm x},s}  \nonumber \\
 &-  t^{\prime}_s c^{\dagger}_{{\bm x} + {\bm e}_{3},s} c_{{\bm x},s}
- t^{\prime}_p c^{\dagger}_{{\bm x} + {\bm e}_{3},p} c_{{\bm x},p} + {\rm H.c.} \Big\}. \label{tb1}
\end{align}
$\epsilon_{s}$, $\epsilon_{p}$ and $\upsilon_{s}(\bm{x})$, $\upsilon_{p}(\bm{x})$, 
are atomic energies for the $s$, $p$ orbital and on-site disorder potential of 
the $s$, $p$ orbital, respectively. The disorder potentials are uniformly 
distributed within $[-W/2,W/2]$ with identical probability distribution. 
$t_{s}$, $t_{p}$, and $t_{sp}$ are intralayer transfer integrals between $s$ orbitals of 
nearest neighboring two sites, that between $p$ orbitals, and that 
between $s$ and $p$ orbitals, respectively, while $t_{s}^{\prime}$ and $t_{p}^{\prime}$ are 
interlayer transfer integrals. ${\bm e}_{\mu}$ ($\mu=1,2$) are primitive translational vectors 
within a square-lattice plane. ${\bm e}_{3}$ is a primitive translational vector 
connecting neighboring square-lattice layers. Without the interlayer transfers, 
the electron model in the clean limit at half electron filling (one electron filling per cubic lattice site) 
is in either one of the two (layered) Chern band insulator phases ($0<\epsilon_s-\epsilon_p<4(t_s+t_p)$, 
$-4(t_s+t_p)<\epsilon_s-\epsilon_p<0$) or in the ordinary insulator phase 
($4(t_s+t_p)<|\epsilon_s-\epsilon_p|$). 
In this paper, we take $\epsilon_{s}-\epsilon_{p}=-2(t_{s}+t_{p})$, $t_{s}=t_{p}>0$, and 
$t_{sp}=4t_{s}/3$. 
A larger interlayer coupling induces a phase transition from the layered Chern band 
insulator phase to Weyl semimetal phase. In this paper, we take $t_{s}^{\prime}=-t_{p}^{\prime}>0$. For 
$\beta\equiv(t_{s}^{\prime}-t_{p}^{\prime})/[2(t_{s}+t_{p})]$ greater than $1/2$, a fourier-transformed tight-binding 
Hamiltonian has three pairs of Weyl nodes at ${\bm k}=(\pi,\pi,\pm k_0)$, $(0,\pi,\pi\pm k_0)$ and 
$(\pi,0,\pi\pm k_0)$ with $\cos k_0 \equiv \frac{1}{2\beta}$. 
In a clean-limit phase diagram subtended by $\beta$, a quantum critical point ($\beta=1/2$) intervenes 
the (Chern) band insulator phase ($\beta<1/2$) and a Weyl semimetal phase with three pairs of 
Weyl nodes ($\beta>1/2$).   
\begin{figure*}[t]
	\centering
	\includegraphics[width=0.65\linewidth]{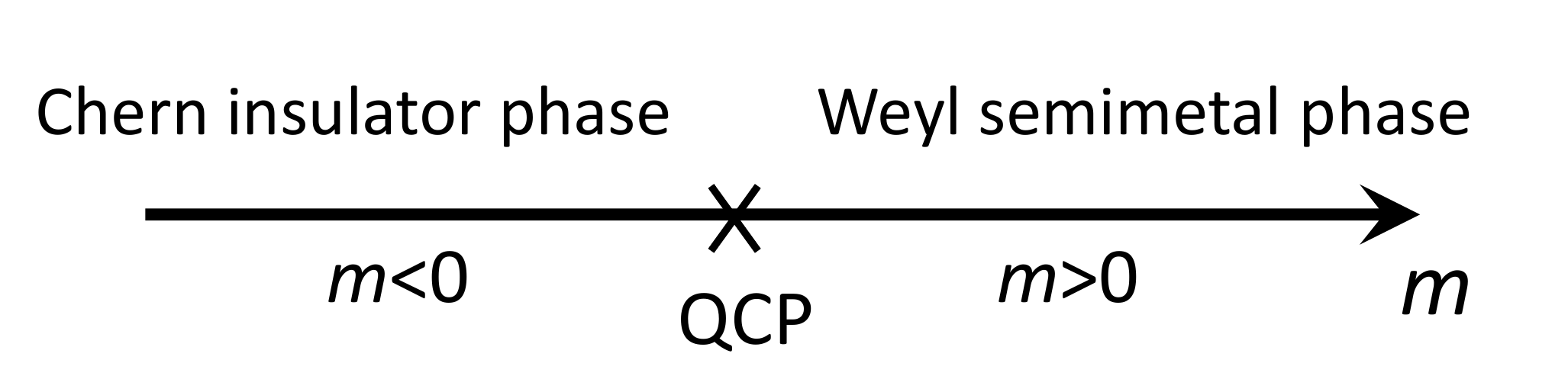}\\
	\caption{Schematic phase diagram for a quantum critical point 
intervening layered (Chern) band insulator phase ($m<0$) and Weyl semimetal phase ($m>0$). $m<0$ 
corresponds to $\beta<\frac{1}{2}$ and $m>0$ to $\beta>\frac{1}{2}$ in the LCI model in this paper.}\label{sfig:1}
\end{figure*}
\section{Renormalization Group Analyses for Low-Energy Effective Electron Hamiltonian with Disorders}\label{sec:B}
At the quantum critical point, three pairs of Weyl nodes annihilate at the high symmetric ${\bm k}$ points;  
${\bm k}=(\pi,\pi,0)$, 
$(\pi,0,\pi)$ and $(0,\pi,\pi)$, respectively. Each pair annihilation is described by the following two by 
two effective Hamiltonian, which is obtained from the ${\bm k}\cdot{\bm p}$ expansion around these 
high symmetric ${\bm k}$ points;
\begin{eqnarray}
{\bm H}_{\rm eff}({\bm p}) = v(p_1 {\bm \sigma}_1 + p_2 {\bm \sigma}_2) + (b_2 p^2_3-m) {\bm \sigma}_3.   \label{eff1}
\end{eqnarray}
Here ${\bm p}\equiv (p_1,p_2,p_3)$ denotes a crystal momentum measured from the symmetric ${\bm k}$ points. 
Velocities along the $x_1$ and $x_2$ directions are same due to the $C_4$ rotational symmetry. For $m<0$, 
the effective Hamiltonian is gapped for the half-filling case (Chern band 
insulator phase), while, for $m>0$, the Hamiltonian has a pair of Weyl 
nodes at $(p_1,p_2,p_3)=(0,0,\pm \sqrt{m/b_2})$.

Effect of disorders on the quantum critical point can be studied by the following replicated effective action~\cite{37,44},
\begin{align}
Z &= \int d\psi^{\dagger}_{\alpha} \psi_{\alpha} e^{-S}, \ \ S  = S_{0} + S_{1}, \nonumber \\
S_0 & \equiv \int d\tau \int d^2{\bm x}_{\perp} dx_3 \!\ \psi^{\dagger}_{\alpha}({\bm x},\tau) \Big\{ a \partial_{\tau} - iv 
(\partial_1 {\bm \sigma}_1 + \partial_2 {\bm \sigma}_2)      \nonumber \\ 
&+ \big( (-i)^2 b_2 \partial^2_3 - m \big) {\bm \sigma}_3 \Big\} \psi_{\alpha}({\bm x},\tau) \label{s0} \\
S_1& \equiv - \frac{\Delta_0}{2} \int d\tau \int d\tau^{\prime} \int d^3{\bm x} \!\ 
\big(\psi^{\dagger}_{\alpha} \psi_{\alpha}\big)_{{\bm x},\tau} 
\big(\psi^{\dagger}_{\beta}\psi_{\beta}\big)_{{\bm x},\tau^{\prime}}  \nonumber \\    
&- \frac{\Delta_3}{2} \int d\tau \int d\tau^{\prime} \int d^3{\bm x} 
\big(\psi^{\dagger}_{\alpha} {\bm \sigma}_3 \psi_{\alpha}\big)_{{\bm x},\tau} 
\big(\psi^{\dagger}_{\beta} {\bm \sigma}_3 \psi_{\beta}\big)_{{\bm x},\tau^{\prime}}  \nonumber  \\  
%
&- \frac{\Delta_\perp}{2} \sum_{\mu=1,2} \!\ \int d\tau \int d\tau^{\prime} \int d^3{\bm x} 
\big(\psi^{\dagger}_{\alpha} {\bm \sigma}_{\mu} \psi_{\alpha}\big)_{{\bm x},\tau} 
\big(\psi^{\dagger}_{\beta} {\bm \sigma}_{\mu} \psi_{\beta}\big)_{{\bm x},\tau^{\prime}}  \label{s1}
\end{align}  
where a coefficient in front of $\partial_{\tau}$ in $S_0$ is for a bookkeeping purpose; $a=1$. 
Here we have considered only short-ranged correlated on-site disorders in the low-energy effective Hamiltonian,
\begin{align}
{\cal H}_{\rm eff}&=\int d^3{\bm x} \!\ \psi^{\dagger}({\bm x}) \!\ \Big\{ 
- iv (\partial_1 {\bm \sigma}_1 + 
\partial_2 {\bm \sigma}_2)+  \nonumber  \\ 
&\big( (-i)^2 b_2 \partial^2_3 - m \big) {\bm \sigma}_3 
 + \sum_{\mu=0,1,2,3} V_{\mu}({\bm x}) {\bm \sigma}_\mu  
\Big\} \psi({\bm x}), 
\label{eff2}
\end{align}    
with $\langle V_{\mu}({\bm x}) V_{\nu}({\bm x}^{\prime}) \rangle_{\rm imp} 
= \Delta_{\mu} \delta_{\mu\nu}\delta^3({\bm x}-{\bm x}')$, and $\Delta_{1} = \Delta_2 \equiv \Delta_{\perp}$. 
The summation over the replica indices $\alpha$ and $\beta$ are omitted ($\alpha, \beta =1,\cdots,N$). 
In the perturbative derivation of renormalization 
group equations, we only keep those Feynman diagram contributions which survive in the limit of $N \rightarrow 0$. 
Following Roy et. al~\cite{31}, we employ a momentum-shell renormalization method, to decompose 
the fermion field $\psi_{\alpha}$ into a slow mode 
($\psi_{\alpha,<}$) and a fast mode ($\psi_{\alpha,>}$) in the momentum space, 
\begin{align}
\psi_{\alpha}({\bm x},\tau) &= \psi_{\alpha,<}({\bm x},\tau) + \psi_{\alpha,>}({\bm x},\tau) \nonumber \\
&= \frac{1}{\sqrt{\beta V}} \sum^{p_{\perp}<\Lambda e^{-dl}}_{i{\cal E}_n,{\bm p}} e^{i{\bm p}{\bm x} - i{\cal E}_{n}\tau} 
\!\  \psi_{\alpha}({\bm p},{\cal E}_n)   \nonumber \\
& \ \ + \frac{1}{\sqrt{\beta V}} \sum^{\Lambda e^{-dl}<p_{\perp}<\Lambda}_{i{\cal E}_n,{\bm p}} 
e^{i{\bm p}{\bm x} - i{\cal E}_{n}\tau} \!\ \psi_{\alpha}({\bm p},{\cal E}_n).\nonumber
\end{align}
with $p_{\perp} \equiv \{p^2_1 + p^2_2\}^{\frac{1}{2}}$.
The integration of the fast mode ($>$) in the partition function leads to a renormalization of 
effective action for the slow mode ($<$),  
\begin{align}
Z &= \int d\psi^{\dagger}_{\alpha,<} d\psi_{\alpha,<} \!\ e^{-S_{0,<}-S_{1,<}}  
\int d\psi^{\dagger}_{\alpha,>} d\psi_{\alpha,>} e^{-S_{0,>}} e^{-S_{1,>}} \nonumber \\
&\equiv Z_{0,>}  \int d\psi^{\dagger}_{\alpha,<} d\psi_{\alpha,<} \!\ e^{-S_{0,<}-S_{1,<}} \!\ 
\big\langle e^{-S_{1,>}} \big\rangle_{0,>} \nonumber \\ 
& = Z_{0,>}  \int d\psi^{\dagger}_{\alpha,<} d\psi_{\alpha,<} \!\ e^{-S_{0,<}-S_{1,<}}  \nonumber \\
& \hspace{2cm}  e^{-\langle S_{1,>} \rangle_{0,>} + \frac{1}{2} \big(\langle S^2_{1,>} \rangle_{0,>} 
- \langle S_{1,>}\rangle^2_{0,>} \big)+\cdots } \label{rg1} 
\end{align}
where 
\begin{align}
\big\langle \cdots \big\rangle_{0,>} &\equiv \frac{1}{Z_{0,>}} \!\ \int d\psi^{\dagger}_{\alpha,>} d\psi_{\alpha,>} 
e^{-S_{0,>}} \cdots, \nonumber \\
Z_{0,>}  &\equiv \int d\psi^{\dagger}_{\alpha,>} d\psi_{\alpha,>} e^{-S_{0,>}} \nonumber    
\end{align}
and 
\begin{widetext}
\begin{align}
S_{1,>} &= - \frac{\Delta_0}{2} \int d\tau \int d\tau' \int d^3{\bm x} \Big\{ 2
\big(\psi^{\dagger}_{\alpha,>}\psi_{\alpha,>}\big)_{{\bm x},\tau}   \big(\psi^{\dagger}_{\beta,<}\psi_{\beta,<}\big)_{{\bm x},\tau'} 
+ 2\big(\psi^{\dagger}_{\alpha,<}\psi_{\alpha,>}\big)_{{\bm x},\tau} \big(\psi^{\dagger}_{\beta,>}\psi_{\beta,<}\big)_{{\bm x},\tau'} \nonumber  \\ 
&\hspace{-0.2cm} + \big(\psi^{\dagger}_{\alpha,>}\psi_{\alpha,<}\big)_{{\bm x},\tau} \big(\psi^{\dagger}_{\beta,>}\psi_{\beta,<}\big)_{{\bm x},\tau'}  
+ \big(\psi^{\dagger}_{\alpha,<}\psi_{\alpha,>}\big)_{{\bm x},\tau} \big(\psi^{\dagger}_{\beta,<}\psi_{\beta,>}\big)_{{\bm x},\tau'}  
\Big\}  
- \frac{\Delta_3}{2} \int \cdots - \frac{\Delta_{\perp}}{2}  \sum_{\mu=1,2} \!\ \int \cdots \label{rg2} \\
S_{0,>(<)} &= \sum_{i{\cal E}_n} \sum_{p_3} {\sum_{p_{\perp}}}^{\prime} 
\psi^{\dagger}_{\alpha}({\bm p},{\cal E}_n) \Big\{ -i{\cal E}_n \!\ a \!\ {\bm \sigma}_0 
+ v(p_1 {\bm \sigma}_1 + p_2 {\bm \sigma}_2) + (b_2 p^2_3 - m) {\bm \sigma}_3 \Big\} \psi_{\alpha}({\bm p},{\cal E}_n). \label{rg3} 
\end{align} 
The integral region over ${\bm p}$ in $S_{0,>(<)}$ is restricted within the fast mode (slow mode).  $S_{1,<}$ comprises 
only of the slow modes. We keep only those terms in Eq.~(\ref{rg1}) which survive in the zero-replica 
limit ($N\rightarrow 0$)~\cite{10,12,20,21,31}; 
\begin{align}
\big\langle S_{1,>}\big\rangle_{0,>}  &= -\big(\Delta_0+2\Delta_{\perp} + \Delta_{3} \big) 
\sum_{i{\cal E}_{n}} {\sum_{\bm p}}^{\prime} h_0 \!\ i{\cal E}_n \psi^{\dagger}_{\alpha,<}({\bm p},i{\cal E}_n) {\bm \sigma}_0 
\psi_{\alpha,<}({\bm p},i{\cal E}_n) \nonumber \\ 
&\ \ - \big(\Delta_0-2\Delta_{\perp} + \Delta_{3} \big) \sum_{i{\cal E}_{n}} {\sum_{\bm p}}^{\prime} h_1 \!\ 
 \psi^{\dagger}_{\alpha,<}({\bm p},i{\cal E}_n) {\bm \sigma}_3 \psi_{\alpha,<}({\bm p},i{\cal E}_n) \label{rg4} 
\end{align} 
with 
\begin{align}
h_0 &= \frac{1}{V} \sum_{p_3} \sum_{\Lambda e^{-dl}<p_{\perp}<\Lambda} 
\frac{1}{{\cal E}^2_n + v^2 p^2_{\perp} + (b_2 p^2_3 - m)^2} = \frac{\Lambda^2 dl}{4\pi^2} 
\int^{+\infty}_{-\infty} \frac{dp_3}{(v\Lambda)^2 
+ (b_2 p^2_3 - m)^2} \equiv H_0 dl \label{h0} \\ 
h_1 &=  \frac{1}{V} \sum_{p_3} \sum_{\Lambda e^{-dl}<p_{\perp}<\Lambda} 
\frac{b_2 p^2_3 - m}{{\cal E}^2_n + v^2 p^2_{\perp} + (b_2 p^2_3 - m)^2} 
= \frac{\Lambda^2 dl}{4\pi^2} \int^{+\infty}_{-\infty} dp_3 \frac{b_2 p^2_3 -m}{(v\Lambda)^2 
+ (b_2 p^2_3 - m)^2}  \equiv H_1 dl \label{h1} 
\end{align}
and 
\begin{align}
&\frac{1}{2} \Big(\big\langle S^2_{1,>} \big\rangle_{0,>} - \big\langle S_{1,>}\big\rangle^2_{0,>} \Big) \nonumber \\
& \ \ = \frac{1}{V} \sum_{\mu=0,1,2,3} \sum_{i{\cal E}}\sum_{i{\cal E}^{\prime}}  {\sum_{{\bm p},{\bm p}^{\prime},{\bm q}}}^{\prime} \!\ 
F_{\mu} \!\ \big(\psi^{\dagger}_{\alpha,<}({\bm p}+{\bm q},i{\cal E}) {\bm \sigma}_{\mu} \psi_{\alpha,<}({\bm p},i{\cal E}) \big)
\big(\psi^{\dagger}_{\alpha,<}({\bm p}^{\prime}-{\bm q},i{\cal E}^{\prime}) 
{\bm \sigma}_{\mu} \psi_{\alpha,<}({\bm p}^{\prime},i{\cal E}^{\prime}) \big)  \label{rg5}
\end{align}
with 
\begin{align}
F_{0} &= (h_2+h_3) \Delta^2_{0} + (h_2 + 3 h_3)\Delta_0 \Delta_3 + 2(h_2 + h_3) \Delta_0 \Delta_{\perp} 
+ 2h_2 \Delta_3 \Delta_{\perp} \label{f0} \\ 
F_{3} &= h_3 \Delta^2_{0} + (-h_2+2h_3) \Delta^2_{3} + 2h_3 \Delta^2_{\perp} + (-h_2+h_3) \Delta_{0}\Delta_3 
+ 2h_2 \Delta_0 \Delta_{\perp} + 2 (h_2 - h_3) \Delta_{3} \Delta_{\perp} \label{f3} \\
F_{1} &= F_{2} =h_2 \Delta_{0} \Delta_{3} - h_3 \Delta_{0} \Delta_{\perp} + 3 h_3 \Delta_{3} \Delta_{\perp} \label{f12} 
\end{align}
\end{widetext}
and 
\begin{align}
h_{2} = \frac{\Lambda^2dl}{4\pi^2} \int^{+\infty}_{-\infty} \frac{(v\Lambda)^2 dp_{3}}{\big[(v\Lambda)^2 + 
(b_2 p^2_3 - m)^2\big]^2} \equiv H_2 dl,  \nonumber\\
h_3  = \frac{\Lambda^2dl}{4\pi^2} \int^{+\infty}_{-\infty} \frac{(b_2 p^2_3 - m)^2 dp_{3}}{\big[(v\Lambda)^2 + 
(b_2 p^2_3 - m)^2\big]^2} \equiv H_3 dl. \label{h23} 
\end{align}
To evaluate $h_0,h_1,h_2,h_3$, we have set the frequencies and momenta for the slow modes (external lines) 
to zero~\cite{10,12,20,21,31}. After the integration over the fast modes, we scale spatial 
coordinate, time and field operators as   
\begin{align}
\psi_{\alpha} &= Z^{-\frac{1}{2}}_{\psi} \psi^{\prime}_{\alpha} \equiv e^{-\frac{gdl}{2}} \psi^{\prime}_{\alpha}, \label{scale1} \\
\tau &= e^{zdl}\tau^{\prime},  \label{scale2} \\ 
{\bm x}_{\perp} &= e^{dl}{\bm x}^{\prime}_{\perp}, \label{scale3} \\ 
x_{3} &= e^{\frac{dl}{2}} x^{\prime}_{3}. \label{scale4}     
\end{align}  
This leads to one-loop renormalization for $a$, $v$, $b_2$, $\Delta_0$, $\Delta_3$ and $\Delta_{\perp}$, 
\begin{align}
a^{\prime} &= a + \big(2+ \frac{1}{2} - g\big)a \!\ dl  + \big(\Delta_0+2\Delta_{\perp} + \Delta_{3}\big) h_0, \label{rg6} \\ 
v^{\prime} & = v + \big(1+ \frac{1}{2} + z - g\big) v\!\ dl,  \label{rg7} \\ 
b^{\prime}_2 &= b_{2} + \big(1 + \frac{1}{2} + z - g\big) b_2\!\ dl, \label{rg8} \\ 
m^{\prime} & =  m + \big(2+ \frac{1}{2} + z - g\big) m \!\ dl  + \big(\Delta_0-2\Delta_{\perp} + \Delta_{3}\big) h_1, \label{rg9} \\
\Delta^{\prime}_0 & = \Delta_0 + \big(2 + \frac{1}{2} + 2z -2g \big) \Delta_{0} \!\ dl  + F_0, \label{rg10} \\ 
\Delta^{\prime}_3 & = \Delta_3 + \big(2 + \frac{1}{2} + 2z -2g \big) \Delta_{3} \!\ dl  + F_3, \label{rg11} \\ 
\Delta^{\prime}_{\perp} & = \Delta_{\perp} + \big(2 + \frac{1}{2} + 2z -2g \big) \Delta_{\perp} \!\ dl  + F_{\perp}. \label{rg12} 
\end{align}
We choose a scaling of the field operator $g$ and dynamical exponent $z$ in a way that a gapless part 
of the effective action ($a$, $v$ and $b_2$ in Eq.~(\ref{s0})) is invariant under the renormalization~\cite{10,12,20,21,31};
\begin{align}
g &= \big(2+\frac{1}{2}\big) + \big(\Delta_{0} + 2\Delta_{\perp} + \Delta_3 \big) H_0, \label{scale5} \\ 
z &= 1 + \big(\Delta_{0} + 2\Delta_{\perp} + \Delta_3 \big) H_0. \label{dynamical} 
\end{align} 
Using this, we obtain from Eqs.~(\ref{rg9})-(\ref{rg12}) one-loop renormalization group equations 
for mass ($m$) and three kinds of disorder strength;
\begin{widetext}
\begin{align}
\frac{dm}{dl} &= m + \big(\Delta_0 - 2\Delta_{\perp} + \Delta_3 \big) H_1, \label{rg13} \\
\frac{d\Delta_0}{dl} &= -\frac{1}{2} \Delta_{0} + 
(H_2+H_3) \big(\Delta_0 + 2\Delta_{\perp} + \Delta_3 \big)\Delta_0 + 
2 H_3\Delta_0 \Delta_3  + 2H_2 \Delta_3 \Delta_{\perp}, \label{rg14} \\ 
\frac{d\Delta_{\perp}}{dl} & = -\frac{1}{2} \Delta_{\perp} + H_2 \Delta_{0} \Delta_{3} - H_3 \Delta_{0} \Delta_{\perp}
 + 3 H_3 \Delta_{3} \Delta_{\perp}, \label{rg15} \\ 
\frac{d\Delta_{3}}{dl} & = -\frac{1}{2} \Delta_{3} + (-H_2+H_3)  
\big(\Delta_0 - 2\Delta_{\perp} + \Delta_{3} \big) \Delta_{3}  
+ H_3 \big(\Delta^2_{0} +  
2 \Delta^2_{\perp} + \Delta^2_{3}\big) + 2H_2 \Delta_0 \Delta_{\perp},  \label{rg16}
\end{align}

To make the RG analysis to be a controlled analysis, Roy et.al. proposed to replace $b_2 (-i)^2\partial^2_{3}$ 
in Eqs.~(\ref{eff1}) and (\ref{eff2}) by $b_{n} (-i)^{n} \partial^{n}_{3}$~\cite{31}. 
This leads to replacements of $b_2 p^2_3$ in Eqs.~(\ref{h0}), (\ref{h1}), and (\ref{h23}) by $b_{n} p^{n}_{3}$ and $\frac{1}{2}$ in 
Eqs.~(\ref{scale4})-(\ref{scale5}) and (\ref{rg14})-(\ref{rg16}) by $\frac{1}{n}$. Thereby, disorder strength of any 
non-trivial fixed point 
becomes on the order of $1/n$ in a large $n$ limit, which always justifies the above perturbative derivation 
of the RG equation in the leading order of $1/n$. All the scaling analyses based on the RG equation around 
such fixed points become controlled in the large $n$ limit. For general even integer $n$, we have 
\begin{align}
H_0 &= \frac{\Lambda^2}{4\pi^2} \int^{+\infty}_{-\infty} dX \frac{1}{(v\Lambda)^2 + (b_n X^{n} - m)^2} 
= \frac{\Lambda^2 R^{\frac{1}{n}-2}}{4\pi^2 b^{\frac{1}{n}}_n}  \frac{2\pi}{n} \frac{\sin\big[(n-1)\phi_n\big]}{\sin \big[n\phi_n\big]} 
\frac{1}{\sin\big[\frac{\pi}{n}\big]}, \label{H0} \\
H_{1} &= \frac{\Lambda^2}{4\pi^2} \int^{+\infty}_{-\infty} dX \frac{b_n X^n -m}{(v\Lambda)^2 + (b_n X^{n} - m)^2} 
= \frac{\Lambda^2 R^{\frac{1}{n}-1}}{4\pi^2 b^{\frac{1}{n}}_n}  \frac{2\pi}{n} \frac{\sin\big[\phi_n\big] + \cos\big[n\phi_n\big] \sin\big[(n-1)\phi_n\big]}{\sin \big[n\phi_n\big]} 
\frac{1}{\sin\big[\frac{\pi}{n}\big]}, \label{H1} \\
H_2 &= \frac{\Lambda^2}{4\pi^2} \int^{+\infty}_{-\infty} dX \frac{(v\Lambda)^2}{[(v\Lambda)^2 + (b_n X^{n} - m)^2]^2}  
= \frac{\Lambda^2 R^{\frac{1}{n}-2}}{4\pi^2 b^{\frac{1}{n}}_n} \frac{\pi}{n} \frac{1}{\sin\big[\frac{\pi}{n}\big]}
\Big(\frac{\sin\big[(n-1)\phi_n\big]}{\sin \big[n\phi_n\big]} - \frac{n-1}{n} \cos\big[(2n-1)\phi_n\big] \Big), \label{H2} \\
H_3 &= \frac{\Lambda^2}{4\pi^2} \int^{+\infty}_{-\infty} dX \frac{(b_n X^n -m)^2}{[(v\Lambda)^2 + (b_n X^{n} - m)^2]^2} 
= \frac{\Lambda^2 R^{\frac{1}{n}-2}}{4\pi^2 b^{\frac{1}{n}}_n}  \frac{\pi}{n} \frac{1}{\sin\big[\frac{\pi}{n}\big]}
\Big(\frac{\sin\big[(n-1)\phi_n\big]}{\sin \big[n\phi_n\big]} + \frac{n-1}{n} \cos\big[(2n-1)\phi_n\big] \Big).  \label{H3} 
\end{align}  
with $H_{2}+H_3 \equiv H_0$. 
$n\phi_n$ measures an angle subtended by the mass $m$ and cutoff energy $v\Lambda$ (see Fig.~\ref{sfig:2}). 
\begin{figure}[b]
	\centering
	\includegraphics[width=0.6\linewidth]{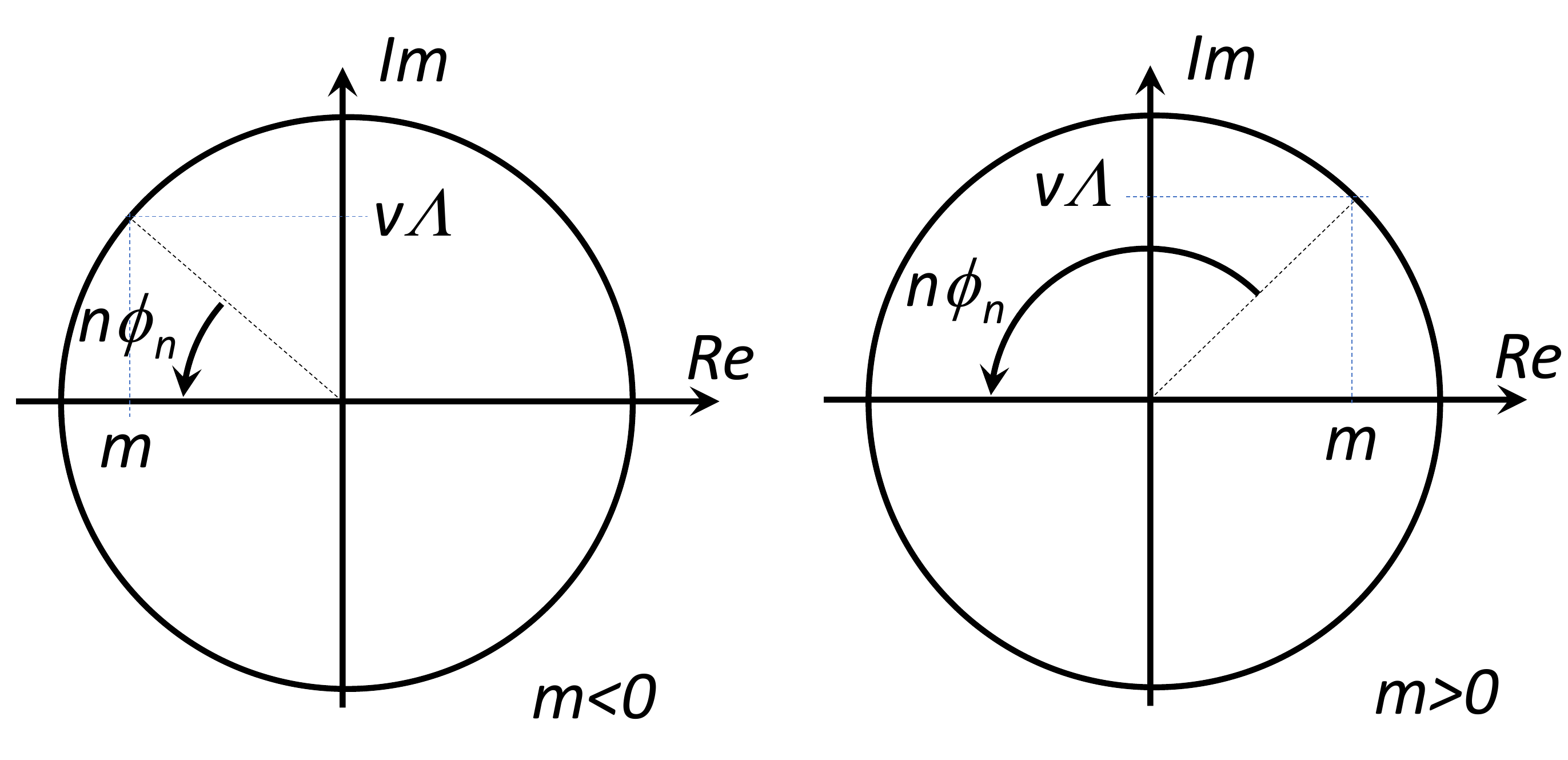}\\
	\caption{$n\phi_n$ is an angle subtended by the mass $m$ 
		and cutoff energy $v\Lambda$ with $\cos[n\phi_n]=-\frac{m}{R}$ and $R=\{m^2+(v\Lambda)^2\}^{\frac{1}{2}}$. 
		(Left) $m<0$, (Right) $m>0$. }\label{sfig:2}
\end{figure}

\subsection{large-$n$ limit}
In the large $n$ limit, the leading order contribution of $H_0$, $H_1$, $H_2$, $H_3$ are as follows,
\begin{align}
H_0 &= \frac{\Lambda^2 R^{\frac{1}{n}-2}}{4\pi^2 b^{\frac{1}{n}}_n} \Big\{ 2 + {\cal O}\big(n^{-1}\big) \Big\}, 
H_1 = \frac{\Lambda^2 R^{\frac{1}{n}-1}}{4\pi^2 b^{\frac{1}{n}}_n} \Big\{ - \frac{2m}{R} + {\cal O}\big(n^{-1}\big) \Big\}, \nonumber \nonumber\\
H_2 &= \frac{\Lambda^2 R^{\frac{1}{n}-2}}{4\pi^2 b^{\frac{1}{n}}_n} \Big\{ \frac{2(v\Lambda)^2}{R^2} + {\cal O}\big(n^{-1}\big) \Big\}, 
H_3 =  \frac{\Lambda^2 R^{\frac{1}{n}-2}}{4\pi^2 b^{\frac{1}{n}}_n} \Big\{ \frac{2m^2}{R^2} + {\cal O}\big(n^{-1}\big) \Big\}. \nonumber
\end{align} 
The RG equations in the large $n$ limit take form of 
\begin{align}
z &= 1 + 2\big(\overline{\Delta}_0 + 2\overline{\Delta}_{\perp} + \overline{\Delta}_3 \big), \label{dynamical2} \\ 
\frac{dm}{dl} &= m - 2m\big(\overline{\Delta}_0 - 2\overline{\Delta}_{\perp} + \overline{\Delta}_3 \big) \Big(\frac{v\Lambda}{R}\Big)^{2-\frac{1}{n}}, 
\label{rg17} \\
\frac{d\overline{\Delta}_0}{dl} &= -\frac{1}{n} \overline{\Delta}_{0} + 
2 \big(\overline{\Delta}_0 + 2\overline{\Delta}_{\perp} + \overline{\Delta}_3 \big)\overline{\Delta}_0 
\Big(\frac{v\Lambda}{R}\Big)^{2-\frac{1}{n}}  
+ 4 \overline{\Delta}_0 \overline{\Delta}_3 \Big(\frac{m}{R}\Big)^{2} \Big(\frac{v\Lambda}{R}\Big)^{2-\frac{1}{n}} + 4 \overline{\Delta}_3 \overline{\Delta}_{\perp} \Big(\frac{v\Lambda}{R}\Big)^{4-\frac{1}{n}}, \label{rg18} \\ 
\frac{d\overline{\Delta}_{\perp}}{dl} & = -\frac{1}{n} \overline{\Delta}_{\perp} + 2\overline{\Delta}_{0} \overline{\Delta}_{3}  
\Big(\frac{v\Lambda}{R}\Big)^{4-\frac{1}{n}} 
 - 2 \overline{\Delta}_{0} \overline{\Delta}_{\perp}  \Big(\frac{m}{R}\Big)^{2} 
\Big(\frac{v\Lambda}{R}\Big)^{2-\frac{1}{n}} 
 + 6 \overline{\Delta}_{3} \overline{\Delta}_{\perp}  \Big(\frac{m}{R}\Big)^{2} 
\Big(\frac{v\Lambda}{R}\Big)^{2-\frac{1}{n}},  \label{rg19} \\ 
\frac{d\overline{\Delta}_{3}}{dl} & = -\frac{1}{n} \overline{\Delta}_{3} +   
2\big(\overline{\Delta}_0 - 2\overline{\Delta}_{\perp} + \overline{\Delta}_{3} \big) \overline{\Delta}_{3} \frac{m^2-(v\Lambda)^2}{R^2} 
\Big(\frac{v\Lambda}{R}\Big)^{2-\frac{1}{n}} \nonumber\\
&\ \ \ + 2 \big(\overline{\Delta}^2_{0} +   
2 \overline{\Delta}^2_{\perp} + \overline{\Delta}^2_{3}\big)  \Big(\frac{m}{R}\Big)^{2} 
\Big(\frac{v\Lambda}{R}\Big)^{2-\frac{1}{n}} 
 + 4 \overline{\Delta}_0 \overline{\Delta}_{\perp} \Big(\frac{v\Lambda}{R}\Big)^{4-\frac{1}{n}},   
  \label{rg20} 
\end{align} 
with $R^2\equiv m^2+(v\Lambda)^2$ and 
\begin{eqnarray}
\overline{\Delta}_{\mu} \equiv \Delta_{\mu} \cdot  \frac{(v\Lambda)^{\frac{1}{n}}}{4\pi^2 v^2 b^{\frac{1}{n}}_n}. \label{scale20}
\end{eqnarray}
In the leading order expansion in large $n$, we can replace 
$(v\Lambda/R)^{2-\frac{1}{n}}$ and $(v\Lambda/R)^{4-\frac{1}{n}}$ in the right hand sides of 
Eqs.~(\ref{rg17})-(\ref{rg20}) by $(v\Lambda/R)^{2}$ and $(v\Lambda/R)^{4}$, 
respectively.  Note also that the RG equations differ from RG equations obtained by 
Roy et. al.~\cite{31} in several aspects. First of all, (i) $-2m$ in Eq.~(\ref{rg17}) is $+m$ in the 
RG equations by Roy et. al.~\cite{31}, (ii) $2(\overline{\Delta}_0 + 
2\overline{\Delta}_{\perp} + \overline{\Delta}_3)$ 
in Eq.~(\ref{dynamical2}) is $\overline{\Delta}_0 + 
2\overline{\Delta}_{\perp} + \overline{\Delta}_3$ in Roy et. al.~\cite{31} 
These differences change the scaling exponent of $m$ and an dynamical exponent $z$, respectively  
as well as the DOS, diffusion constant and conductivities scalings around the quantum multicritical point (see the next 
section). Besides, we keep $m/R$ to be finite, instead of setting $m/R$ and 
$\Lambda/R$ to be zero and unit, respectively~\cite{31}.

The set of RG equations above have only one non-trivial fixed point and a trivial fixed point;
\begin{align}
&{\rm FP0} : \ \ \ (z,m,\overline{\Delta}_{0},\overline{\Delta}_{\perp},\overline{\Delta}_3) = \Big(1,0,0,0,0\Big),  \label{FP0-a} \\
&{\rm FP1} : \ \ \ (z,m,\overline{\Delta}_{0},\overline{\Delta}_{\perp},\overline{\Delta}_3) = \Big(1+\frac{1}{n},0,\frac{1}{2n},0,0\Big).  \label{FP1-a} 
\end{align}
These two fixed points are on a plane of $\overline{\Delta}_{3}=\overline{\Delta}_{\perp}=0$. Besides, any points in the plane of 
$\Delta_{3}=\Delta_{\perp}=0$ flows within the same plane. The former fixed point FP0 is a saddle point of the RG flow around which the RG equations are linearized as 
\begin{eqnarray}
\frac{dm}{dl} = m, \ \ \frac{d\Delta_0}{dl} = - \frac{1}{n} \Delta_0. \label{FP0-rg} 
\end{eqnarray}
The latter fixed point FP1 is an unstable point of the RG flow, around which linearized equations 
take forms of 
\begin{eqnarray}
\frac{dm}{dl} = \big(1-\frac{1}{n}\big) m, \ \ \frac{d\Delta_0}{dl} = \frac{1}{n} \big(\Delta_0 - \frac{1}{2n}\big). \label{FP1-rg} 
\end{eqnarray}         

\begin{figure}
	\centering
	\includegraphics[width=0.4\linewidth]{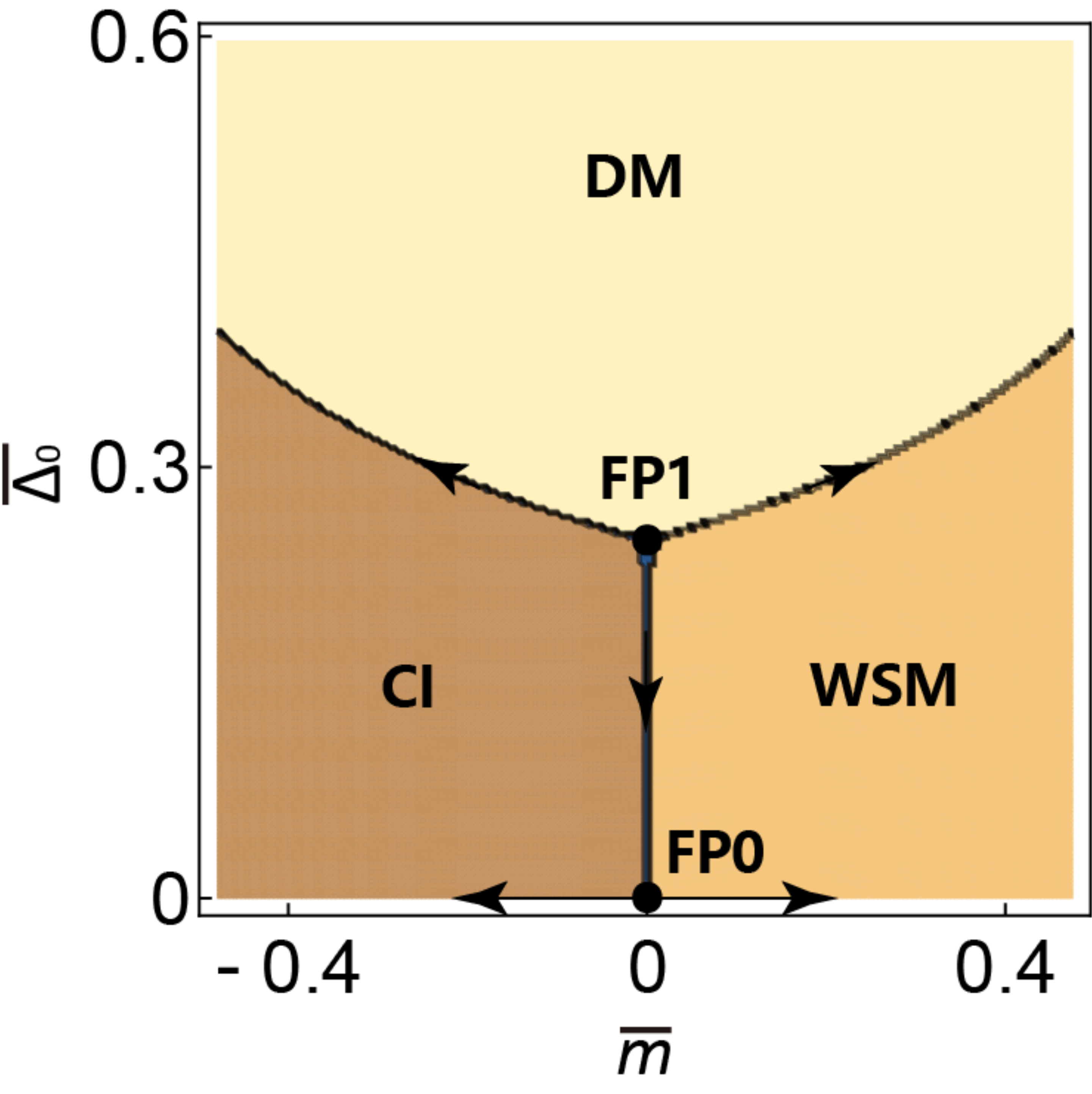}\\
	\caption{(color online) Phase diagram in $\overline{\Delta}_0$-$\overline{m}$ plane determined by 
the RG equations in the large $n$ limit.}\label{sfig:3}
\end{figure}

\subsection{in the case of $n=2$} 
For complementary information, we also analysize the RG equations in the case of $n=2$, where 
\begin{align}
H_0 &= \frac{\Lambda^2 R^{\frac{1}{2}-2}}{4\pi^2 b^{\frac{1}{2}}_2} \frac{\pi}{2} \frac{1}{\cos[\phi_2]}, \ \ 
H_1 = \frac{\Lambda^2 R^{\frac{1}{2}-1}}{4\pi^2 b^{\frac{1}{2}}_2} \pi \cos[\phi_2], \nonumber \\
H_2 &= \frac{\Lambda^2 R^{\frac{1}{2}-2}}{4\pi^2 b^{\frac{1}{2}}_2} \frac{\pi}{4} \bigg\{ \frac{1}{\cos[\phi_2]} - \cos[3\phi_2] \bigg\}, \nonumber\\
H_3 &=   \frac{\Lambda^2 R^{\frac{1}{2}-2}}{4\pi^2 b^{\frac{1}{2}}_2} \frac{\pi}{4} \bigg\{ \frac{1}{\cos[\phi_2]} + \cos[3\phi_2] \bigg\}. \nonumber
\end{align}
A set of the RG equations is simplified as follows,  
\begin{align}
z & = 1 + 2 (\overline{\Delta}_0 + 2\overline{\Delta}_{\perp} + \overline{\Delta}_3) \Big(\frac{v\Lambda}{R}\Big)^{2-\frac{1}{2}} x, \label{dynamical3} \\ 
\frac{dm}{dl} &= m + 4 R \big(\overline{\Delta}_0 - 2\overline{\Delta}_{\perp} + \overline{\Delta}_3 \big) 
\Big(\frac{v\Lambda}{R}\Big)^{2-\frac{1}{2}} y, 
\label{rg21} \\
\frac{d\overline{\Delta}_0}{dl} &= -\frac{1}{2} \overline{\Delta}_{0} + 2 \big(\overline{\Delta}_0 
+ 2\overline{\Delta}_{\perp} + \overline{\Delta}_3 \big)\overline{\Delta}_0 
\Big(\frac{v\Lambda}{R}\Big)^{2-\frac{1}{2}}  x  
+ 2 \overline{\Delta}_0 \overline{\Delta}_3  \Big(\frac{v\Lambda}{R}\Big)^{2-\frac{1}{2}} (x+z)  
+ 2 \overline{\Delta}_3 \overline{\Delta}_{\perp}  \Big(\frac{v\Lambda}{R}\Big)^{2-\frac{1}{2}}(x-z), \label{rg22} \\ 
\frac{d\overline{\Delta}_{\perp}}{dl} & = -\frac{1}{2} \overline{\Delta}_{\perp} 
+ \overline{\Delta}_{0} \overline{\Delta}_{3}  
\Big(\frac{v\Lambda}{R}\Big)^{2-\frac{1}{2}} (x-z)  
- \overline{\Delta}_{0} \overline{\Delta}_{\perp}  \Big(\frac{v\Lambda}{R}\Big)^{2-\frac{1}{2}} (x+z)
 + 3 \overline{\Delta}_{3} \overline{\Delta}_{\perp}   \Big(\frac{v\Lambda}{R}\Big)^{2-\frac{1}{2}} (x+z), \label{rg23} \\ 
\frac{d\overline{\Delta}_{3}}{dl} & = -\frac{1}{2} \overline{\Delta}_{3} +   
2\big(\overline{\Delta}_0 - 2\overline{\Delta}_{\perp} + \overline{\Delta}_{3} \big) \overline{\Delta}_{3} 
\Big(\frac{v\Lambda}{R}\Big)^{2-\frac{1}{2}} z \nonumber \\ 
& \ \ \ + \big(\overline{\Delta}^2_{0} +  2 \overline{\Delta}^2_{\perp} 
+ \overline{\Delta}^2_{3}\big) \Big(\frac{v\Lambda}{R}\Big)^{2-\frac{1}{2}} (x+z) 
 + 2 \overline{\Delta}_0 \overline{\Delta}_{\perp} \Big(\frac{v\Lambda}{R}\Big)^{2-\frac{1}{2}} (x-z),
  \label{rg24}
\end{align}  
with $R^2\equiv m^2+(v\Lambda)^2$, $\overline{m} \equiv 
m/R = -\cos[2\phi_2]$, $v\Lambda/R = \sin[2\phi_2]$ and,  
\begin{align}
\overline{\Delta}_{\mu} \equiv \Delta_{\mu} \frac{\pi}{4} \frac{(v\Lambda)^{\frac{1}{2}}}{4\pi^2 v^2 b^{\frac{1}{2}}_2}, \ \ 
x \equiv \frac{1}{\cos[\phi_2]}, \ \ y \equiv \cos[\phi_2], \ \ z \equiv \cos[3\phi_2]. 
\end{align}
The set of equations have two non-trivial fixed points and one trivial fixed points;
\begin{align}
& {\rm FP0}: \ \ \Big(z,\frac{m}{R},\overline{\Delta}_0,\overline{\Delta}_{\perp},\overline{\Delta}_3\Big) 
= (1,0,0,0,0.0)  \label{FP0-b} \\ 
&{\rm FP1}: \ \ \Big(z,\frac{m}{R},\overline{\Delta}_0,\overline{\Delta}_{\perp},\overline{\Delta}_3\Big) 
= (1.41,-0.447,0.137,0.0128,0.0440)  \label{FP1-b} \\ 
&{\rm FP2}: \ \ \Big(z,\frac{m}{R},\overline{\Delta}_0,\overline{\Delta}_{\perp},\overline{\Delta}_3\Big) 
= (1.43,-0.753,0.0,0.0,0.376).  \label{FP2-b}
\end{align}  
\end{widetext} 
Here FP0 and FP1 correspond to FP0 and FP1 in the large-$n$ limit, respectively. 
The trivial fixed point is a saddle point of the RG flows, 
around which the linearized equations take forms of 
\begin{eqnarray}
\frac{dm}{dl} = m, \frac{d\overline{\Delta}_{\mu}}{dl} = -\frac{1}{2}\overline{\Delta}_{\mu}, \label{FP0-rg-b}
\end{eqnarray}  
with $\mu=0,\perp,3$. 
The fixed point FP1 is a saddle point, which 
determines the quantum multicriticality among the DM, WSM, and CI phases. 
Namely, linearized RG equations around FP1 have two linear-independent 
directions in a four-dimensional parameter space, 
$m$-$\overline{\Delta}_{0}$-$\overline{\Delta}_{\perp}$-$\overline{\Delta}_{3}$ space, 
along which the RG flow always goes into FP1, and has two other linear-independent directions, along 
which the RG flow always runs away from FP1. Let us call the 
latter two directions as ${\bm e}_1$ and ${\bm e}_2$. These two  
belong to eigenvalues $\lambda_1=+1.14\cdots$ and $\lambda_{2}=+0.287\cdots$, respectively. 
When an initial parameter set of 
$(m,\overline{\Delta}_0,\overline{\Delta}_{\perp},\overline{\Delta}_3)$ slightly 
deviates from FP1 along $\pm {\bm e}_1$, the corresponding 
RG flow goes to the DM and CI phase fixed points, respectively (see Fig.~\ref{sfig:4}(b)). 
When an initial set deviates from 
FP1 along $\pm {\bm e}_2$, the flow goes to 
CI and WSM phase fixed points, respectively (see Fig.~\ref{sfig:4}(b)). 
FP2 is an unstable point of the RG flow; 
linearized RG equations around FP2 have four 
linear-independent directions, along which the RG flow  always runs away from FP2.  
 
\begin{figure}[t]
	\centering
	\includegraphics[width=1\linewidth]{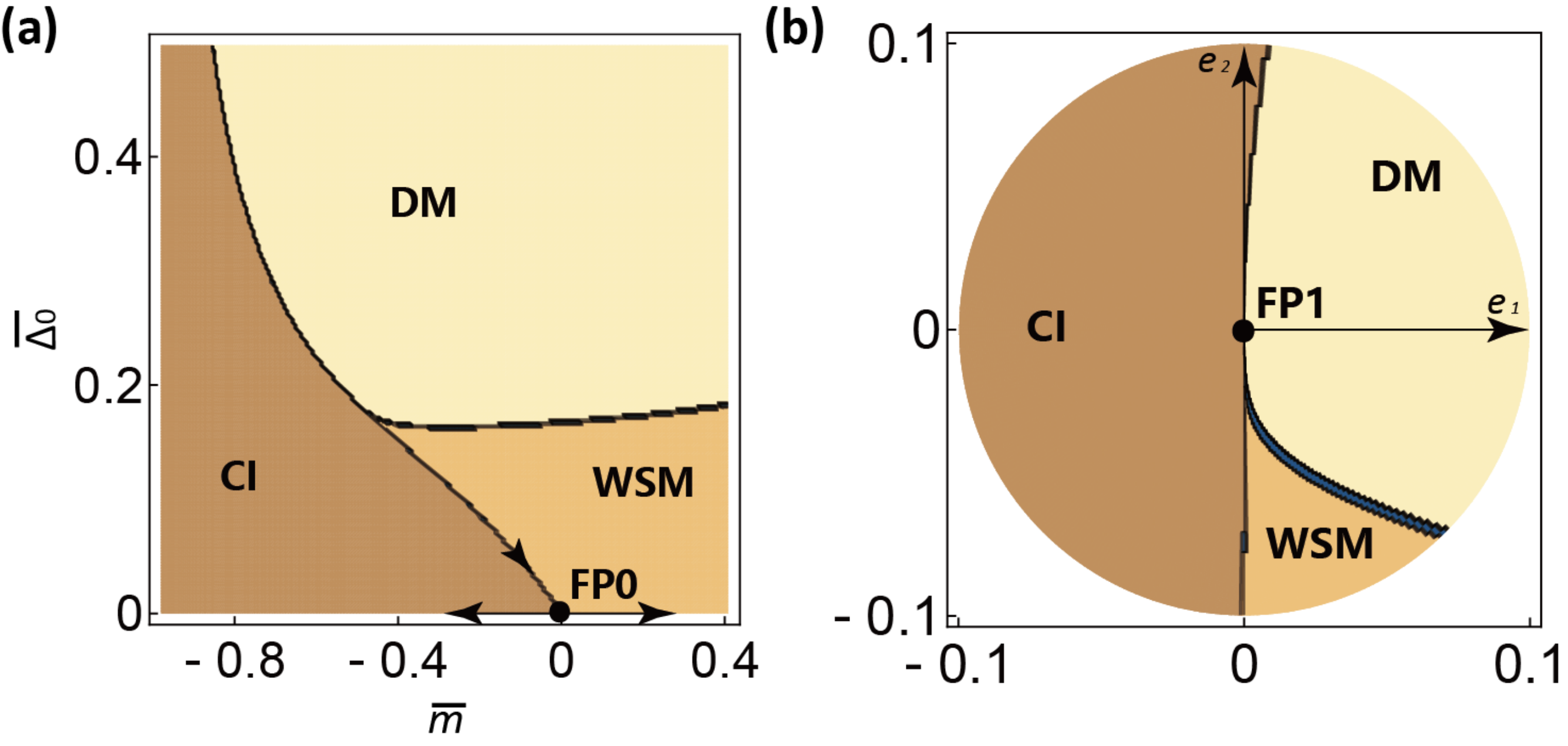}\\
	\caption{(color online) (a) Phase diagram in the $\overline{\Delta}_0$-$\overline{m}$ plane determined by 
the RG equations in the case of $n=2$. (b) Phase diagram around FP1 in a plane subtended by ${\bm e}_1$ and 
${\bm e}_2$.}\label{sfig:4}
\end{figure}

\section{Density of States and Diffusion Constant Scaling in Crossover Regimes}\label{sec:C}
A scaling theory of the density of states (DOS) and diffusion constant 
proposed by Kobayashi et.al.~\cite{15} is  characterized by dynamical exponent $z$, 
critical exponent $\nu$ and spatial dimension $d$. As above, there exist a number of fixed 
points in a phase diagram of disordered Weyl semimetal. Every fixed point has 
different critical and dynamical exponents. To see how these scalings change along 
a quantum critical line which connects two fixed points with 
different exponents, let us argue the DOS and diffusion constant scalings in the  
framework of the renormalization group language.    

\begin{figure}[t]
	\centering
	\includegraphics[width=1\linewidth]{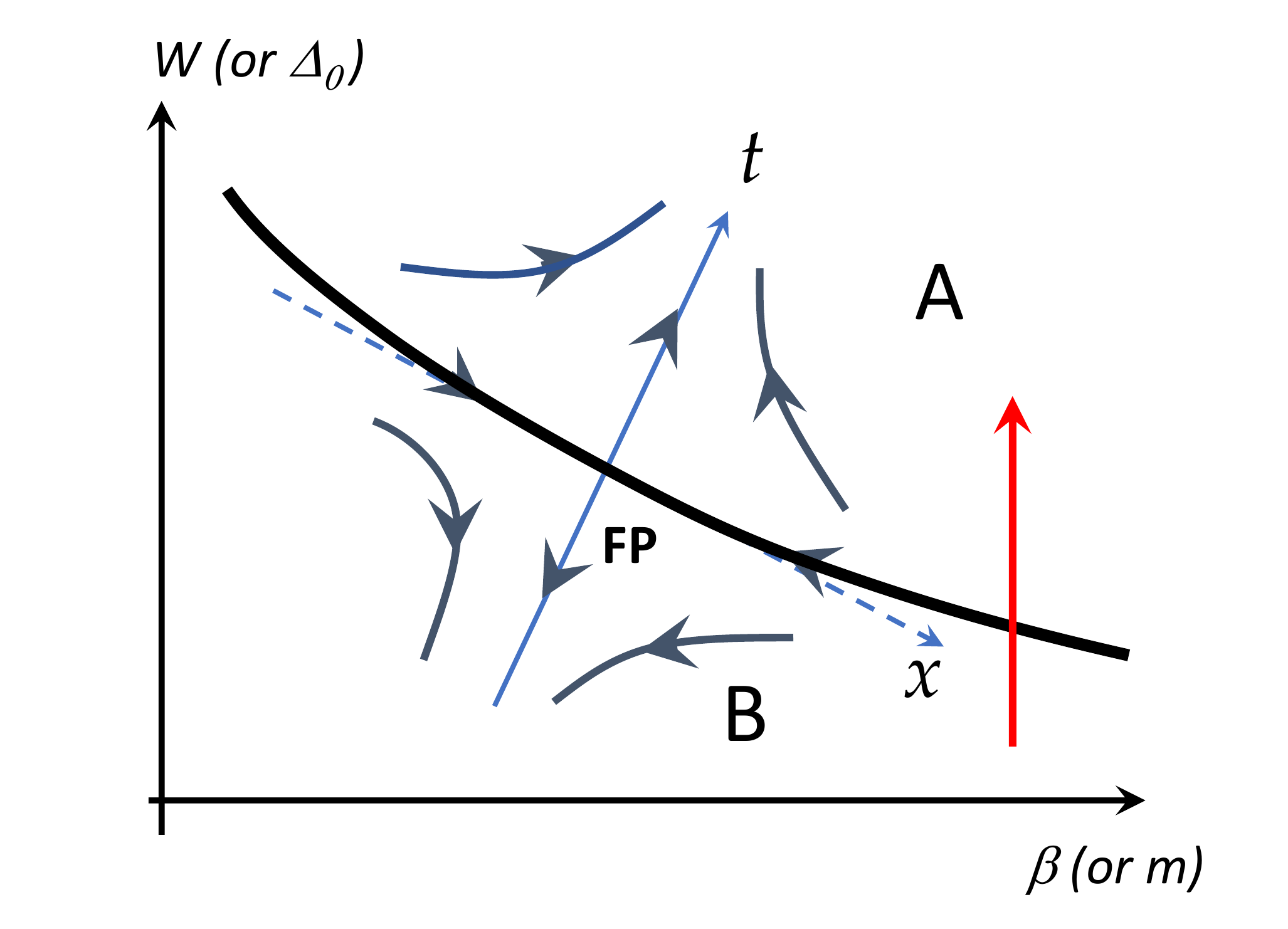}\\
	\caption{(color online) Schematic RG phase diagram around a saddle-point 
fixed point with one relevant scaling variable $t$ and one irrelevant scaling variable $x$ 
in a two-dimensional parameter space subtended by microscopic parameters such as $W$ and $\beta$. 
The bold solid line is a phase boundary between A phase and B phase. The fixed point controls 
critical properties of the continuous phase transition between these two.}\label{sfig:4a}
\end{figure}

\subsection{DOS scaling around Saddle-Point Fixed Point}\label{sec:C-1}
Consider first a quantum phase transition between two phases, where a finite density of states 
at ${\cal E}=0$ in one of the two phases or in both of these two phases become zero 
or infinite at the critical point. Suppose that critical properties of this continuous phase transition is 
controlled by a saddle-point fixed point, which has one relevant scaling variable $t$ and 
one irrelevant scaling variable $x$ (see Fig.~\ref{sfig:4a}). 
Around the fixed point, these two scaling variables, single-particle energy and spatial 
length are scaled as follows,  
\begin{align}
t^{\prime} &= b^{-y_t} t, \label{scale-1} \\
x^{\prime} &= b^{-y_x} x = b^{|y_x|} x, \label{scale-2} \\
\xi^{\prime} &= b \xi, \label{scale-3} \\ 
{\cal E}^{\prime} & = b^{-z} {\cal E}, \label{scale-4} 
\end{align}   
with $y_t>0$ and $y_x<0$. Variables with prime in the  left hand 
side are those after a renormalization, and those 
without prime in the right hand side are variables before the 
renormalization; $b=e^{-dl}<1$. When the scaling 
is spatially isotropic, a volume is scaled by $b^{d}$ ($d$ is a spatial dimension). Thereby, a total number of 
single-particle states below a given energy ${\cal E}$ per unit volume $N({\cal E},t,x)$ is 
scaled by $b^{-d}$ under the renormalization;
\begin{eqnarray}
N^{\prime}({\cal E}^{\prime},t^{\prime},x^{\prime}) = b^{-d} N({\cal E},t,x). \nonumber 
\end{eqnarray} 
Taking a derivative with respect to the energy ${\cal E}$, we obtain a scaling of the density of states under the 
renormalization,
\begin{eqnarray}
b^{d-z} \rho^{\prime}(b^{-z}{\cal E},b^{-y_t}t,b^{-y_x}x) = \rho({\cal E},t,x), \label{important}
\end{eqnarray} 
with $\rho \equiv dN/d{\cal E}$. 

To address ourselves to a scaling property of DOS 
near a critical point, take $t$ to be very small while $x$ is not. 
Suppose that we renormalize many times, such that $b^{-y_t} t=1$ ($b<1$, $y_t>0$). Solving $b$ 
in favor for $t$, we have, 
\begin{eqnarray}
\rho({\cal E},t,x) = t^{-\frac{z-d}{y_t}} \rho^{\prime}(t^{-\frac{z}{y_t}}{\cal E}, 1, t^{\frac{|y_x|}{y_t}} x). \nonumber 
\end{eqnarray}
$t$ is sufficient small, so that the third argument is tiny. Thus, we have, 
\begin{align}
\rho({\cal E},t,x)&\simeq t^{-\frac{z-d}{y_t}} \rho^{\prime}(t^{-\frac{z}{y_t}}{\cal E}, 1, 0)  \nonumber\\
&\equiv   t^{-\frac{z-d}{y_t}} \Psi(t^{-\frac{z}{y_t}}{\cal E}). \label{important1} 
\end{align}
Noting that $\xi = at^{-\frac{1}{y_t}}$ (where $a$ is an atomic length scale), we reach 
the same DOS scaling by Kobayashi et. al.~\cite{15},  
\begin{eqnarray}
\rho({\cal E},t,x) = \xi^{z-d} \Psi(\xi^z {\cal E}). \label{important2}
\end{eqnarray} 
In the presence of finite strength of the irrelevant scaling variable ($x \ne 0$), the above DOS scaling is valid 
in a critical regime specified by  
\begin{eqnarray}
x \ll t^{-\frac{|y_x|}{y_t}}. \label{important3}
\end{eqnarray}    

Our lattice model is characterized by microscopic parameters such as the disorder strength 
$W$ and an interlayer coupling strength $\beta$. In the two-dimensional parameter 
space subtended by $W$ and $\beta$, 
the fixed point has one relevant scaling variable $t$ and one irrelevant scaling variable $x$, 
both of which are functions of $W$ and $\beta$. The phase boundary between the two phases, 
$W_{c}(\beta)$, is determined by 
\begin{eqnarray}
t(W_{c}(\beta),\beta) = 0. \label{important4}
\end{eqnarray}  
For a given $\beta$, we can expand $t(W,\beta)$ in small $W-W_{c}(\beta)$;
\begin{align}
t(W,\beta)&= \Big(\frac{\partial t}{\partial W}\Big)_{|W=W_{c}(\beta)} \cdot (W-W_{c}(\beta)) + \cdots  \nonumber\\
&\equiv a(\beta) \cdot (W-W_{c}(\beta)) + \cdots. \label{important5}
\end{align}
Substituting this expansion into Eq.~(\ref{important1}), we obtain the 
following scaling form for the density of  states as a function of $W$ and ${\cal E}$, 
\begin{align}
\rho({\cal E},W,\beta) &= \big(a(\beta) |W-W_{c}(\beta)|\big)^{-\frac{z-d}{y_t}} \nonumber \\
 & \hspace{0.5cm} \times \Phi 
\Big(\big(a(\beta) |W-W_{c}(\beta)|\big)^{-\frac{z}{y_t}} 
{\cal E}\Big), \label{important6}
\end{align} 
for $W\simeq W_{c}(\beta)$ and arbitrary $\beta$.

\subsection{Diffusion Constant Scaling around the Saddle-Point Fixed Point}\label{sec:C-2}
Consider next a quantum phase transition between two phases, where a finite 
diffusion constant at ${\cal E}=0$ in one of the two phases or both phases  
diverges or vanishes at its transition point. Such a phase transition includes 
not only disorder-driven semimetal-metal quantum phase transition 
in single Weyl node, but also conventional localization-delocalization 
transition. We suppose that critical properties of this continuous phase transition 
is controlled by a saddle-point fixed point which has one relevant scaling variable 
$t$ and one irrelvant scaling variable $x$ (Fig.~\ref{sfig:4b}). 
Around this fixed point, these two variables 
as well as characteristic length scale $\xi$ and the single-particle energy ${\cal E}$ 
are scaled as in Eqs.~(\ref{scale-1}-\ref{scale-4}). To see scaling properties of 
the diffusion constant, we begin with a mean-square 
displacement of single-particle states of energy ${\cal E}$ at time $s$ as a function of 
the two scaling variables;
\begin{eqnarray}
g_{{\bm r}}({\cal E},s,t,x) \equiv \langle {\bm r}^2 ({\cal E},s,t,x) \rangle. \label{diffusion1}
\end{eqnarray}  
Under the spatially isotropic scaling, the mean square displacement is normalized by 
$b^2$ under the renormalization,
\begin{eqnarray}
b^{-2} g^{\prime}_{\bm r}({\cal E}',s',t',x') = g_{\bm r}({\cal E},s,t,x)
\end{eqnarray}
with ${\cal E}'=b^{-z} {\cal E}$, $s'=b^{z}s$, $t' = b^{-y_t} t$, $x'=b^{|y_x|}x$, and $b<1$. 
We consider that the system is close to the critical point, such that $t$ is 
tiny while $x$ is not. Solving $b$ in favor for $t$ with $b^{-y_t} t=1$, we obtain 
a universal scaling form for the mean square distance;
\begin{eqnarray}
g_{\bm r}({\cal E},s,t,x) = t^{-\frac{2}{y_t}} \Psi_{\bm r}(t^{-\frac{z}{y_t}} {\cal E}, t^{\frac{z}{y_t}}s). \nonumber 
\end{eqnarray}
Since $t$ is tiny and $y_t>0$, we Taylor-expand the second argument for arbitrary 
time $s$;
\begin{eqnarray}
g_{\bm r}({\cal E},s,t,x) = t^{\frac{z-2}{y_t}} f(t^{-\frac{z}{y_t}}{\cal E}) s + {\cal O}(s^2). \nonumber 
\end{eqnarray}
This gives us a scaling form of the diffusion constant as 
\begin{eqnarray}
D({\cal E},t,x) = t^{\frac{z-2}{y_t}} f(t^{-\frac{z}{y_t}}{\cal E}). \label{diffusion}
\end{eqnarray}
In the parameter space subtended by microscopic parameters such as $W$ and $\beta$, the scaling 
takes a form of 
\begin{align}
D({\cal E},W,\beta) &= \big(a(\beta) |W - W_{c}(\beta)| \big)^{\frac{z-2}{y_t}} \nonumber \\ 
& \ \times f\Big(\big(a(\beta) |W - W_{c}(\beta)| \big)^{-\frac{z}{y_t}} {\cal E}\Big). \label{diffusion2}
\end{align}

\section{DOS, Diffusion Constant, and Conductivity Scalings around the Quantum Multicritical Point (QMCP)}
In the previous section, we have discussed the DOS and diffusion constant scalings around a saddle-point 
fixed point, which has only one relevant scaling variable and (more than) one irrelvant scaling variables. In this 
section, we generalize the RG scaling argument into an unstable fixed point which has two relevant scaling variables. 
Such a fixed point generally plays a role of a quantum multicritical point (QMCP). In the present study, 
FP1 in Fig.~\ref{sfig:3} corresponds to this QMCP. 

In the next subsection, we first postulate a global RG phase diagram of disordered 
Weyl semimetal. The RG phase diagram includes not only the saddle-point fixed point (FP0) and unstable fixed point 
(FP1), which are derived from the RG analyses of disordered `magnetic dipole' model in the 
previous section, but also all the other saddle-point 
fixed points that surround the unstable fixed point. 
In Appendix.~\ref{sec:D-2}, we derive scaling forms of DOS and diffusion constants 
around and on the QMCP. 
In Appendices.~\ref{sec:D-3}, \ref{sec:D-4}, and \ref{sec:D-5}, 
we derive scaling forms around all the other saddle-point fixed points 
that encompass QMCP.  Combining these knowledge together and 
using the Einstein relation, we complete in Table~\ref{table:scaling} DOS, 
diffusion constant and conductivity scalings around QMCP as well as all the 
quantum critical lines which branch out from QMCP. 

\begin{table*}[ht]
  \centering
    \begin{tabular}{c|c|cc|cc|c}
\hline 
& $\rho(0)$ or $\rho({\cal E})$ & $D_{3}(0)$ or $D_3({\cal E})$ & $D_{\perp}(0)$ or $D_{\perp}({\cal E})$ 
&$v_3(0)$  or $v_3({\cal E})$ & $v_{\perp}(0)$ or $v_{\perp}({\cal E})$ & $\tau(0)$ or $\tau({\cal E})$  \\ \hline \hline 
(i) & $|\delta \overline{\Delta}_0|^{\frac{2d-1-2z}{2y_{\Delta}}}$ 
& $|\delta \overline{\Delta}_0|^{\frac{z-1}{y_{\Delta}}}$ & $|\delta \overline{\Delta}_0|^{\frac{z-2}{y_{\Delta}}}$ 
& --- &  --- & $\rho^{-1}(0)$ \\
(ii) & ${\cal E}^{d-\frac{3}{2}}$ & $\ne 0$ & ${\cal E}^{-1}$ & ${\cal E}^{\frac{1}{2}}$ & $\ne 0$ & ${\cal E}^{-1}$ \\ 
(ii)$^{\prime}$ & $|\delta \overline{\Delta}_0|^{\frac{2d-1}{2} \frac{1-z}{y_{\Delta}}} 
{\cal E}^{d-\frac{3}{2}}$ & $|\delta \overline{\Delta}_0|^{\frac{z-1}{y_{\Delta}}}$ 
& $|\delta \overline{\Delta}_0|^{\frac{2(z-1)}{y_{\Delta}}}  {\cal E}^{-1}$ & 
$|\delta \overline{\Delta}_0|^{\frac{1}{2}\frac{z-1}{y_{\Delta}}} {\cal E}^{\frac{1}{2}}$
&$|\delta \overline{\Delta}_0|^{\frac{z-1}{y_{\Delta}}}$ & ${\cal E}^{-1}$ \\  
(iii) & ${\cal E}^{\frac{d-z'}{z'}}$ & \multicolumn{2}{c|}{${\cal E}^{\frac{z'-2}{z'}}$} 
& \multicolumn{2}{c|}{${\cal E}^{\frac{z'-1}{z'}}$} & ${\cal E}^{-1}$ \\ 
(iii)$^{\prime}$ & $|m|^{\frac{2d(z'-z)-z'}{2z'y_m}} {\cal E}^{\frac{d-z'}{z'}}$ 
& $|m|^{\frac{2z-z'}{z' y_m}} {\cal E}^{\frac{z'-2}{z'}}$ 
& $|m|^{\frac{2(z-z')}{z' y_m}} {\cal E}^{\frac{z'-2}{z'}}$ 
& $|m|^{\frac{1}{2}\frac{2z-z'}{z'y_m}}{\cal E}^{\frac{z'-1}{z'}}$ &
$|m|^{\frac{z-z'}{z'y_m}}{\cal E}^{\frac{z'-1}{z'}}$& ${\cal E}^{-1}$\\ 
(iv) & ${\cal E}^{\frac{2d-1-2z}{2z}}$ & ${\cal E}^{\frac{z-1}{z}}$ & ${\cal E}^{\frac{z-2}{z}}$ 
& ${\cal E}^{\frac{2z-1}{2z}}$ &${\cal E}^{\frac{z-1}{z}}$ &${\cal E}^{-1}$ \\
(v) & ${\cal E}^{\frac{d-z''}{z''}}$ & \multicolumn{2}{c|}{0}  & --- & --- & --- \\
(v)$^{\prime}$ & $|m|^{\frac{2d(z^{\prime\prime}-z)-z^{\prime\prime}}{2z^{\prime\prime}y_m}}{\cal E}^{\frac{d-z''}{z''}}$ 
& \multicolumn{2}{c|}{0}   & --- & --- & --- \\
(vi) & $\ne 0$ & \multicolumn{2}{c|}{${\cal E}^{\frac{z_{\rm 3d,u}-2}{z_{\rm 3d,u}}}$} 
& --- & --- & --- \\
(vii) & $|m|^{-\frac{1}{2}} {\cal E}^{d-1}$ & $|m|^{d-1} {\cal E}^{-(d-1)}$ & $|m|^{d-2} {\cal E}^{-(d-1)}$ 
& $|m|^{\frac{1}{2}}$ & $\ne 0$ &  $m^{d-2} {\cal E}^{-(d-1)}$  \\ 
(viii) & $|\delta \overline{\Delta}_0|^{\frac{d-z'}{y^{\prime}_{\Delta}}}$ 
& \multicolumn{2}{c|}{$|\delta \overline{\Delta}_0|^{\frac{z'-2}{y^{\prime}_{\Delta}}}$} & --- & --- & $\rho^{-1}(0)$ \\ 
(ix) & $|\delta \overline{\Delta}_0|^{-\frac{dz'-d}{y^{\prime}_{\Delta}}} {\cal E}^{d-1}$
& \multicolumn{2}{c|}{ $|\delta \overline{\Delta}_0|^{\frac{dz'-2}{y^{\prime}_{\Delta}}}{\cal E}^{-(d-1)}$} 
&\multicolumn{2}{c|}{$|\delta \overline{\Delta}_0|^{\frac{z'-1}{y^{\prime}_{\Delta}}}$}  
& $|\delta\overline{\Delta}_0|^{\frac{z'(d-2)}{y^{\prime}_{\Delta}}}{\cal  E}^{-(d-1)}$ \\ 
(x) & $|\delta \overline{\Delta}_0|^{\frac{d-z''}{y^{\prime\prime}_{\Delta}}}$ & \multicolumn{2}{c|}{0} & --- & --- & 
--- \\ 
(xi) & $\ne 0$ & \multicolumn{2}{c|}{$|\delta \overline{\Delta}_{0}|^{\frac{z_{\rm 3d,u}-2}{y_{\rm 3d,u}}}$}  
& --- & --- &  --- \\ \hline 
 \end{tabular} 
 \caption{Scalings of DOS, diffusion constants, effective velocities, and life time near 
Weyl nodes. ${\cal E}$ denotes a single-particle energy. $\rho(0)$ 
and $\rho({\cal E})$ are DOS for the zero-energy states and for finite (but small) energy states, 
respectively. $D_{\mu}(0)$ and $D_{\mu}({\cal E})$ are the diffusion 
constants along the $\mu$ direction ($\mu=3, \perp$) for 
the zero-energy and the finite energy states, respectively. $v_{\mu}(0)$ and $v_{\mu}({\cal E})$ are the  effective velocities along the $\mu$-direction at the zero-energy and finite-energy state in the WSM phase 
or in the boundary between the WSM and CI phases. 
$\tau(0)$ and $\tau({\cal E})$ are the life times for the zero-energy and finite-energy 
states. 
The Roman number such as (i), (ii), $\cdots$ in the first column specifies either 
a route along which the system approaches the quantum multicritical point or quantum critical lines. The routes with 
the same Roman number are depicted in Fig.~\ref{xunlong:2}. In (iv), the system is  on the quantum multicritical 
point. In (ii), (iii), (v) and (vi), the system is on the quantum critical line between CI and WSM phases, that between 
WSM and DM phases, that between CI and CI$^{\prime}$ phases, and that between CI$^{\prime}$ and DM phases, respectively. 
In (ii)$^{\prime}$, (iii)$^{\prime}$ and (v)$^{\prime}$, the system approaches QMCP along 
(ii), (iii) and (v), respectively. $\delta\overline{\Delta}_0$ denotes the disorder strength measured 
from a critical disorder strength of respective phase transitions (see also the text). $m$ denotes the mass term. 
$y_{\Delta}$, $y_m$ and $z$ are scaling dimensions of the disorder strength $\delta \overline{\Delta}_0$ and the mass 
term $m$  and the dynamical exponent at the quantum multicritical point (QMCP), respectively. 
$y^{\prime}_{\Delta}$ and $z^{\prime}$ are scaling dimension of the disorder 
strength and dynamical exponent around FP2, respectively. 
$y^{\prime\prime}_{\Delta}$ and $z^{\prime\prime}$ are those for FP3. $y_{\rm 3d,u}$ and $z_{\rm 3d,u}$ are those for 
FP4, which belongs to the conventional 3D unitary class. See also Table~\ref{table:scaling0}.}\label{table:scaling}
\end{table*}

\subsection{postulated Global RG phase diagram}\label{sec:D-1}
The previous RG analyses of the disordered magnetic-dipole model gives the saddle-point fixed point in the clean limit 
(FP0) and unstable fixed point at finite disorder strength (FP1). The critical line connecting these two intervenes the insulator phase 
with finite gap ($m<0$) and Weyl semimetal phase with a pair of Weyl nodes ($m>0$). Near the critical line including 
the two fixed points, the mass term ($m$) is a relevant scaling variable. When the mass term blows up into 
a larger value (either positive or negative), the magnetic dipole model is no longer an appropriate low-energy model. 

\begin{figure}[t] 
	\centering
	\includegraphics[width=0.975\linewidth]{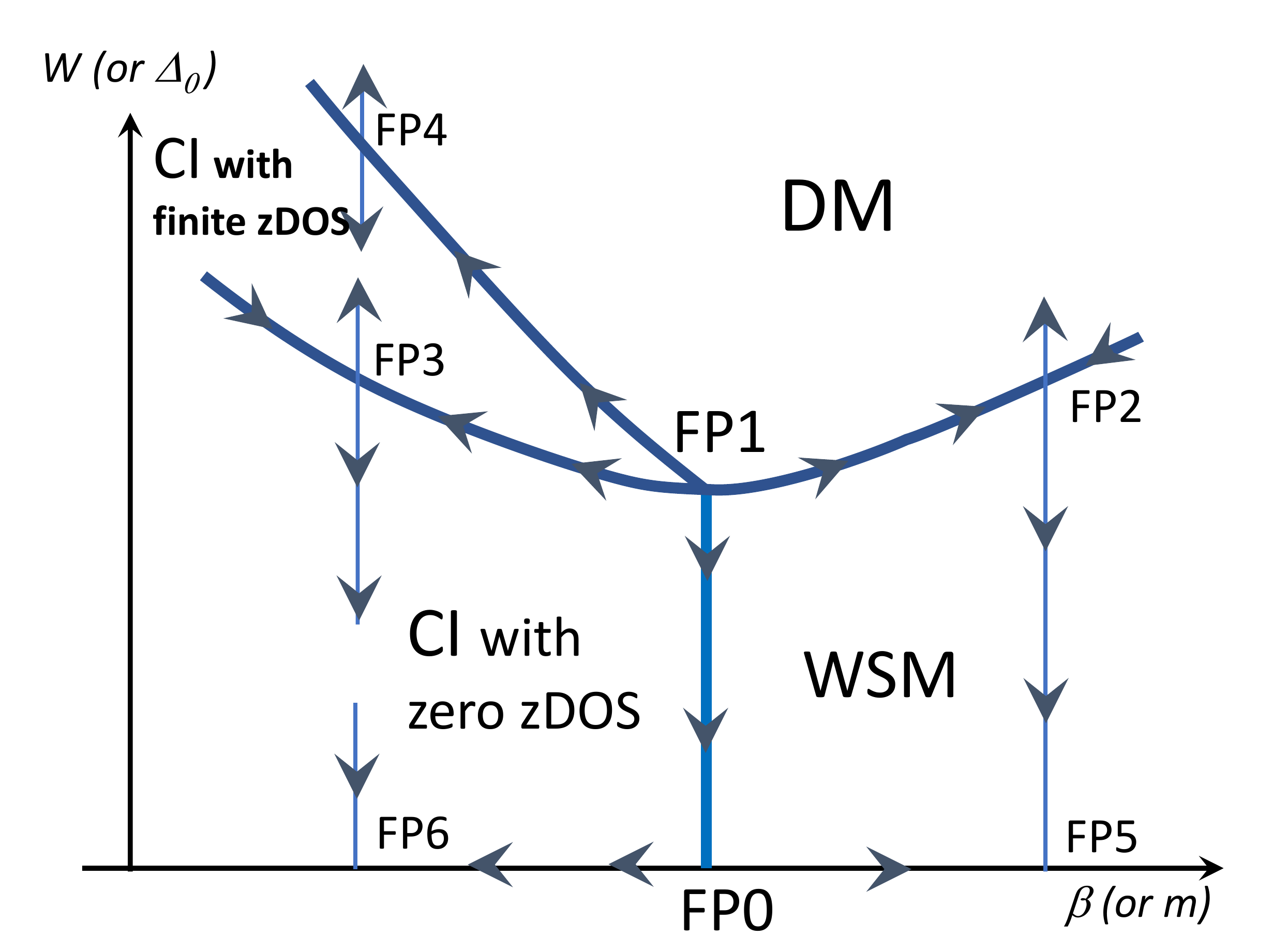} \\
	\caption{(color online) Schematic picture of a global RG phase diagram with fixed points. Blue 
arrow specifies a direction of the RG flow. Bold blue lines are phase boundaries between WSM phase, 
DM phase, CI phase with zero zDOS and CI phase with finite zDOS.}\label{xunlong:1}
\end{figure}

For positively large $m$, low-energy physics is well described by a single Weyl node model. Previous 
RG analyses of the disordered single Weyl node model shows the presence of the stable fixed point in the clean limit 
and a saddle-point fixed point at finite disorder strength. Let us call them FP5 and FP2, respectively. The numerical 
studies of the LCI model indicate that critical properties of the semimetal-metal quantum phase boundary  between 
DM and WSM phases are controlled by the latter saddle-point (FP2). This indicates a structure of the 
global RG phase diagram with positive mass term side as in Fig.~\ref{xunlong:1}. 

For negatively large $m$, the situation is rather more involved. Our scaling analysis of localization length and DOS in 
the main paper suggests that there exist two saddle-point fixed points between QMCP at $m=0$ and the two-dimensional 
limit with $m$ being negatively infinite. One saddle-point fixed point controls DOS scaling around a transition 
between CI phase with zero zDOS and CI phase with finite zDOS. Let us call the CI phase with finite zDOS as 
CI$^{\prime}$ phase and the fixed point as FP3 as in Fig.~\ref{xunlong:1}. 
The other saddle-point fixed point controls critical properties around a localization-delocalization 
transition between CI$^{\prime}$ and DM phases, which belongs to conventional 3D unitary class. 
Let us call this fixed point as FP4 as in Fig.~\ref{xunlong:1}. 
The diffusion constant is always zero in both of the two CI phases as well as the transition point 
between these two. On the one hand, DOS does not show any critical behaviour 
at the transition point between CI$^{\prime}$ and DM phases.

\begin{figure}[t] 
	\centering
	\includegraphics[width=0.975\linewidth]{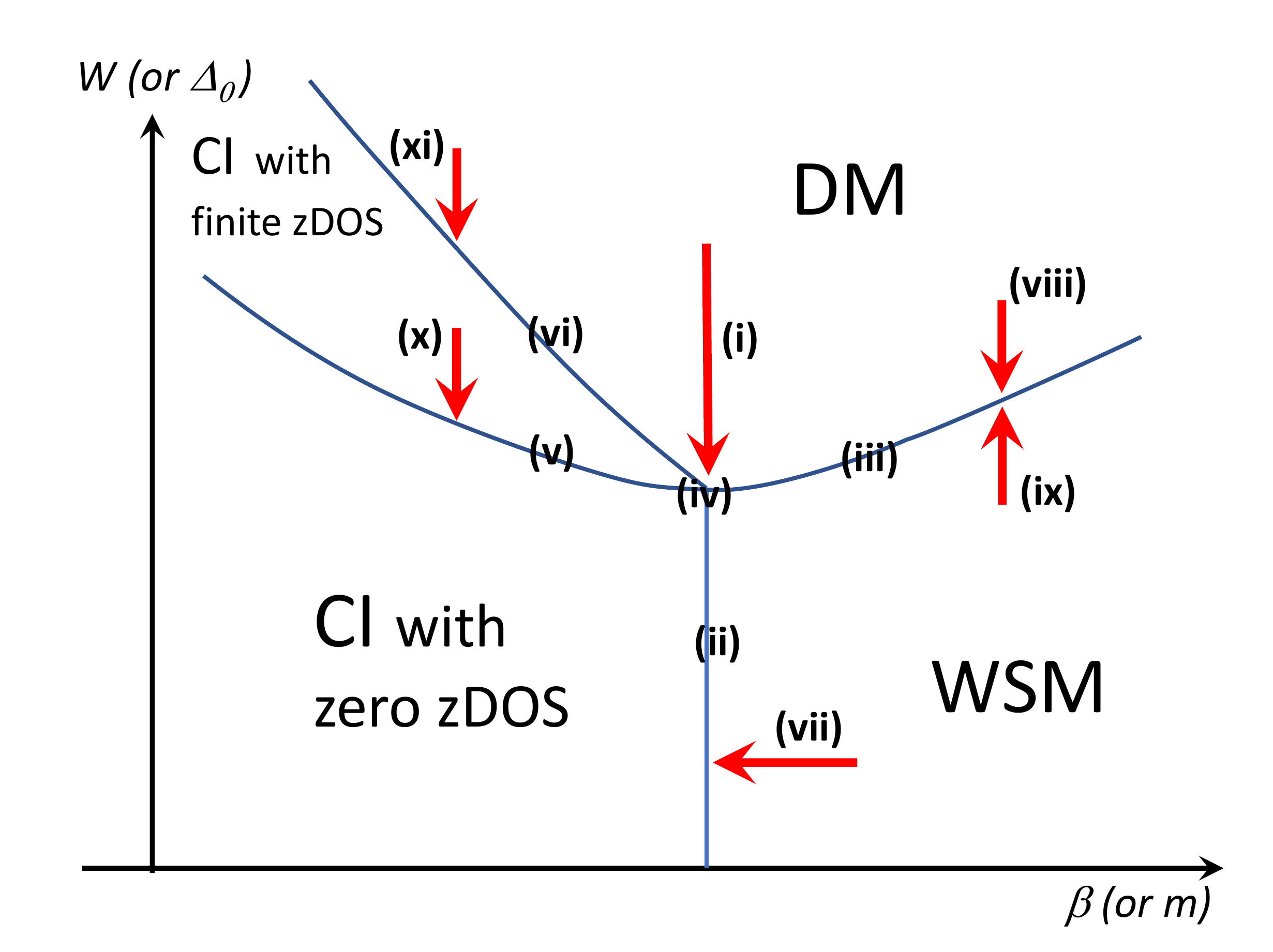} \\
	\caption{(color online) Schematic picture of a phase diagram of disordered Weyl semimetal and 
its critical regions. Roman number (iv) specifies the multicritical point (QMCP). Roman numbers (ii,iii,v,vi) 
specify quantum critical line between CI and WSM phases, that between WSM and DM phases, 
that between CI with finite zDOS and CI with zero zDOS, and that between DM and CI with finite zDOS, respectively.  
Red arrows with roman numbers (i,vii,viii,ix,xi,x) specify a route along which the system approaches quantum 
critical lines or quantum multicritical point. Corresponding scaling forms of DOS, diffusion constants, 
effective velocities and life time are described in Table~\ref{table:scaling}.}\label{xunlong:2}
\end{figure}

\subsection{spatially anisotropic scaling in magnetic dipole model}\label{sec:D-A}
The critical theory described by Eq.~(\ref{eff0}) at $m=0$ has a linear dispersion along the in-plane directions, 
and a quadratic dispersion along the 3-direction. This naturally 
leads to a spatially anisotropic scaling around the quantum critical line between 
CI and WSM phases including its two end points (FP0 and FP1),  
\begin{eqnarray}
\left\{\begin{array}{c}
\xi^{\prime}_{3} = b^{\frac{1}{2}} \xi_3, \\
\xi^{\prime}_{\perp} = b \xi_{\perp}. \\ 
\end{array} \right. \label{anisotropic-scaling} 
\end{eqnarray} 
Here $\xi_{\perp}$ and $\xi_{3}$ denote characteristic length scales within the in-plane directions  
and along the 3-direction, respectively. FP1 (QMCP) has two relevant scaling variables, 
$\delta \overline{\Delta}_0 \equiv \overline{\Delta}_0 - \overline{\Delta}_{0,c}$ and $m$, which 
are scaled by respective scaling dimensions, $y_{\Delta}$ and $y_m$;
\begin{eqnarray}
\left\{\begin{array}{c}
\delta \overline{\Delta}^{\prime}_{0} = b^{-y_{\Delta}} \delta \overline{\Delta}_0, \\ 
m^{\prime} = b^{-y_m} m. \\  
\end{array}\right. \label{scaling-FP1} 
\end{eqnarray} 
with $y_{\Delta}>0$ and $y_m>0$. FP0 has one relevant scaling variable $m$ and an irrelvant 
scaling variable $\overline{\Delta}_0$;  
\begin{eqnarray}
\left\{\begin{array}{c}
\overline{\Delta}^{\prime}_{0} = b^{(d-\frac{5}{2})} \overline{\Delta}_0, \\ 
m^{\prime} = b^{-1} m. \\ 
\end{array}\right. \label{scaling-FP0}
\end{eqnarray} 

\subsection{DOS and Diffusion Constant Scalings around QMCP}\label{sec:D-2}
Consider first the scalings around FP1 (QMCP). Under the anisotropic 
scaling, the volume element is scaled by $b^{d-\frac{1}{2}}$ rather than by $b^d$. 
Thereby, the DOS after and before the renormalization are related by 
\begin{eqnarray}
b^{(d-\frac{1}{2}-z)} \rho^{\prime}({\cal E}',\delta \overline{\Delta}^{\prime}_0,m') 
= \rho({\cal E},\delta \overline{\Delta}_{0},m), \label{D-2-1} 
\end{eqnarray}
where ${\cal E}'=b^{-z}{\cal E}$ and $z$ denotes the dynamical exponent at FP1 (QMCP). 
Besides, the mean square distance along the 3-direction and that within the 1-2 plane 
are scaled differently, 
\begin{eqnarray}
\left\{\begin{array}{c}
b^{-1} g^{\prime}_{3}({\cal E}',s',\delta \overline{\Delta}^{\prime}_{0},m') = g_{3}({\cal E},s,\delta \overline{\Delta}_{0},m), \\
 b^{-2} g^{\prime}_{\perp}({\cal E}',s',\delta \overline{\Delta}^{\prime}_{0},m') = g_{\perp}({\cal E},s,\delta \overline{\Delta}_{0},m), \\
\end{array}\right. \label{D-2-2} 
\end{eqnarray}
with $s'=b^z s$ and 
\begin{eqnarray}
\left\{\begin{array}{c}
g_{3}({\cal E},s,\delta \overline{\Delta}_{0},m) \equiv \langle x^2_3 ({\cal E},s,\delta \overline{\Delta}_{0},m) \rangle, \\
g_{\perp}({\cal E},s,\delta \overline{\Delta}_{0},m) \equiv \langle {\bm x}^2_{\perp} ({\cal E},s,\delta \overline{\Delta}_{0},m) \rangle. \\ 
\end{array}\right. \label{D-2-3} 
\end{eqnarray}

Eqs.~(\ref{D-2-1}), (\ref{D-2-2}), and (\ref{D-2-3}) lead to the following two-parameter scalings of DOS and diffusion 
constants around QMCP. When $\delta \overline{\Delta}_{0}=0$ 
and the system approaches QMCP ($m=0$), the scalings for small $m$ and ${\cal E}$ are given by 
\begin{eqnarray}
\left\{\begin{array}{c}
\rho({\cal E},m) = |m|^{\frac{d-\frac{1}{2}-z}{y_m}} \Psi(|m|^{-\frac{z}{y_m}} {\cal E}), \\
D_{3}({\cal E},m)  = |m|^{\frac{z-1}{y_m}} f_z (|m|^{-\frac{z}{y_m}} {\cal E}), \\
D_{\perp}({\cal E},m)  = |m|^{\frac{z-2}{y_m}} f_{\perp} (|m|^{-\frac{z}{y_m}} {\cal E}), \\ 
\end{array}\right. \label{D-2-4}
\end{eqnarray} 
respectively. When $m=0$ and the system approaches QMCP ($\delta \overline{\Delta}_0=0$), the scalings for small 
$\delta\overline{\Delta}_{0}$ and ${\cal E}$ are given by 
\begin{eqnarray}
\left\{\begin{array}{c}
\rho({\cal E},\delta \overline{\Delta}_0) = |\delta \overline{\Delta}_0|^{\frac{d-\frac{1}{2}-z}{y_\Delta}} 
\Psi(|\delta \overline{\Delta}_0|^{-\frac{z}{y_\Delta}} {\cal E}), \\
D_{3}({\cal E},\delta \overline{\Delta}_0)  = |\delta \overline{\Delta}_0|^{\frac{z-1}{y_\Delta}} 
f_z (|\delta \overline{\Delta}_0|^{-\frac{z}{y_\Delta}} {\cal E}), \\
D_{\perp}({\cal E},\delta\overline{\Delta}_0)  = |\delta \overline{\Delta}_0|^{\frac{z-2}{y_\Delta}} 
f_{\perp} (|\delta \overline{\Delta}_0|^{-\frac{z}{y_\Delta}} {\cal E}). \\
\end{array}\right. \label{D-2-5} 
\end{eqnarray} 
On QMCP where $\delta \overline{\Delta}_0=m=0$, the scalings at small ${\cal E}$ are determined only 
by the dynamical exponent at QMCP;
\begin{eqnarray}
\left\{\begin{array}{c}
\rho({\cal E}) \propto |{\cal E}|^{\frac{d-\frac{1}{2}-z}{z}},  \\
D_{z}({\cal E}) \propto  |{\cal E}|^{\frac{z-1}{z}}, \\
D_{\perp}({\cal E}) \propto  |{\cal E}|^{\frac{z-2}{z}}. \\ 
\end{array}\right.  \label{D-2-6}
\end{eqnarray} 

When $\delta\overline{\Delta}_0$ is negative and the system approaches the quantum critical line between 
CI and WSM phases, the scalings for larger $m$ still respect Eq.~(\ref{D-2-4}). Meanwhile, 
the scalings for smaller $m$ (what we call `critical' regime) are determined by exponents of the 
saddle-point fixed point in the clean limit (FP0 in Fig.~\ref{xunlong:1}: Appendix.~\ref{sec:D-3}). A 
crossover boundary between these two regimes are given by the scaling 
dimensions of the two relevant scaling variables at FP1;
\begin{eqnarray}
\left\{\begin{array}{cc} 
|\delta \overline{\Delta}_{0}| \ll B |m|^{\frac{y_{\Delta}}{y_m}}  & {\rm :controlled} \  {\rm by} \ {\rm FP1} \\ 
|\delta \overline{\Delta}_{0}| \gg B |m|^{\frac{y_{\Delta}}{y_m}}  & {\rm :controlled} \ {\rm by} \  {\rm FP0} \\ 
\end{array}\right. \label{D-2-7}
\end{eqnarray} 

When $m$ is finite, and the system approaches the quantum critical lines specified by 
$\delta \overline{\Delta}_{0}=0$, the scalings for larger $\delta \overline{\Delta}_0$ 
still follow Eq.~(\ref{D-2-5}). Meanwhile, 
the scalings for smaller $\delta \overline{\Delta}_0$ (`critical' regime) are controlled by 
the saddle-point fixed points such as FP2 ($m>0$), or FP3 and FP4 ($m<0$).  Crossover 
boundaries among these regimes are given by, 
\begin{eqnarray}
\left\{\begin{array}{cc} 
|m| \ll A |\delta\overline{\Delta}_0|^{\frac{y_{m}}{y_\Delta}}  & {\rm :controlled}  \ {\rm by} \ {\rm FP1} \\ 
|m| \gg A |\delta\overline{\Delta}_0|^{\frac{y_{m}}{y_\Delta}}  & {\rm :controlled} \ {\rm by} \  {\rm FP2,3,4} \\ 
\end{array}\right. \label{D-2-8}
\end{eqnarray} 

\subsection{DOS and Diffusion Constant Scalings in CI-WSM branch}\label{sec:D-3}
Let us consider the scalings of the quantum phase transition between CI and WSM phases. 
The scalings in the critical regime are controlled by the saddle-point fixed point (FP0) with the 
anistropic scaling. Thus, the scaling argument goes in the same way as in Appendix.~\ref{sec:C}; 
$\overline{\Delta}_0$ corresponds to the irrelevant scaling variable $x$ and $m$ corresponds to 
the relevant scaling variable $t$ in Appendix.~\ref{sec:C}. Due to the anisotropic scaling,  
DOS and mean square distances after a renormalization and those before the renormalization 
are related by 
\begin{eqnarray}
\left\{\begin{array}{c}
b^{(d-\frac{1}{2}-1)} \rho^{\prime}({\cal E}',\overline{\Delta}^{\prime}_0,m') 
= \rho({\cal E},\overline{\Delta}_{0},m), \\
b^{-1} g^{\prime}_{3}({\cal E}',s',\overline{\Delta}^{\prime}_{0},m') = g_{3}({\cal E},s,\overline{\Delta}_{0},m), \\
b^{-2} g^{\prime}_{\perp}({\cal E}',s',\overline{\Delta}^{\prime}_{0},m') = 
g_{\perp}({\cal E},s,\overline{\Delta}_{0},m), \\
\end{array}\right. \label{D-3-1}
\end{eqnarray}
with ${\cal E}^{\prime}=b^{-1}{\cal E}$, $s^{\prime}=bs$, 
$\overline{\Delta}^{\prime}_{0}=b^{d-\frac{5}{2}}\overline{\Delta}_{0}$, and 
$m^{\prime}=b^{-1}m$. The dynamical exponent at FP0 is 1. When $m$ is tiny, 
DOS and diffusion constants follow universal scaling forms,
\begin{eqnarray}
\left\{\begin{array}{c}
\rho({\cal E},\overline{\Delta}_0,m) = m^{d-\frac{3}{2}} \Psi(m^{-1}{\cal E}), \\
D_{3}({\cal E},\overline{\Delta}_0,m) =  f_3(m^{-1}{\cal E}), \\
D_{\perp}({\cal E},\overline{\Delta}_0,m) = m^{-1} f_{\perp}(m^{-1}{\cal E}). \\ 
\end{array}\right. \label{D-3-2}
\end{eqnarray} 

Suppose that the system is in the WSM phase ($m>0$). A low-energy effective electronic band structure 
comprises of a pair of two Weyl nodes with velocities $v_3$ and $v_{\perp}$ 
(see Eq.~(\ref{eff0})). The DOS near the nodes is 
given by a quadratic function of the energy;  
$\rho({\cal E})=v^{-1}_3 v^{-2}_{\perp}|{\cal E}|^{d-1}$. The diffusion 
constants near nodes $D_{\mu}$ ($\mu=3,\perp$) 
are given by respective velocities $v_{\mu}$ and a life 
time around the node $\tau({\cal E})$; 
\begin{eqnarray}
D_{\mu}({\cal E})=v^2_{\mu}\tau({\cal E}), \label{D-3-3}
\end{eqnarray} 
($\mu=3,\perp$). Note that $v_3$ depends on the single-particle energy 
$\cal E$. The self-consistent Born analysis relates the life time 
with an inverse of the DOS~\cite{25}, 
\begin{eqnarray}
\rho({\cal E}) = \frac{2}{\pi \tau({\cal E}) \overline{\Delta}_0}.  \label{D-3-4}
\end{eqnarray} 
Thus, the diffusion constants near the Weyl nodes are inversely proportional to 
a quadratic function of the energy; $D_{\mu}({\cal E})\propto |{\cal E}|^{-(d-1)}$.  
Combining these observations with Eq.~(\ref{D-3-2}), we obtain 
\begin{eqnarray}
\left\{\begin{array}{c}
\rho({\cal E},\overline{\Delta}_0,m) \propto m^{-\frac{1}{2}} |{\cal E}|^{d-1}, \\
D_{3}({\cal E},\overline{\Delta}_0,m) \propto m^{(d-1)}|{\cal E}|^{-(d-1)}, \\
D_{\perp}({\cal E},\overline{\Delta}_0,m) \propto m^{(d-2)} |{\cal E}|^{-(d-1)}, \\ 
\end{array}\right. \label{D-3-5}
\end{eqnarray}
for small $m$ ($>0$) and ${\cal E}$. 
In terms of the Einstein relation, $\sigma_{\mu} = e^2 \rho D_{\mu}$, we 
conclude that the zero-energy conductivities reduce to zero toward the quantum critical line 
between CI and WSM phases with the following exponents;
\begin{eqnarray}
\left\{\begin{array}{c}
\sigma_3 \propto m^{d-\frac{3}{2}}, \\
\sigma_{\perp}  \propto m^{d-\frac{5}{2}}. \\
\end{array}\right. \label{D-3-7} 
\end{eqnarray}  

Both DOS and diffusion constants at finite energy ${\cal E}$ are 
finite even on the critical line ($m=0$). This requires  
the universal scaling functions in Eq.~(\ref{D-3-2}) to have 
asymptotic forms as, 
\begin{eqnarray}
\left\{\begin{array}{c}
\Psi(x) \propto x^{d-\frac{3}{2}}, \\ 
f_3(x) \propto x^0, \\
f_{\perp}(x) \propto x^{-1}, \\  
\end{array}\right. \label{D-3-8}
\end{eqnarray}
for large $x$. In other words, DOS and diffusion constants on the critical line 
have energy dependences as,
\begin{eqnarray}
\left\{\begin{array}{c}
\rho({\cal E}) \propto |{\cal E}|^{d-\frac{3}{2}},  \\
D_3({\cal E}) \propto |{\cal E}|^0, \\
D_{\perp}({\cal E}) \propto |{\cal E}|^{-1}. \\ 
\end{array}\right. \label{D-3-9}
\end{eqnarray} 
The Einstein relation further gives the energy dependences of the conductivities 
around the Weyl node as,
\begin{eqnarray}
\left\{\begin{array}{c}
\sigma_3({\cal E}) \propto |{\cal E}|^{d-\frac{3}{2}}, \\
\sigma_{\perp}({\cal E}) \propto |{\cal E}|^{d-\frac{5}{2}}. \\
\end{array}\right. \label{D-3-10}
\end{eqnarray}

Finally, let us consider that the system approaches the quantum multicritical point (QMCP) 
along the critical line between CI and WSM phases ($m=0$). Near QMCP, DOS and  
diffusion constants have the universal scaling forms given by Eq.~(\ref{D-2-5}). Combining 
Eqs.~(\ref{D-3-9}) and (\ref{D-3-10}) with these scaling forms, we obtain;
\begin{eqnarray}
\left\{\begin{array}{c}
\rho({\cal E})  \propto |\delta \overline{\Delta}_0|^{\frac{2d-1}{2}\frac{1-z}{y_{\Delta}}} |{\cal E}|^{d-\frac{3}{2}},  \\
D_3({\cal E}) \propto |\delta \overline{\Delta}_0|^{\frac{z-1}{y_{\Delta}}} |{\cal E}|^0,  \\
D_{\perp}({\cal E}) \propto |\delta \overline{\Delta}_0|^{\frac{2(z-1)}{y_{\Delta}}} |{\cal E}|^{-1}, \\
\sigma_3({\cal E}) \propto |\delta \overline{\Delta}_0|^{\frac{2d-3}{2}\frac{1-z}{y_{\Delta}}} |{\cal E}|^{d-\frac{3}{2}}, \\
\sigma_{\perp}({\cal E}) \propto |\delta \overline{\Delta}_0|^{\frac{2d-5}{2}\frac{1-z}{y_{\Delta}}} |{\cal E}|^{d-\frac{5}{2}}. \\  
\end{array}\right. \label{D-3-11}
\end{eqnarray} 
Here $y_{\Delta}$ and $z$ are exponents at QMCP.  

On closing this subsection, it is noteworthy to comment that, near the quantum phase transition line 
between CI and WSM phases, the life time near Weyl nodes respect the following universal scaling 
form,
\begin{align}
{\tau}({\cal E},\overline{\Delta}_0,m) = m^{-1} \Phi(m^{-1}{\cal E}). \label{D-3-12}
\end{align}
To derive this, we use the RG argument in Eq.~(\ref{D-3-4}). The form has a different 
exponent with respect to small $m$, compared to $\rho^{-1}({\cal E},\overline{\Delta}_0,m)$ 
in Eq.~(\ref{D-3-2}). The difference 
is because, in the WSM phase as well as on the phase boundary between CI and WSM phases, 
$\overline{\Delta}_0$ in the right hand side of Eq.~(\ref{D-3-4}) goes to 
zero on the renormalization; $\overline{\Delta}^{\prime}_0=b^{d-\frac{5}{2}}\overline{\Delta}_0$. 
In the WSM phase ($m>0$), the life time is inversely proportional to ${\cal E}^{d-1}$, so that 
it behaves as 
\begin{eqnarray}
\tau({\cal E},\delta\overline{\Delta}_0,m) \propto m^{d-2} |{\cal E}|^{-(d-1)}, \label{D-3-13}
\end{eqnarray}
for small positive $m$. At the critical point ($m=0$), the life time is scaled by the finite 
energy ${\cal E}$ as 
\begin{eqnarray}
\tau({\cal E},\delta\overline{\Delta}_0,m=0) \propto |{\cal E}|^{-1}.  \label{D-3-14}
\end{eqnarray}

\subsection{DOS and Diffusion Constant Scalings in WSM-DM branch}\label{sec:D-4}
Consider the scalings in the quantum phase transition between WSM and DM phases. 
The scaling properties in the critical regime are determined by the saddle-point fixed point 
with the spatially isotropic scaling (FP2 in Fig.~\ref{xunlong:1}). 
Thus, the argument goes exactly in the same way as in 
Appendix.~\ref{sec:C}, where $\delta \overline{\Delta}_0 \equiv \overline{\Delta}_0-\overline{\Delta}_{0,c}$ and 
$m$ correspond to relevant scaling variable $t$ and irrelevant scaling variable $x$ in Appendix.~\ref{sec:C},  respectively 
and $\overline{\Delta}_{0,c}$ generally depends on $m$. For small $\delta \overline{\Delta}_{0}$ and positive 
$m$, DOS and diffusion constant respect the following scaling forms;
\begin{eqnarray}
\left\{\begin{array}{c}
\rho({\cal E},\delta\overline{\Delta}_0,m) = |\delta \overline{\Delta}_{0}|^{-\frac{z'-d}{y^{\prime}_{\Delta}}} 
\Psi(|\delta \overline{\Delta}_{0}|^{-\frac{z'}{y^{\prime}_{\Delta}}}{\cal E}), \\
D({\cal E},\delta\overline{\Delta}_0,m) = |\delta \overline{\Delta}_{0}|^{-\frac{2-z'}{y^{\prime}_{\Delta}}} 
f(|\delta \overline{\Delta}_{0}|^{-\frac{z'}{y^{\prime}_{\Delta}}}{\cal E}). \\
\end{array}\right. \label{D-4-1}
\end{eqnarray}  
$z'$ and $y^{\prime}_{\Delta}$ are dynamical exponent and scaling exponent of the 
relevant scaling variable $\delta \overline{\Delta}_{0}$ at the saddle-point fixed point. These exponents 
have been evaluated both by the renormalization group studies and by 
simulational studies~\cite{10,12,15,19,20,21,22,25,27,29,31}. In DM phase 
($\delta \overline{\Delta}_0>0$), DOS and diffusion constant are finite at the zero energy; 
\begin{eqnarray}
\left\{\begin{array}{c}
\rho({\cal E}=0,\delta\overline{\Delta}_0,m) \propto \delta \overline{\Delta}_{0}^{-\frac{z'-d}{y^{\prime}_{\Delta}}}, \\
D({\cal E}=0,\delta\overline{\Delta}_0,m) \propto \delta \overline{\Delta}_{0}^{-\frac{2-z'}{y^{\prime}_{\Delta}}},\\
\sigma({\cal E}=0,\delta\overline{\Delta}_0,m) \propto \delta \overline{\Delta}_{0}^{-\frac{2-d}{y^{\prime}_{\Delta}}}.\\
\end{array}\right. \label{D-4-2}
\end{eqnarray}
 
In the WSM phase ($\delta \overline{\Delta}_0<0$), DOS and diffusion constant near Weyl nodes behave 
as $|{\cal E}|^{d-1}$ and $|{\cal E}|^{-(d-1)}$, respectively; these scaling forms thus reduce to
\begin{eqnarray}
\left\{\begin{array}{c}
\rho({\cal E},\delta\overline{\Delta}_0,m) \propto |\delta \overline{\Delta}_{0}|^{-\frac{dz'-d}{y^{\prime}_{\Delta}}} 
|{\cal E}|^{d-1}, \\
D({\cal E},\delta\overline{\Delta}_0,m) \propto |\delta \overline{\Delta}_{0}|^{-\frac{2-dz'}{y^{\prime}_{\Delta}}}
|{\cal E}|^{-(d-1)}, \\
\sigma({\cal E},\delta\overline{\Delta}_0,m) \propto |\delta \overline{\Delta}_{0}|^{-\frac{2-d}{y^{\prime}_{\Delta}}}, \\ 
\end{array}\right. \label{D-4-3}
\end{eqnarray} 
for small ${\cal E}$ and $|\delta\overline{\Delta}_0|$. Note that, contrary to the previous study~\cite{25}, 
the zero-energy conductivity follows the Wegner's relation without any assumption of the dynamical 
exponent. 

On the quantum critical line ($\delta \overline{\Delta}_0=0$), both DOS and diffusion constant 
take finite values at a finite energy ${\cal E}$. This requirement in combination with the universal scaling forms 
Eq.~(\ref{D-4-1}) tells us how the DOS, diffusion constant and conductivity on the 
critical line are scaled with respect to small ${\cal E}$;
\begin{eqnarray}
\left\{\begin{array}{c}
\rho({\cal E},\delta\overline{\Delta}_0=0,m) \propto |{\cal E}|^{\frac{d-z'}{z'}}, \\
D({\cal E},\delta\overline{\Delta}_0=0,m) \propto |{\cal E}|^{\frac{z'-2}{z'}},\\
\sigma({\cal E},\delta\overline{\Delta}_0=0,m) \propto |{\cal E}|^{\frac{d-2}{z'}}. \\
\end{array}\right. \label{D-4-4}
\end{eqnarray} 
Combining Eq.~(\ref{D-4-4}) with the other universal scaling forms around QMCP (Eq.~(\ref{D-2-4})), we 
obtain;
\begin{eqnarray}
\left\{\begin{array}{c} 
 \rho({\cal E},\delta\overline{\Delta}_0=0,m) 
\propto m^{\frac{1}{2y_m}\big(2d\frac{z'-z}{z'}-1\big)} |{\cal E}|^{\frac{d-z'}{z'}}, \\
D_3({\cal E},\delta\overline{\Delta}_0=0,m) 
\propto m^{\frac{1}{y_m}\big(\frac{2z}{z'}-1\big)} |{\cal E}|^{\frac{z'-2}{z'}}, \\
D_{\perp}({\cal E},\delta\overline{\Delta}_0=0,m) 
\propto m^{\frac{2}{y_m}\big(\frac{z}{z'}-1\big)} |{\cal E}|^{\frac{z'-2}{z'}}, \\
\sigma_3({\cal E},\delta\overline{\Delta}_0=0,m) 
\propto m^{\frac{1}{y_m}\big(d-\frac{z}{z'}(d-2)-\frac{3}{2}\big)} |{\cal E}|^{\frac{d-2}{z'}}, \\
\sigma_{\perp}({\cal E},\delta\overline{\Delta}_0=0,m) 
\propto  m^{\frac{1}{y_m}\big(d-\frac{z}{z'}(d-2)-\frac{5}{2}\big)}  |{\cal E}|^{\frac{d-2}{z'}}. \\ 
\end{array}\right. \label{D-4-5}
\end{eqnarray}
for small positive $m$. 
  
\begin{figure*}[t] 
	\centering
	\includegraphics[width=0.9\linewidth]{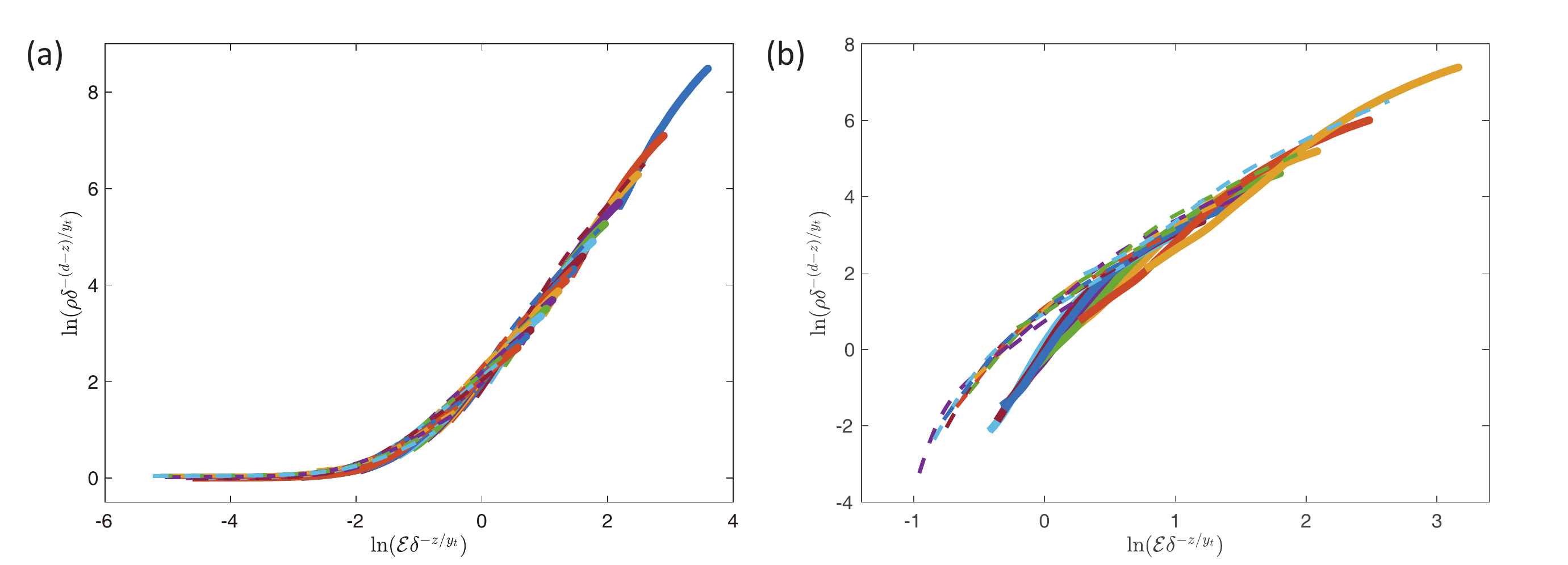} \\
	\caption{(color online) Single-parameter scaling of the density of 
states for $\beta=0.2$ (solid lines) and $\beta=0.3$ (dotted lines) 
for different disorder strength $W$ near the transition point between 
CI phase with zero zDOS and CI phase with finite zDOS; 
(a) $W>W_{c,1}$ $(\delta\overline{\Delta}_{0}>0)$ (b) $W<W_{c,1}$ 
$(\delta\overline{\Delta}_{0}<0)$. Only those data points with $\rho({\cal E})>0.008$$(0.004)$ 
are used for the plot for $\beta=0.2$$(0.3)$ in (a) and $\rho({\cal E})>0.001$ for (b). 
We use Eq.~(\ref{important6}) for the scaling fit with $(d-z)/y_t=2$, $z/y_t=1$ and $\delta\equiv|W-W_{c,1}|/W_{c,1}$. 
In the fitting for $\beta=0.2$ and $\beta=0.3$, we use the same 
$a(\beta)$ in Eq.~(\ref{important6}). For the CI with finite energy gap, Fig.\ref{sfig:4b}~(b), 
the data streams for $\beta=0.3$ and those for $\beta=0.2$ deviate 
from each other especially when ${\cal E}\delta^{-z/y_t}$ 
becomes smaller. These data points are from those single-particle states which are 
proximate (and greater than) the energy gap $\Delta$. We speculate that the deviation 
results from a non-analytic feature of $\rho({\cal E})$ at ${\cal E}=\Delta$,  
and is rather generic in the side of the CI with finite energy gap.}\label{sfig:4b}
\end{figure*}

\subsection{DOS Scaling in CI-CI$^{\prime}$ branch and diffusion constant scaling 
in CI$^{\prime}$-DM branch}\label{sec:D-5}
DOS around ${\cal E}=0$ shows a universal scaling feature at the transition between CI phase 
with zero zDOS and CI phase with finite zDOS,
\begin{align}
\rho({\cal E},\delta \overline{\Delta}_0,m) = |\delta \overline{\Delta}_0|^{\frac{d-z''}{y^{\prime\prime}_\Delta}} 
\Psi(|\delta \overline{\Delta}_0|^{-\frac{z''}{y^{\prime\prime}_\Delta}} {\cal E}), \label{D-5-1} 
\end{align} 
with $\delta \overline{\Delta}_{0} \equiv \overline{\Delta}_0-\overline{\Delta}_{0,c,1}$ (see  Eq.~(\ref{important6}), $y_\Delta''=y_t$ and $z''=z$ ). 
$\overline{\Delta}_{0,c,1}$ is a critical disorder for a transition between 
CI with zero zDOS and CI with finite zDOS, and it generally depends on the mass term 
$m$. $z^{\prime\prime}$ and $y^{\prime\prime}_{\Delta}$ are dynamical exponent and scaling 
dimension of the relevant scaling variable $\delta\overline{\Delta}_{0}$ at the saddle-point 
fixed point (FP3 in Fig.~\ref{xunlong:1}), respectively. From a single-parameter scaling 
fit given in Fig.~\ref{sfig:4b}, we numerically determine these two exponents as 
$z^{\prime\prime} \simeq 1$, and $y^{\prime\prime}_{\Delta} \simeq 1$. The zero-energy 
DOS (zDOS) plays role of an order parameter of the transition; the zDOS behaves as 
\begin{align}
\rho({\cal E}=0,\delta \overline{\Delta}_0,m) \propto 
\delta \overline{\Delta}_0^{\frac{d-z^{\prime\prime}}{y^{\prime\prime}_{\Delta}}},  
\label{D-5-2}
\end{align}
for $\delta \overline{\Delta}_{0}>0$. At the transition point, the zDOS vanishes and DOS has a scaling form 
as a function of small energy ${\cal E}$;
\begin{align}
\rho({\cal E},\delta \overline{\Delta}_0=0,m) \propto |{\cal E}|^{\frac{d-z^{\prime\prime}}{z^{\prime\prime}}}. \label{D-5-3} 
\end{align}
Near QMCP, the other universal scaling forms (Eq.~(\ref{D-2-4})) endows Eq.~(\ref{D-5-3}) 
with an additional exponent with respect 
to small mass term $m$;
\begin{align}
\rho({\cal E},\delta \overline{\Delta}_0=0,m) \propto |m|^{\frac{2d(z^{\prime\prime}-z)-z^{\prime\prime}}{2z^{\prime\prime}y_m}} 
|{\cal E}|^{\frac{d-z^{\prime\prime}}{z^{\prime\prime}}}. \label{D-5-4} 
\end{align}
Here $z$ and $y_{m}$ are dynamical exponent and scaling dimension of the mass term at QMCP,  respectively. 

The diffusion constant near ${\cal E}=0$ also shows a universal scaling at the transition between 
CI with finite zDOS and DM phases. The scaling form is determined by exponents of the conventional 
3D unitary fixed point (FP4 in Fig.~\ref{xunlong:1}). In terms of dynamical exponent 
$z_{\rm 3d,u}$ and critical exponent $\nu_{\rm 3d,u}\equiv 1/y_{\rm 3d,u}$ associated with the fixed point, 
the diffusion constant around ${\cal E}=0$ is given by a scaling function,
\begin{align}
D({\cal E},\delta\overline{\Delta}_0,m) = |\delta \overline{\Delta}_0|^{\frac{z_{\rm 3d,u}-2}{y_{\rm 3d,u}}} 
f(|\delta \overline{\Delta}_{0}|^{-\frac{z_{\rm 3d,u}}{y_{\rm 3d,u}}}{\cal E}), \label{D-5-5} 
\end{align} 
with $\delta \overline{\Delta}_{0}\equiv \overline{\Delta}_0-\overline{\Delta}_{0,c,2}$. $\overline{\Delta}_{0,c,2}$ 
denotes a critical disorder strength for the localization-delocalization transition between CI with finite zDOS and 
DM phases. The zero-energy DOS takes a finite value at the transition point, 
$\overline{\Delta}_0=\overline{\Delta}_{0,c,2}>\overline{\Delta}_{0,c,1}$, so that the DOS scaling (Eq.~(\ref{important1})) 
generally leads to $z_{\rm 3d,u}=d=3$. 

For DM phase side ($\delta \overline{\Delta}_{0}>0$), the diffusion constant 
at the zero energy evolves continuously from the transition point;
\begin{align}
D({\cal E}=0,\delta \overline{\Delta}_{0},m) \propto \delta \overline{\Delta}_{0}^{\frac{z_{\rm 3d,u}-2}{y_{\rm 3d,u}}}. \label{D-5-6} 
\end{align} 
At the critical point ($\delta \overline{\Delta}_0=0$), the diffusion constant evolves continuously from the zero energy;
\begin{align}
D({\cal E},\delta \overline{\Delta}_{0}=0,m) \propto |{\cal E}|^{\frac{z_{\rm 3d,u}-2}{z_{\rm 3d,u}}}. \label{D-5-7} 
\end{align} 
Combining Eq.~(\ref{D-5-7}) with the other universal scalings around QMCP (Eq.~(\ref{D-2-4})), we obtain 
\begin{eqnarray}
\left\{\begin{array}{c}
D_3({\cal E},\delta \overline{\Delta}_{0}=0,m) \propto 
|m|^{\frac{1}{y_m}\big(\frac{2z}{z_{\rm 3d,u}}-1\big)}|{\cal E}|^{\frac{z_{\rm 3d,u}-2}{z_{\rm 3d,u}}}, \\
D_{\perp}({\cal E},\delta \overline{\Delta}_{0}=0,m) 
\propto |m|^{\frac{2}{y_m}\big(\frac{z}{z_{\rm 3d,u}}-1\big)} 
|{\cal E}|^{\frac{z_{\rm 3d,u}-2}{z_{\rm 3d,u}}}, \\
\end{array}\right. \label{D-5-8}
\end{eqnarray}
for small mass term ($m$). The Einstein relation gives the same scalings for the 
conductivities as Eq.~(\ref{D-5-8}).

\section{polynomial fitting of CI$^{\prime}$-DM branch} 
\begin{figure}[t]
	\centering
	\includegraphics[width=1\linewidth]{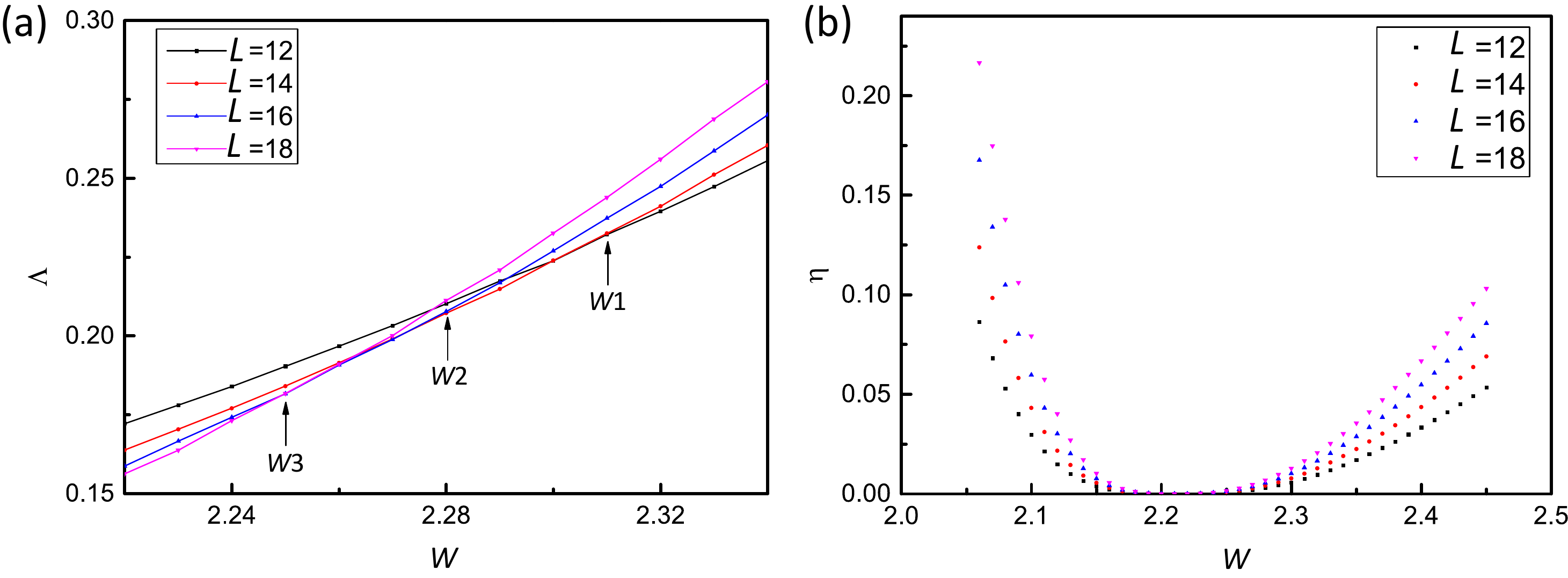}\\
	\caption{(color online) (a) Enlarged figure of Fig.~\ref{fig:2} around $W_c$. $W1$, $W2$, $W3$ denote an intersection point  
between $L$=12, and 14, that between $L$ = 14 and 16, that between $L$ = 16 and 18, respectively. (b) Ratio between 
the 3rd order term and a sum of the lower order terms in the polynomial fitting with $(m_1,m_2,n_1,n_2)=(2,0,3,1)$, 
$\eta$ given in Eq.~(\ref{eta}), as a function of the disorder strength $W$ for different system size $L$. }\label{sfig:56}
\end{figure}

{\label{sec: E}}
For those data points in Fig.~\ref{fig:2}, the polynomial fitting results give $W_c=2.21$, while we could see 
by eye that the intersection of curves with different $L$ occur at $W_c=2.27$ in Fig.~\ref{fig:2}. Fig.~\ref{sfig:56}(a) 
is an enlarged figure of Fig.~\ref{fig:2}. From this, one can see that 
the intersection of curves with larger size moves in the direction of smaller $W$, being 
consistent with the polynomial fitting result ($W_c=2.21$).   

A convergence of the polynomial fitting results with $(m_1,m_2,n_1,n_2)=(2,0,3,1)$ for the data points 
in Fig.~\ref{fig:2} is demonstrated in Fig.~\ref{sfig:56}(b). The figure shows a ratio between the 3rd order 
term and a sum of the lower order terms:
\begin{eqnarray}
\eta \equiv \frac{a_{3,0}\phi_{1}^{3}}{a_{0,0}+a_{1,0}\phi_{1}+a_{2,0}\phi_{1}^{2}}. \label{eta}
\end{eqnarray}
One can see that the ratio is already tiny (below 8\%) near the critical disorder $W_c$ 
($2.1<W<2.4$). This verifies the polynomial fitting with $(m_1,m_2,n_1,n_2)=(2,0,3,1)$.

\begin{acknowledgements}
This work (XL, BX, RS) was supported by NBRP of China Grants No.~2014CB920901, 
No.~2015CB921104, and No.~2017A040215. TO was supported by  
JSPS KAKENHI Grant No. JP15H03700 and JP17K18763. 
\end{acknowledgements}

\end{document}